\documentclass[12pt]{article}
    \usepackage{fullpage,epsfig,graphics,amsbsy,amssymb,cancel,slashed,mathrsfs,booktabs}
    \usepackage{psfrag,hyperref}
    \usepackage{graphicx,color}
    \usepackage{wrapfig}
    \usepackage{subfigure}
    \usepackage{booktabs}

    \newcommand{\be}{\begin{equation}}
    \newcommand{\ee}{\end{equation}}
    \newcommand{\bes}{\begin{split}}
    \newcommand{\ees}{\end{split}}
    \usepackage{amsmath}
    \numberwithin{equation}{section}
    \usepackage{caption}
    \usepackage[nosort]{cite}

    \def\L{{\cal L}}

    \def\al{\alpha}
    \def\eps{\epsilon}
    
    \newcommand{\bea}{\begin{eqnarray}}
    \newcommand{\eea}{\end{eqnarray}}  
    
    \newcommand{\Tr}{\textrm{Tr}}

    \newcommand{\NN}{\mathcal{N}}
    
    \newcommand{\sfrac}[2]{\mbox{$\frac{#1}{#2}$}}
    \newcommand{\ms}{\!-\!}
    \newcommand{\ps}{\!+\!}
    
    \newcommand{\pint}{\int\!\!\!\!\!\!-}

    \newcommand{\LL}{{\mathcal L}}
    \newcommand{\DD}{{\mathcal D}}
    
    \newcommand{\ZZ}{{\mathcal Z}}
    \newcommand{\OO}{\mathcal{O}}
   \newcommand{\QQ}{\mathcal{Q}}

    \newcommand{\Zm}{{Z_{\rm mass}}}
    
    \newcommand{\vareps}{\varepsilon}
    
    \def\s{\sigma}

    \newcommand{\sbkt}[1]{\left[#1\right]}
    \newcommand{\bkt}[1]{\left(#1\right)}
    
    \newcommand{\Zv}{Z_{{1-\text{loop}}}^{\text{vec}}}
    \newcommand{\Zh}{Z_{{1-\text{loop}}}^{\text{hyp}}}
     \newcommand{\Zc}{Z_{{1-\text{loop}}}^{\text{chi}}}
    \newcommand{\tv}{\langle \alpha,\sigma\rangle^2}
    
    \newcommand{\alphi}{\langle\alpha,\sigma\rangle}
    \usepackage{amsmath}

    \usepackage[nameinlink]{cleveref}
\begin{document}

\thispagestyle{empty}
\begin{flushright} \small
	UUITP-41/17\\
	MIT-CTP/4962
\end{flushright}
\smallskip
\begin{center} \LARGE
	{\bf Analytic continuation of dimensions in supersymmetric localization}
	\\[12mm] \normalsize
	{\bf  Anastasios Gorantis\,${}^a$, Joseph A. Minahan\,${}^a$ and Usman Naseer\,${}^b$} \\[8mm]
	{\small\it
	${}^a$\  Department of Physics and Astronomy,
	Uppsala University,\\
	Box 516,
	SE-751 20 Uppsala,
	Sweden
	\\

	${}^b$\ Center for Theoretical Physics,\\
	Massachusetts Institute of Technology,\\
	Cambridge, MA 02139, USA.
	}

	\medskip
	\texttt{anastasios.gorantis, joseph.minahan@physics.uu.se}\\
	\texttt{unaseer@mit.edu}
\end{center}
\vspace{7mm}
\begin{abstract}
	\noindent

\end{abstract}
We compute the perturbative partition functions for gauge theories with eight supersymmetries on spheres of dimension $d\le5$, proving a conjecture by 
the second author.  
  We apply similar methods to gauge theories with four supersymmetries on spheres with $d\le3$.
   The results are valid for non-integer $d$ as well.
    We further propose an analytic continuation from $d=3$ to $d=4$ that gives the perturbative partition function for an $\NN=1$ gauge theory.  The results are consistent with the free multiplets and the one-loop $\beta$-functions for general $\NN=1$ gauge theories.  We also consider the analytic continuation of an $\NN=1$ preserving mass deformation of the maximally supersymmetric gauge theory and compare to recent holographic results 
    for $\NN=1^*$ super Yang-Mills.  We find that the general structure 
for the real part of the free energy coming from the analytic continuation is consistent with the holographic results.
\eject
\normalsize
\tableofcontents

\section{Introduction}
\label{intro}

Localization has proven to be a powerful tool for investigating supersymmetric gauge theories on compact spaces with isometries (for a recent review  see \cite{Pestun:2016zxk}).  Localizing a gauge  theory reduces its partition function to a  sum\footnote{The sum may include integrals over continuous parameters which parametrize the localization loci} over the various localization loci, with a structure of the form
\be
\ZZ=\sum_{k\in \rm{loci}}e^{-S_k} \mbox{Det}_k\,,
\ee
where $S_k$ refers to the Euclidean action evaluated at the $k^{\rm{th}}$ localization locus and $\mbox{Det}_k$ is the contribution from the Gaussian fluctuations about that locus.

Evaluating the $\mbox{Det}_k$ is subtle as there are contributions from both fermions and bosons and they almost completely cancel out against each other.  One possible way to compute it is to evaluate the fluctuations from bosons and fermions separately and combine the results, as  was done in  \cite{Kapustin:2009kz} for $d=3$ and  in \cite{Kim:2012ava} for $d=5$.  In both cases one observes a very large cancellation.

Alternatively, one can use index theorems to find the determinant factors, as was done by Pestun for $d=4$ in his groundbreaking paper \cite{Pestun:2007rz}.  Generalizations to $d=3$ \cite{Kallen:2011ny}, $d=5$ \cite{Kallen:2012cs,Kallen:2012va}, and $d=6,7$ \cite{Minahan:2015jta} followed thereafter (for a further list of references see \cite{Pestun:2016zxk}). 
 In computing the determinants via index theorems there was a difference in approach for odd and even dimensional spheres.  In the odd case one takes advantage of an everywhere nonvanishing vector field.  In the even case a vector field necessarily has fixed points and one adjusts the methods accordingly\footnote{ The differences are spelled out more thoroughly in \cite{Pestun:2016jze}, where separate subsections are devoted to the odd and even  case.}.

However, even though the methods used were different, the final results were strikingly similar.    In \cite{Minahan:2015any} a conjecture was given for the partition function of supersymmetric gauge theories in the zero instanton sector on round spheres with eight supersymmetries, for general dimension $d$. 
 The conjecture passes many tests.  As was observed in \cite{Minahan:2015any} one could combine the partition function for a vector multiplet and an adjoint hypermultiplet with appropriate mass such that the number of supersymmetries is enhanced to the maximal number of 16, and analytically continue the result up to six and seven dimensions to obtain the result found previously in \cite{Minahan:2015jta}.  Other tests were performed in \cite{Minahan:2017wkz}, where it was shown that the analytically continued result for a vector multiplet in six-dimensions is consistent with the one-loop runnings of the coupling in flat space.  A similar story is true for  maximal supersymmetry in eight and nine dimensions.

In this paper we will verify this conjecture by calculating explicitly the determinants for general dimensions.  Our methods do not use index theorems but are instead generalizations of the procedures used in\cite{Kapustin:2009kz} and \cite{Kim:2012ava}.  When localizing with eight supersymmetries on $S^d$, we will choose a spinor whose vector bilinear leaves an $S^{4-d}$ sphere fixed.  So for example,  on $S^5$ it acts freely, on $S^4$ there is a fixed $S^0$, namely the north and south poles, while on $S^3$ there is a fixed $S^1$.  In the last case this is a different choice than the one used in \cite{Kapustin:2009kz}, where the vector bilinear acts freely on $S^3$.  Of course, the two procedures must give the same result. The determinant factors for the vector multiplet and hypermultiplet are given in~\cref{eq:Zvec8,eq:Zhyp8} respectively.

We then consider theories with four supersymmetries.   
Actions for gauge theories  on $S^4$  preserving four supersymmetries have been constructed \cite{Festuccia:2011ws},  but a direct localization procedure has not yet been found.  Hence, our starting point is on $S^3$.  Here we follow the prescription in \cite{Kapustin:2009kz} to generate a vector field that acts freely.  We show how to generalize the construction to $d\le3$  
and write down an explicit expression for the determinant factors given in~\cref{4ssvec,4sschi}.  In the generalization the fixed point set for the vector field is $S^{2-d}$, hence $S^2$ will have fixed points at the poles.  

We then make a proposal for analytically continuing gauge theories with four supersymmetries up to $d=4$.  The pitfalls of dimensionally regularizing supersymmetric gauge theories have been known for a long time \cite{Siegel:1980qs,Siegel:1979wq}.  However, except perhaps for  anomalies, it appears to work in one- and two-loop calculations~\cite{Avdeev:1981vf}.  Analytical continuation of the dimension has also been successfully applied to conformal field theories \cite{Fei:2014yja,Giombi:2014xxa,Fei:2014xta,Fei:2015kta,Fei:2015oha}.   
With this proposal for minimal supersymmetry on $S^4$ we test it against various cases.    We first show that the continuation is consistent with the partition functions for a  $U(1)$ vector multiplet or a free massless chiral multiplet.  Both of these situations are conformal and so can be mapped from flat space onto $S^4$.  Since they are free, their partition functions on the sphere  are calculable.  We next consider a general gauge theory with $\NN=1$ supersymmetry.   We show that in the limit of large radius we can extract the correct one-loop $\beta$-function. 

Lastly,  we investigate a mass deformation of $\NN=4$ super Yang-Mills.  Here we concentrate on $\NN=1^*$ theories with three chiral multiplets in the adjoint representation and masses $m_i$, with $i=1,2,3$.  The superpotential also has a term cubic  in the chiral fields that stays fixed as the mass parameters are varied.  A straightforward dimensional reduction of $\NN=1^*$ gives a three dimensional gauge theory with complex masses for chiral multiplets. In our analytic continuation we start with a vector multiplet and three chiral multiplets.  However, the three dimensional mass deformed gauge theory that we can analytically continue requires real masses.  Such terms appear explicitly as central charges in the superalgebra.  The presence of the cubic term in the superpotential forces the sum of the three real masses to be zero in order to maintain supersymmetry.

Despite these subtleties, one can compare the general structure of the analytically continued partition function with the $\NN=1^*$ partition function. We make a straightforward identification, up to a sign,  of the real masses of the analytically continued theory with the masses, up to a sign,  that appear in the $\NN=1^*$ superpotential. 
 $\NN=1$ superconformal theories on $S^4$ are scheme dependent \cite{Gerchkovitz:2014gta}. 
 However, in \cite{Bobev:2016nua} it was argued that the the fourth derivatives of the free energy with respect to the mass parameters are scheme independent.  This is in line with our observations here.  We compute the corrections to the free energy to sixth order in the chiral masses at strong coupling.   At least for the real part of the free energy we find no inconsistencies with the holographic results in \cite{Bobev:2016nua}.
   In fact, having the sum of the real masses be zero turns out to play a crucial role.

The rest of this paper is structured as follows.  In section 2 we review and extend the results in \cite{Minahan:2015jta} for constructing gauge theories on round spheres for eight and four supersymmetries.  In section 3 we compute the fluctuations about the perturbative localization locus.  In section 4 we explicitly construct the determinant factors for theories with eight supersymmetries.  In section 5 we do the same for theories with four supersymmetries.  In section 6 we use the analytically continued result for four supersymmetries to compute the free energy of the mass deformed $\NN=1$ theory to quartic order in the  masses of the chiral multiplet.  In section 7 we present our conclusions and discuss further issues.  The appendices contain our conventions and numerous technical details.

\section{Supersymmetric gauge theories on $S^d$ by dimensional reduction}
\label{s-dimred}

In this section we review and extend the procedure in  \cite{Minahan:2015jta} to construct supersymmetric gauge theories on $S^d$. This is a generalization of Pestun's study in four dimensions \cite{Pestun:2016zxk}, and includes further details to reduce the number of supersymmetries to eight and four respectively.

As in \cite{Pestun:2016zxk} our starting point is the 10 dimensional $\NN=1$ SYM Lagrangian\footnote{As in \cite{Pestun:2016zxk} we consider the real form of the gauge group so that the group generators are anti-Hermitian and independent generators satisfy $\Tr(T^aT^b)=-\delta^{ab}$.}
\begin{equation}\label{LL}
\LL= -\frac{1}{g_{10}^2}\Tr\left(\sfrac12F_{MN}F^{MN}-\Psi\slashed{D}\Psi\right)\,,
\end{equation}
The space-time indices  $M,N$  run from $0$ to $9$ and $\Psi^a$ is a Majorana-Weyl spinor in the adjoint representation.  Properties of  $\Gamma^{M}_{ab}$ and $\tilde\Gamma^{M\,ab}$ are given in appendix \ref{s-useful}.  The 16 independent supersymmetry transformations that leave \cref{LL} invariant are
\begin{equation}\label{susy}
\begin{split}
\delta_\eps A_M&=\eps\,\Gamma_M\Psi\,,\\
\delta_\eps \Psi&=\sfrac12 \Gamma^{MN}F_{MN}\,\eps\,,
\end{split}
\end{equation}
where $\eps$ is a constant bosonic  real spinor, but is otherwise arbitrary.

We next dimensionally reduce this theory to $d$ dimensions by choosing Euclidean spatial indices $\mu=1,\dots d$ with gauge fields $A_\mu$ and scalars $\phi_I$ with $I=0,d+1,\dots 9$.  The field strengths with  scalar indices become $F_{\mu I}=D_\mu \phi_I$ and $F_{I J}=[\phi_I,\phi_J]$.  As in \cite{Pestun:2016zxk} we are choosing one scalar component to come from dimensionally reducing the time direction, leading to a wrong-sign kinetic term for this field.

We take the  $d$-dimensional Euclidean space to be the round sphere $S^d$ with radius $r$ with the metric
\begin{equation}\label{metric}
ds^2=\frac{1}{(1+\beta^2x^2)^2}\, dx_\mu dx^\mu\,,
\end{equation}
where $\beta=\frac{1}{2r}$.
The supersymmetry parameters are modified to be conformal Killing spinors on the sphere, satisfying
\begin{equation}\label{KS1}
\nabla_\mu\eps=\tilde\Gamma_\mu\tilde\eps\,,\qquad\qquad \nabla_\mu\tilde\eps=-\beta^2\Gamma_\mu\eps\,.
\end{equation}
We impose the further condition
\begin{equation}\label{KS}
\nabla_\mu\eps=\beta\,\tilde\Gamma_\mu\Lambda\, \eps\,,
\end{equation}
leaving 16 independent supersymmetry transformations. To be consistent with \cref{KS1} $\Lambda$ must satisfy $\tilde\Gamma^\mu\Lambda=-\tilde\Lambda\Gamma^\mu$,  $\tilde\Lambda\Lambda=1$, $\Lambda^T=-\Lambda$.  The simplest choice has $\Lambda=\Gamma^0\tilde\Gamma^8\Gamma^9$.  The solution to \cref{KS1} and \cref{KS} is \begin{equation}\label{spinsol}
\eps=\frac{1}{(1+\beta^2 x^2)^{1/2}}\left(1+\beta\, x\cdot\tilde\Gamma\,\Lambda\right)\eps_s\,,
\end{equation}
where $\eps_s$ is  constant.
On the sphere the supersymmetry transformations for the bosons are unchanged, but those for the  fermions are modified to
\begin{equation}\label{susysp}
\begin{split}
\delta_\eps \Psi=&\sfrac12 \Gamma^{MN}F_{MN}\eps+\frac{\alpha_I}{2}\Gamma^{\mu I}\phi_I\nabla_\mu\,\eps\,,
\end{split}
\end{equation}
where the constants $\al_I$ are given by
\begin{equation}
\label{alrel}
\begin{split}
\alpha_I=&\frac{4(d-3)}{d}\,,\qquad I=8,9,0,\\
\alpha_I=&\frac{4}{d}\,,\qquad I=d+1,\dots 7\, .
\end{split}
\end{equation}
The index $I$ in \cref{susysp} is summed over. This particular choice  preserves all 16 supersymmetries. One needs to add following extra terms to get a supersymmetric Lagrangian:
\begin{equation}\label{eq:extra}
\begin{split}
\LL_{\Psi \Psi} & =\ -\frac{1}{g_{YM}^2} \Tr \frac{(d-4)}{2r}\Psi\Lambda\Psi, \\
 \LL_{\phi\phi}&=- \frac{1}{g_{YM}^2}\left(\frac{d\,\Delta_I}{2\,r^2}\,\Tr \phi_I\phi^I\right)\,, \\ 
 \LL_{\phi \phi \phi }\ & = \frac{1}{g_{YM}^2}\frac{2}{3r}(d-4) 
 \varepsilon_{ABC} \Tr\bkt{  [\phi^A,\phi^B]\phi^C}.
\end{split}
\end{equation}
Here $\Delta_I$ is defined as 
\begin{equation}
\Delta_I= \alpha_I, \qquad \text{for}\qquad I=8,9,0, \qquad \Delta_I\ =\ 2\frac{d-2}{d}\qquad \text{for}\qquad I=d+1,\cdots 7.
\end{equation}
The scalars  split into two groups, $\phi^A$, $A=0,8,9$ and $\phi^i$, $i=d+1,\cdots 7$ and the $R$-symmetry is manifestly  broken from $SO(1,9-d)$ to 
$SO(1,7-d)$. 
The full supersymmetric Lagrangian is the dimensionally reduced version of~\cref{LL} supplemented with $\LL_{\Psi \Psi}, \LL_{\phi \phi}$ and $\LL_{\phi\phi\phi}$. 
\subsection{Eight supersymmetries}
In this paper we are interested in theories with less supersymmetry.  To construct theories with eight supersymmetries when $d\le 5$ we put a further condition on $\eps$.
\begin{equation}
 \Gamma\eps=+\eps, \qquad \Gamma\equiv \Gamma^{6789}.
\end{equation}
This reduces the number of  independent supersymmetry transformations to eight.  We divide the spinor $\Psi$ as
\begin{equation}
\Psi\ =\ \psi\ + \chi, \qquad \Gamma\psi=+\psi, \qquad \Gamma \chi\ =\ -\chi.
\end{equation} 
$\psi$  and   $\chi$ fields will be the fermionic components of the vector multiplet and the hypermultiplet respectively.  The scalars $\phi^I$, $I=6,7,8,9$ are in the hypermultiplet, while the remaining scalars belong to the vector multiplet.  Given a hypermultiplet mass $m$, the constants in \cref{alrel} paired with the hypermultiplet scalars are modified to
\begin{equation}\label{al8}
\begin{split}
\al_I&=\frac{2(d-2)}{d}+\frac{4i\sigma_I \,m\,r}{d}, \qquad I=6\dots 9, \\
\sigma_6 &=\sigma_7=-\sigma_8=-\sigma_9=1.
\end{split}
\end{equation}
To preserve supersymmetry we must modify  the cubic scalar  terms in the  Lagrangian to
\begin{equation}\label{3phi}
\LL_{\phi\phi\phi}=-\frac{4}{g_{YM}^2}\left(\left(\beta (d-4)+im\right)\Tr(\phi^0[\phi^6,\phi^7])-\left(\beta (d-4)-im\right)\Tr(\phi^0[\phi^8,\phi^9])\right)\,.
\end{equation}
We also need to change   the   quadratic term  for the hypermultiplet fermion to
\begin{equation}\label{Lchi}
\begin{split}
\LL_{\chi\chi} &=-\frac{1}{g_{YM}^2}\left(- im\Tr \chi\Lambda\chi\right)\,.
\end{split}
\end{equation}
The quadratic term for the hypermultiplet scalars is modified by changing the  value of the constant $\Delta_I$ 
\begin{equation}\label{Delta8}
\Delta_I=\frac{2}{d}\left(mr(mr+i\sigma_I)+\frac{d(d-2)}{4}\right) , \qquad \text{for} \qquad I=6,7,8,9.
\end{equation}
The quadratic term for the vector multiplet fermion is the same as in the case of 16 supersymmetries with $\Psi$ replaced by $\psi$. 
The full supersymmetric Lagrangian is then the dimensional reduction of~\cref{LL} supplemented with $\LL_{\phi\phi}+\LL_{\psi\psi}+\L_{\chi\chi}+\LL_{\phi\phi\phi}$. 
\subsection{Four supersymmetries}
If $d\le3$ then we can further reduce the number of supersymmetries by imposing the extra condition 
\begin{equation}
\Gamma'\eps=+\eps, \qquad\Gamma'\equiv\Gamma^{4589}.
\end{equation}
Now we decompose the spinor $\Psi$ into four parts
\begin{equation}
\Psi\ =\ \psi\ + \sum_{\ell=1}^3 \chi_\ell,
\end{equation}
where $\psi$ belongs to the vector multiplet and the $\chi_\ell$ belong to three different types of chiral multiplets. If we write $\ell$ in binary form as $\ell= 2 \beta_2(\ell)+\beta_1(\ell)$, where $\beta_s(\ell)$ are the binary digits for $\ell$, then we can write the chirality conditions as
\begin{equation}
 \Gamma\chi_{\ell}=(-1)^{\beta_1(\ell)}\chi_{\ell},\qquad   \Gamma'\chi_{\ell}=(-1)^{\beta_2(\ell)}\chi_{\ell}, \qquad \Gamma'\psi\ =\ \Gamma \psi\ =\ +\psi.
\end{equation} We also split the scalar fields into 4 groups. The fields $\phi^0$ and $\phi^i$, $i=d+1,\dots 3$  belong to the vector multiplet.  Each chiral multiplet contains two scalar fields $\phi_{I_\ell}$, where the index $I_\ell$ takes two values $I_\ell\ =\ 2\ell+2, 2\ell+3$.  {Given the chiral multiplet masses $m_{\ell}$, the constants in \cref{alrel} are further split into
\begin{equation}\label{al9}
\begin{split}
\al_{I_\ell}=&\frac{2(d-2)}{d}+\frac{4i\sigma_{I_\ell} \,m_{\ell}\,r}{d}\equiv \al_{(\ell)}\,,\qquad
\sigma_{I_\ell}=(-1)^{\beta_2(\ell)\beta_1(\ell)}\equiv \sigma_{(\ell)}.
\end{split}
\ee

It is instructive to look at the individual supersymmetry transformations of the fermions in  the vector and chiral multiplets.  For the fermion $\psi$ in the vector multiplet the transformations in  \cref{susysp} reduces to
\begin{equation}\label{susyspv}
\begin{split}
\delta_\eps \psi=&\sfrac12 F_{ M'  N'}\Gamma^{  M'   N'}\eps+\sfrac12 \sum_{\ell=1}^3[\phi_{I_\ell},\phi_{J_\ell}]\Gamma^{I_\ell J_\ell}\eps+\frac{\alpha_a}{2}\Gamma^{\mu a}\phi_a\nabla_\mu\,\eps\,,
\end{split}
\ee
where $ M', N'=0,\dots,3$ and $a=0,d+1\dots 3$.
Likewise, for the chiral multiplet fermions we have
\begin{equation}\label{susyspch}
\begin{split}
\delta_\eps \chi_{\ell}=&D_\mu\phi_{I_\ell}\Gamma^{\mu I_\ell}\eps+[\phi_a,\phi_{I_\ell}]\Gamma^{a I_\ell}\eps+\sfrac12\vareps^{\ell m n}[\phi_{I_m},\phi_{J_n}]\Gamma^{I_mJ_n}\eps+\frac{\alpha_{(\ell)}}{2}\Gamma^{\mu I_\ell}\phi_{I_\ell}\nabla_\mu\,\eps\,.
\end{split}
\ee
Notice that \cref{susyspv} and \cref{susyspch} have terms that contain fields outside of their respective multiplets.  In the usual construction for four supersymmetries, the transformations of the fermions would contain the auxiliary fields $D$ and $F_\ell$.  The terms outside the multiplets arise from evaluating the auxiliary fields on-shell\footnote{We thank Guido Festuccia for a helpful discussion on this point.}.  In our construction we will still use auxiliary fields, but in this case they equal zero on-shell.

With the modification in \cref{al9} the Lagrangian is {\it almost} supersymmetric under four supersymmetries if the  mass terms have the form
\begin{equation}\label{Lchi4}
\begin{split}
\LL_{\chi\chi}&=-\frac{1}{g_{YM}^2}\sum_{\ell=1}^3\left(- im_{\ell}\Tr \chi_{\ell}\Lambda\chi_{\ell}\right) , \\
\LL_{\phi\phi}&=- \frac{1}{g_{YM}^2}\sum_{\ell=1}^3\left(\frac{d\,\Delta_{(\ell)}}{2\,r^2}\,\Tr \phi_{I_\ell}{\phi^{I_\ell}}\right)\,,
\end{split}
\ee
where
\begin{equation}\label{Delta4}
\Delta_{(\ell)}\equiv \Delta_{I_{\ell}}=\frac{2}{d}\left(m_{\ell}r\big(m_{\ell}r+i\sigma_{(\ell)}\big)+\frac{d(d-2)}{4}\right)\,,
\ee
and we include the cubic terms
\begin{equation}\label{3phi4}
\LL_{\phi\phi\phi}=-\frac{4}{g_{YM}^2}\sum_{\ell=1}^3 \left(\left( i m_{\ell}  +\beta \sigma_{(\ell)} \bkt{d-4}\right)\Tr(\phi^0[\phi_{2\ell+2},\phi_{2\ell+3}])\right)\,.\ee
 However,  under a supersymmetry transformation  the Lagrangian changes by
\begin{equation}\label{susybr}
\delta_\eps\LL=\frac{1}{2g_{YM}^2}\left(\beta({d-4})+i\sum_{\ell=1}^3\sigma_{(\ell)}m_{\ell}\right)\Tr\left(\eps\Lambda\Gamma^{I_m I_n}\chi_\ell [\phi_{I_m},\phi_{J_n}]\right)\vareps^{\ell m n}\,.
\ee
The only  way  to get rid of this term is to set
\begin{equation}\label{masscond}
\beta({d-4})+i\sum_{\ell=1}^3\sigma_{(\ell)}m_{(\ell)}=0\,.
\ee
One might have expected that the leftover term in $\delta_\eps \LL$ could have been cancelled by modifying the Lagrangian with a cubic term of the form $\sim \phi_{I_m} \phi_{I_n} \phi_{I_l}$.  However, one can quickly check that this will not work because of the reality conditions imposed on the original spinor $\Psi$.

Another way to understand the origin of (\ref{masscond})  is to consider the reduction of $\NN=4$ in four dimensions down to three dimensions.  To avoid unnecessary complications we assume  the space is flat.   In three dimensions, $\NN=2$ SYM can have two types of mass terms, real and complex \cite{deBoer:1997kr,Aharony:1997bx}.  Complex masses descend directly from an $\NN=1$ superpotential in four dimensions.   However, a real mass arises from a Wilson line of a background $U(1)$ gauge field \cite{Aharony:1997bx}\footnote{In Euclidean space the real masses do not have to be real, but we will continue to use this term.}.  Writing the $4$-dimensional Lagrangian in terms of $\NN=1$ superfields, one has the term
\be\label{4dN=1}
\int d^2\theta d^2\bar\theta\exp(q_i U)\Tr( Q_i^\dag e^V Q_i e^{-V})\,,
\ee
 where V is the vector superfield for the $SU(N)$ gauge theory and $U$ is the superfield for the background $U(1)$.  The $q_i$'s are the charges of the chiral multiplets under this $U(1)$.    If we then compactify down to three dimensions, turn on the background Wilson line and integrate around the compactified dimension, (\ref{4dN=1})  becomes
\be\label{3dN=2}
R \int d^2\theta d^2\bar\theta\Tr( Q_i^\dag e^V Q_i e^{-V}) + \int d^2\theta'(q_i\Delta \Phi )\Tr( Q_i^\dag e^V Q_i e^{-V})
\ee
where $R$ is the size of the compactified circle, which can be absorbed into the gauge coupling.  The three-dimensional Grassmann variables are of the form $\theta_\alpha$ and $\bar\theta_\alpha$, while $d^2\theta'\equiv(d\theta+d\bar\theta)^2$. 
For the Wilson line we assume that $U_\mu=\nabla_\mu\Phi$ along the compactified direction. The second term in (\ref{3dN=2}) is the contribution for a real mass, $m^R_i=q_i\Delta \Phi/R$.  In the large $r$ limit,  (\ref{Lchi4}) and (\ref{3phi4}) arise from such a term, with $m^R_{\ell}=\sigma_{(\ell)}m_{(\ell)}$.  

However, the four-dimensional $\NN=4$ Lagrangian has a term in the superpotential proportional to $\Tr(Q_iQ_jQ_k)\vareps^{ijk}$ which descends directly  to the three-dimensional superpotential.  In order to couple the background $U(1)$ field to the theory, this term in the superpotential needs to be gauge invariant.  This requires setting $q_1+q_2+q_3=0$, which immediately means that the sum of the real masses is zero.  Putting the theory on the sphere modifies this condition to (\ref{masscond}).  

We can also understand (\ref{masscond}) using the three-dimensional $\NN=2$ superalgebra \cite{deBoer:1997kr,Aharony:1997bx},
\be
\{\QQ_\alpha,\bar\QQ_\beta\}=i\,\s^\mu_{\alpha\beta}\,P_\mu+i\,m^R\,\vareps_{\alpha\beta}\,,
\ee
where the real mass appears explicitly in the algebra as a central charge.   The  contribution of the superpotential to the action is
\be\label{superpot}
\int d^3x\, d^2\theta\, W + {\rm c.c.}\,.
\ee
If the superpotential has the term  $\Tr(Q_iQ_jQ_k)\vareps^{ijk}$   then acting with  $\{\QQ_\alpha,\bar \QQ_\beta\}$ on (\ref{superpot}) gives a term proportional to $m^R_1+m^R_2+m^R_3$.  Hence, supersymmetry  requires the sum to be zero.
\subsection{Off-shell supersymmetry}

We  need an off-shell formulation of supersymmetry in order to localize.  One must also ensure that the supersymmetry transformations close in the algebra.  To this end we select a particular Killing spinor $\eps$ and introduce seven auxiliary fields $K_m$ and bosonic  pure spinors $\nu_m$ with $m=1\dots 7$. These pure spinors satisfy the orthonormality conditions~\eqref{purespinors}. The off-shell Lagrangian  has  the additional term
  \begin{equation}\label{Laux}
 \LL_{aux}=  \frac{1}{g_{YM}^2}\,\Tr K^mK_m\,.
 \ee
 
 When reducing the number of supersymmetries we split the pure spinors accordingly.
With 16 supersymmetries the full set of transformations are \cite{Minahan:2015jta}
 \begin{equation}\label{susyos}
 \begin{split}
 \delta_\eps A_M=&\eps\,\Gamma_M\Psi\,,\\
  \delta_\eps \Psi=&\sfrac12 \Gamma^{MN}F_{MN}\eps+\frac{\alpha_I}{2}\Gamma^{\mu I}\phi_I\nabla_\mu\,\eps+K^m\nu_m\, ,\\
\delta_\eps K^m=&-\nu^m\slashed{D}\Psi+\beta(d-4) \nu^m\Lambda\Psi\, .
\end{split}
 \end{equation}
 Acting twice with the supersymmetry transformation on the gauge fields one finds
  \begin{equation}\label{At}
  \delta^2_\eps A_\mu
 =-v^\nu F_{\nu\mu}+[D_\mu,v^I\phi_I]\,,
 \ee
 which is the  Lie derivative of $A_\mu$ along the $-v^\nu$ direction, plus a gauge transformation.  Likewise, the action on the scalar fields is
 \begin{equation}\label{phit}
 \delta^2_\eps\phi_I=-v^\nu D_\nu\phi_I-[v^J\phi_J,\phi_I]-\frac{1}{2}\al_I\beta d\, \eps\tilde\Gamma_{IJ}\Lambda\eps\,\phi^J\,,
 \ee
 where again we have a Lie derivative plus a gauge transformation.  The last term in \cref{phit} is an $R$-symmetry transformation.  The transformation on the fermions is
 \begin{equation}\label{Psit}
 \begin{split}
\delta^2_\eps\Psi=&-v^N D_N\Psi-\frac{1}{4}(\nabla_{[\mu}v_{\nu]})\Gamma^{\mu\nu}\Psi \\
&\qquad\qquad - \frac{1}{2}\beta(\eps\tilde\Gamma^{ij}\Lambda\eps)\Gamma_{ij}\Psi-\frac{1}{2}(d-3)\beta(\eps\tilde\Gamma^{AB}\Lambda\eps)\Gamma_{AB}\Psi\,,
\end{split}
  \ee
  where the terms in the last line are $R$-symmetry transformations.  Finally, the transformation on the auxiliary fields is
   \begin{equation}\label{Kt}
 \delta^2_\eps K^m=-v^MD_M K^m-(\nu^{[m}\Gamma^\mu\nabla_\mu \nu^{n]})K_n+(d-4)\beta (\nu^{[m}\Lambda\nu^{n]})K_n\,,
 \ee
 where the last two terms are generators of an internal $SO(7)$ symmetry.
 
 With fewer supersymmetries the fields divide up into vector, hyper or chiral multiplets along with the accompanying 
 modifications to the $\al_I$.   For the case of eight supersymmetries, we split the pure spinors such that $\Gamma\nu_m=+\nu_m$ for $m=1,2,3$, while $\Gamma\nu_m=-\nu_m$ for $m=4,5,6,7$.   
The associated auxiliary fields $K^m$ belong to the vector and hypermultiplet respectively.
Their transformations are 
 \begin{equation}
 \begin{split}
 \delta_\eps K^m\ =&\  -\nu^m \slashed{D}\psi + \beta \bkt{d-4} \nu^m \Lambda \psi, \qquad \text{for} \qquad m=1,2,3,  \\
 \delta_\eps K^m\ =& \ -\nu^m \slashed{D} \chi-2 i \mu \beta \nu^m \Lambda \chi, \qquad \hspace{7 pt}\qquad \text{for} \qquad m=4,5,\cdots, 7. 
 \end{split}
 \ee 
 Here $\mu\equiv m r$ is a dimensionless parameter.

 With reduced supersymmetry, the transformations in \cref{At} are unchanged while those in \cref{phit} are modified by the change in the $\al_I$.  For fermions in the vector multiplet \cref{Psit} holds with $\Psi$ replaced by $\psi$. For fermions in the hypermultiplet  \cref{Psit} becomes
  \begin{equation}\label{Chit}
  \begin{split}
\delta^2_\eps\chi=&-v^N D_N\chi-\frac{1}{4}(\nabla_{[\mu}v_{\nu]})\Gamma^{\mu\nu}\chi\\
&\qquad\qquad - \frac{1}{2}\beta(\eps\tilde\Gamma^{IJ}\Lambda\eps)\Gamma_{IJ}\chi-2i\mu\beta(\eps\tilde\Gamma^{A}\Lambda\eps)\tilde\Gamma_{A}\Lambda\chi\, .
\end{split}
  \ee
 For the auxiliary fields,  equation~(\ref{Kt}) splits into two:
 \begin{equation}
 \begin{split}
 \label{K8t}
  \delta^2_\eps K^m =&-v^MD_M K^m-(\nu^{[m}\Gamma^\mu\nabla_\mu \nu^{n]})K_n+(d-4)\beta (\nu^{[m}\Lambda\nu^{n]})K_n\,    \\
    \delta^2_\eps K^m =&-v^MD_M K^m-(\nu^{[m}\Gamma^\mu\nabla_\mu \nu^{n]})K_n -2i\mu \beta (\nu^{[m}\Lambda\nu^{n]})K_n\, ,
    \end{split}
 \ee
 where the first equation is for $m=1,2,3$ and the second is for $m=4,5,6,7$.
  Invariance under  off-shell supersymmetry for the Lagrangian supplemented with $\LL_{aux}$  can be shown by a computation  that is almost identical to the one in \cite{Minahan:2015jta} for 16 supersymmetries.
 
 Reducing the number of supersymmetries to four, we split the pure spinors further as follows.
 \begin{equation}\Gamma'\nu_m=+\nu_m \qquad \text{for} \qquad m=1,4,5, \qquad
 \Gamma'\nu_m=-\nu_m \qquad \text{for}\qquad  m=2,3,6,7.
 \ee
 The transformations of the auxiliary fields are 
  \begin{equation}
  \begin{split}
 \delta_\eps K^m\ &=\  -\nu^m \slashed{D}\psi + \beta \bkt{d-4} \nu^m \Lambda \psi, \qquad \hspace{10 pt}\text{for} \qquad m=1,   \\
 \delta_\eps K^m\ &= \ -\nu^m \slashed{D} \chi_1-2 i \mu_1 \beta\nu^m \Lambda \chi_1, \qquad \hspace{19 pt}\text{for} \qquad m=2,3,   \\
 \delta_\eps K^m\ &= \ -\nu^m \slashed{D} \chi_2-2 i \mu_2 \beta\nu^m \Lambda \chi_2, \qquad \hspace{19 pt} \text{for} \qquad m=4,5,  \\
 \delta_\eps K^m\ &= \ -\nu^m \slashed{D} \chi_3-2 i \mu_3 \beta\nu^m \Lambda \chi_3, \qquad \hspace{19 pt}\text{for} \qquad m=6, 7,
 \end{split}
 \ee 
 with $\mu_\ell\equiv m_\ell r$ being dimensionless parameters.
 As before, equation~\cref{At} is unchanged and ~\cref{phit} is modified by the change in $\alpha_I$. For two supersymmetry variations of the auxiliary field we have a straightforward generalization of~\cref{K8t}, where we split the auxiliary fields into four different types. 
 Two supersymmetry variations of the chiral multiplet fermions take the following form
 \begin{equation}
 \bes
\delta^2_\eps\chi_{\ell}&=-v^N D_N\chi_\ell-\frac{1}{4}(\nabla_{[\mu}v_{\nu]})\Gamma^{\mu\nu}\chi_\ell \\
&\qquad\qquad - \frac{1}{2}\beta(\eps\tilde\Gamma^{IJ}\Lambda\eps)\Gamma_{IJ}\chi_\ell-2i\mu_\ell\beta(\eps\tilde\Gamma^{A}\Lambda\eps)\tilde\Gamma_{A}\Lambda\chi_\ell\,.
\end{split}
  \ee  
Invariance of the Lagrangian under off-shell supersymmetry follows just as in the case of eight and 16 supersymmetries.
\section{The localization Lagrangian}
\label{sec:BPAAction}
In this section we present the localization argument and  compute the quadratic fluctuations about the fixed point locus.  We also add a gauge fixing term in the Lagrangian and give the precise form of the partition function in terms of the  determinants of the quadratic fluctuations around the fixed point locus. 
 We only consider contributions in the zero instanton sector where the fixed point locus has a vanishing gauge field.
\subsection{Fixed point locus}
  Let us  modify the partition function path integral as follows:
\begin{equation}
    \ZZ \sbkt{t} \equiv \int \mathcal{D}\Phi\, e^{-S-tQV},
\end{equation}
where $\mathcal{D}\Phi$ denotes the integration measure for all the fields, $Q$ is a fermionic symmetry of both the integration measure and the action and $QV$ is positive semi-definite. The partition function is then independent of the parameter $t$. This allows us to evaluate the partition function at  $t\to\infty$, where it only receives contributions from quadratic fluctuations of the fields about the locus of the zeros of $QV$.

For our purposes we choose $Q$ to be the supersymmetry transformation generated by $\epsilon$,  and $V$ to be
\begin{equation}
        V =   \int d^dx \sqrt{g}\, \, \Tr' \bkt{\Psi \overline{\delta_\epsilon\Psi}},
\end{equation}
where $\Tr'$   is a positive definite inner product on the Lie algebra, which can be different than the product used in the original action. 
 We will  drop the $\Tr'$ sign  henceforth for notational simplicity.
 $\overline{\delta_\epsilon\Psi}$ is given by
\begin{equation}
        \overline{\delta_\epsilon\Psi} = 
        \frac{1}{2}\tilde{\Gamma}^{MN}F_{MN}\Gamma^0\epsilon
        +\frac{\alpha_I}{2}\,\tilde{\Gamma}^{\mu I}\phi_I\Gamma^0\nabla_\mu\epsilon
        -K^m\Gamma^0 \nu_m.
\end{equation}
So, $QV$ will be
\be\begin{split}
	Q V & =
	\int d^d x\,\sqrt{g}\, \delta_\varepsilon\Psi\overline{\delta_\varepsilon \Psi}
	-\int d^d x\,\sqrt{g}\, \Psi\delta_\varepsilon\left(\overline{\delta_\varepsilon \Psi} \right)\ \equiv\ \int d^dx \sqrt{g}\ \mathcal{L}^{\text{b}}\ + \int d^d x \sqrt{g} \mathcal{L}^{\text{f}}. 
\end{split}\ee
The first and second terms in the above equation contain the bosonic and fermionic part of the localization Lagrangian  respectively. Let us now find the locus where the path integral  localizes when $t\to\infty$. The bosonic part is~\cite{Minahan:2015jta}
\begin{equation}
\label{eq:LocLagBos}
\begin{split}
     \LL^{\text{b}}\ 
      =\  &
        \frac{1}{2}F_{MN}F^{MN}
        -\frac{1}{4}F_{MN}F_{M'N'} \left(\epsilon\Gamma^{MNM'N'0}\epsilon\right)\\ &
        +\frac{\beta d\alpha_I}{4}\,F_{MN}\phi_I\left(\epsilon\Lambda
        (\tilde{\Gamma}^I\tilde{\Gamma}^{MN}\Gamma^0 -\tilde{\Gamma}^0\Gamma^I\Gamma^{MN})\epsilon\right) \\
        &-K^m K_m v^0 -\beta d \alpha_0\phi_0 K^m\left(\nu_m\Lambda\epsilon\right)
        +\frac{\beta^2d^2}{4}\sum_I(\alpha_I)^2\phi_I\phi^Iv^0 . \end{split}
        \ee
We choose the spinor $\epsilon$ such that $v^0=1$ and $v^8=v^9=0$. Then the fixed point condition in the zero instanton sector can be written as 
\begin{equation}
        \nabla_\mu\phi^I\nabla^\mu\phi^I
        -\left(K^m +2\beta(d-3)\phi_0\left(\nu_m\Lambda\epsilon\right)\right)^2
        +\frac{\beta^2 d^2}{4}\sum_{I\neq0}(\alpha_I)^2\phi_I\phi^I = 0,
\end{equation}
All terms on the left hand side of the above equation are positive definite if fields $K^m$ and $\phi_0$ are imaginary. So the fixed point locus is given by
\begin{equation}
        K^m = -2\beta(d-3)\phi_0\left(\nu_m\Lambda\epsilon\right), \qquad
       \phi_0\ =\text{const}= \phi_0^{\text{cl}}\ \equiv \   {\frac{ \sigma}{r}} ,\qquad
        \phi_J = 0 \quad (J\neq0)\,.
\end{equation}
The dimensionless variable $\sigma$ is an element of the Lie algebra and parameterizes the fixed point locus. 
The action evaluated at the fixed point becomes
\begin{equation}\label{fpaction}
        S_{\text{fp}} = \frac{V_d}{g^2_{\text{YM}}}
        \frac{(d-1)(d-3)}{r^2}\text{Tr}\left(\phi_0^{\text{cl}}\phi_0^{\text{cl}}\right)
        =   \frac{8\pi^{\frac{d+1}{2}}r^{d-4}}{g^2_\text{YM}\Gamma\left(\frac{d-3}{2}\right)} \text{Tr } \sigma^2,
\end{equation}
where $V_d$ is the volume of the $d$-dimensional sphere.

\subsection{Quadratic fluctuations}
The next step is to move away from the localization locus by perturbing the fields about their fixed point values. We write
\be\begin{split}
\Phi' =\ \Phi^{\text{cl}}+\frac{1}{\sqrt{t}} \Phi,
\end{split}\ee
for all fields $\Phi'$ in $Q V$, with $\Phi^{\text{cl}}$ being their value at the fixed point. In the $t\rightarrow \infty$ limit, the only terms that survive in the localization Lagrangian are quadratic in the perturbations $\Phi$. Details of the computation of quadratic fluctuations about the fixed point locus are given in Appendix~\ref{ap:fluct}. Here we briefly summarize our results.

The bosonic  
fluctuations for the vector multiplet takes the following form
\begin{equation}\label{eq:LvmBos}
\begin{split}
	\mathcal{L}_{\text{v.m}}^{\text{b}}\ &=\ 
	A^{\tilde{M}}
	\ \mathcal{O}_{\tilde{M}}{}^{\tilde{N}} \
	A_{\tilde{N}}
	\ -[A_{\tilde{M}},\phi^{\text{cl}}_0][A^{\tilde{M}},\phi^{\text{cl}}_0]
	\\&
	-K^m K_m -4\beta (d-3) \phi_{\,0} K^m (\nu_m\Lambda\epsilon)
	-\phi_{\,0}\left(
	-\nabla^2
	+4\beta^2 (d-3)^2
	\right)\phi_0.
\end{split}
\ee
The indices with a tilde take the values as defined below
\begin{equation}
\tilde{M}\ =\ \{\mu, i\}, \qquad \mu=1,2,\cdots, d, \qquad i=d+1,\cdots D,
\ee
where $D=5(3)$ for theories with eight(four) supersymmetries.
$A_\mu$ is the usual vector field, while fields $A_{i}$ denote scalars in the vector multiplet other than $\phi_0$. 
The operator $\mathcal{O}_{\tilde{M}}{}^{\tilde{N}} $ is defined as
\begin{equation}\label{eq:OpO}
	\mathcal{O}_{\tilde{M}}{}^{\tilde{N}} =
	-\delta_{\tilde{M}}{}^{\tilde{N}} \nabla^2
	+\alpha_{\tilde{M}}{}^{\tilde{N}}
	-2\beta(d-3)\epsilon\Gamma_{\tilde{M}}{}^{\nu\tilde{N}89}\epsilon\nabla_\nu.
\end{equation} 
 $\alpha_{\tilde{M}}{}^{\tilde{N}} $ is a diagonal matrix given by
\begin{equation}\label{eq:MatAlph}
\alpha_{\tilde{M}}{}^{\tilde{N}} \ =\  4\beta^2
\begin{pmatrix}
	\bkt{d-1} \delta_\mu^\nu & 0          \\
	0                        & \delta_i^j
\end{pmatrix}.
\ee

The fermionic fluctuations for the vector multiplet  can be written as 
\be
\label{eq:LvmFermFinal}
\begin{split}
	\mathcal{L}_{\text{v.m}}^{\text{f}} =&
	\left(\psi\slashed{\nabla}\psi\right)
	+ \left( \psi \Gamma^0 \sbkt{\phi^{\text{cl}}_0, \psi} \right)
	-\frac{1}{2}(d-3)\beta v^{\tilde{M}}\left(\psi\Gamma^0\tilde{\Gamma}_{\tilde{M}}\Lambda\psi\right)
	 \\	
	&-\frac{1}{4}(d-3)\beta\left(\epsilon\tilde{\Gamma}^{\tilde{M}\tilde{N}}\Lambda\epsilon\right)\left( \psi \Gamma^0 \Gamma_{\tilde{M}\tilde{N}}\psi \right)
	+ m_{\psi} \left(\psi\Lambda\psi\right).
\end{split}
\ee
Here $m_{\psi} = \frac{d-1}{2}$ for eight supersymmetries and $m_\psi=\bkt{d-2} $ for four supersymmetries.

 For theories with eight supersymmetries  we have one hypermultiplet. The bosonic part contains four scalars. Their contribution to the quadratic fluctuations can be written as
\be\label{eq:Lhm8bos}
\begin{split}
\mathcal{L}_{\text{h.m}}^{\text{b}}=& \ \sum_{i=6}^{9} \sbkt{\phi_i\left(
	-\nabla^2
	+\beta^2 (d-2+2i \sigma_i \mu)^2
	\right)\phi_i
	-[\phi^{\text{cl}}_{0},\phi_i][\phi^{\text{cl}}_{0},\phi_i]} \\ &
	+ 4 \beta \bkt{2i\mu-1}\phi_6  v^\mu \nabla_\mu \phi_7 
	+4 \beta \bkt{2i\mu+1}\phi_8  v^\mu \nabla_\mu \phi_9.
\end{split}
\ee
For the hypermultiplet fermions we have
\be
\label{eq:Lfermhyp8final}
	\mathcal{L}^{\text{f}}_{\text{h.m}} = 
	\bkt{\chi \slashed{\nabla} \chi}	+ \bkt{\chi \Gamma^0  [\phi^{\text{cl}}_{0},\chi]} -\frac{1}{2}\beta\left(\epsilon\tilde{\Gamma}^{\tilde{M}\tilde{N}}\Lambda\epsilon\right)\left( \chi \Gamma^0 \Gamma_{\tilde{M}\tilde{N}}\chi \right)
	+2  i \mu \beta v^{\tilde{N}} \bkt{\chi \Gamma^0 \tilde{\Gamma}_{\tilde{N}} \Lambda \chi}. 
\ee

For the case of four supersymmetries we have three chiral multiplets. The chiral multiplet part contains six scalars. Their contribution to the quadratic fluctuations is given by
\be
\begin{split}
	\mathcal{L}_{\text{c.m}}^{\text{b}}
	\  =
	 \ \sum_{\ell=1}^{3}  & \sbkt{\phi_{I_\ell}\left(
	-\nabla^2
	+\beta^2 (d-2+2i \sigma_{(\ell)}\mu_\ell)^2
	\right)\phi^{I_\ell}
	-[\phi^{\text{cl}}_{0},\phi_{I_\ell}][\phi^{\text{cl}}_{0}\phi^{I_\ell}]} \\
	& +4 \beta \bkt{2i\mu_{\ell}-\sigma_{(\ell)}} \phi_{2\ell+2}  v^\mu \nabla_\mu \phi_{2\ell+3}.
\end{split}
\ee
Finally the contribution from the chiral multiplet fermions is
\be
\begin{split}
\mathcal{L}^{\text{f}}_{\text{c.m}}\ & =\sum_{\ell=1}^3 \bkt{\chi_{\ell} \slashed{\nabla} \chi_{\ell}}	
+ \bkt{\chi_{\ell} \Gamma^0  [\phi^{\text{cl}}_{0},\chi_{\ell}]} 
-\frac{1}{2}\beta\left(\epsilon\tilde{\Gamma}^{\tilde{M}\tilde{N}}\Lambda\epsilon\right)\left( \chi_{\ell} \Gamma^0 \Gamma_{\tilde{M}\tilde{N}}\chi_{\ell} \right)
\\
	&
	+\sigma_{(\ell)} \beta \bkt{2  i \mu_{\ell} v^{\tilde{N}} \bkt{\chi_i \Gamma^0 \tilde{\Gamma}_{\tilde{N}} \Lambda \chi_\ell}+ \chi_i\Lambda \chi_\ell} .
	\end{split}
\ee

\subsection{Gauge fixing}
With the expressions for quadratic fluctuations in hand, let us give the precise form of the partition function in terms of quadratic fluctuations. 
To compute the partition function we need to add a gauge fixing term. In the computation of the quadratic fluctuations, we employed the Lorenz gauge, so we need to use the following gauge fixing term
\begin{equation}
S_{\text{g.f}}\ =\ -\int d^dx \sqrt{g} \Tr\bkt{ b \nabla_\mu A'^\mu - \bar{c}\nabla^2 c}.
\ee
Here $b$ is the Lagrange multiplier which enforces the Lorenz gauge condition in the path integral. $c, \bar{c}$ are the usual Fadeev-Popov ghosts. $A'_\mu$ denotes the off-shell gauge field which can be decomposed as
\begin{equation}
A'_\mu\ =\ A_\mu\ + \nabla_\mu \phi,
\ee
where $A_\mu$ is divergenceless and $\phi$ encodes the pure divergence part. 

To compute the partition function one now has to integrate over the following set of fields:
\begin{equation}
b, c, \bar{c}, \phi, K_m, \phi_0, A_\mu, \phi_{I\neq 0}, \Psi.
\ee
The  first six give the following contributions:
\begin{itemize}
\item The $b$ ghosts give a factor of $\delta \bkt{\nabla_\mu A'^{\mu}}= \delta\bkt{\nabla^2 \phi}$.
\item The $c$ and $\bar c$ ghosts give a factor of $\det\bkt{\nabla^2}$. 
\item The gauge parameter $\phi$ has two contributions. There is a Jacobian factor $\sqrt{\det \nabla^2}$ coming from the change of integration measure $\DD \nabla_\mu \phi\to \DD \phi$, while the integration over $\phi$ gives a factor of $ \bkt{\det \nabla^2}^{-1} $ coming from the delta function $\delta\bkt{\nabla^2 \phi}$. 
\item The contribution of the auxiliary fields $K_m$ is trivial. It gets rid of the mass term for the scalar field $\phi_0$ in the quadratic fluctuations.
\item The  scalar $\phi_0$ gives a factor of $(\sqrt{\det \nabla^2})^{-1}$.
\end{itemize}
These factors  cancel and the partition function reduces to
\begin{equation}
\ZZ\ =\ \int d\sigma e^{-S_{\text{fp}}\bkt{\sigma}} \int \DD A_{\mu} \DD\phi_{I\neq 0} \DD \Psi  e^{-S_{\text{quad}}\bkt{\phi_0=2 \beta \sigma}}.
\ee
 
 Since the integrand is invariant under the adjoint action of the gauge group, we can replace the integral over the entire Lie algebra with an integral over a Cartan subalgebra. This introduces a Vandermonde determinant and we can write the partition function, with some convenient normalization as follows:
\begin{equation}
\ZZ\ =\ \int \sbkt{d\sigma}_{\text{Cartan}} e^{-S_{\text{fp}}\bkt{\sigma}} \prod_\alpha i\alphi  \int \DD A_{\mu} \DD\phi_{I\neq 0} \DD \Psi  e^{- S_{\text{quad}}\bkt{\phi_0=2 \beta \sigma}}.
\ee
Now, what is left to be computed is the integral over the fields $A_\mu, \Phi_{I\neq 0}$ and $\Psi$. Before doing that,  let us comment on the decomposition of the fields and quadratic fluctuations in terms of the root vectors of the Lie algebra. Schematically, bosonic quadratic fluctuations are given by
\be
\LL^{\text{b}}\ =\ \Tr' \bkt{ \Phi\cdot \OO^{\text{b}} \cdot \Phi - \sbkt{\Phi,\phi_0^{\text{cl}}} \sbkt{\Phi,\phi_0^{\text{cl}}}}\,.
\ee
Let us expand the field $\Phi$ in the Cartan-Weyl basis. The component of $\Phi$ along the Cartan generators only contributes an uninteresting $\phi_0^{\text{cl}}$ independent overall constant to the partition function, and so we do not need to focus on that part. Next, we can write $\Phi$ as: 
\begin{equation}
\Phi\ =\ \sum_{\alpha} \Phi^{\alpha} E_{\alpha},
\ee
where $E_\alpha$ are the root vectors of the Lie algebra. They are normalized so that $\Tr'\bkt{E_\alpha E_\beta}\ =\ \delta_{\alpha+\beta}$. Using $\sbkt{\sigma , E_{\alpha}}\ =\ \langle \alpha , \sigma\rangle E_{\alpha}$,   the quadratic fluctuations can be written as
\begin{equation}
\LL^{\text{b}}\ =\ \sum_{\alpha} \Phi^{-\alpha} \cdot \bkt{\OO^{\text{b}} + 4 \beta^2 \alphi^2} \cdot \Phi^{\alpha}. 
\ee
Similarly the fermionic quadratic fluctuations can be decomposed as
\begin{equation}
\LL^{\text{f}}\ =\ \Tr'\bkt{ \Psi \Gamma^0 \OO^{\text{f}} \Psi + \Psi \Gamma^0 \sbkt{\phi_0^{\text{cl}}, \Psi }} 
=\ \sum_\alpha \Psi^{-\alpha} \Gamma^0 \bkt{\OO^{\text{f}} + 2\beta \alphi} \Psi^{\alpha}.
\ee
After integrating over the quadratic fluctuations in $\LL^{\text{b,f}}$ one gets:
\be
\int \DD \Phi \DD \Psi  e^{-\int d^dx \sqrt{g} \bkt{\LL^{\text{b}}+ \LL^{\text{f}}}}\ =\ \prod_\alpha \frac{ \det \bkt{\OO^{\text{f}} + 2 \beta \alphi}_{\Psi}}{\sqrt{\det\bkt{ \OO^{\text{b}} + 4 \beta^2 \alphi^2}}_\Phi}.
\ee
Hence to compute the one-loop determinants one  needs to diagonalize the action of the ``quadratic'' operators $\OO^{\text{f,b}}$ appearing in the quadratic fluctuations. We turn to this computation in the next section.

\section{Determinants for eight supersymmetries}
\label{s-Det8}
In this section we compute 
the determinants for theories with eight supersymmetries. We compute the determinants for bosons and fermions separately and then combine them to see that after a large cancellation the results match exactly with the conjectured form in \cite{Minahan:2015any}. 
\subsection{Vector multiplet}
\label{s-1loopvec8}
Let us first compute the  
 determinant for  the vector multiplet. We start by introducing a complete set of basis elements that span spinor and vector harmonics on $S^d$. Then we diagonalize the action of the quadratic operator on these basis elements.
\subsubsection{Complete set of basis elements}
\label{s-basis}
To compute  
the determinants we need to diagonalize the action of the quadratic operator. This can be done by using a suitable set of basis elements. To this end, we define spinors
\begin{equation}
	\label{eq:etadef}
	\begin{split}
		\eta_\pm &\equiv \left( 1 \pm i \Gamma^{67} \right) \epsilon = \left( 1 \mp i \Gamma^{89}\right)\epsilon, \qquad		\tilde{\eta}_\pm \equiv \left(\Gamma^{68} \pm i \Gamma^{69}\right)\epsilon,
	\end{split}
\end{equation}
which satisfy
\begin{align}
	\Gamma^{89}\eta_\pm                                                & =  \pm i \eta_\pm,       & \tilde{\Gamma}_{0}v^{\tilde{M}} \Gamma_{\tilde{M}} \eta_\pm & = \eta_\pm,  \\
	\Gamma^{89} \tilde{\eta}_\pm                                       & = \pm i \tilde{\eta}_\pm, &
	\tilde{\Gamma}_0 v^{\tilde{M}} \Gamma_{\tilde{M}} \tilde{\eta}_\pm & = \tilde{\eta}_\pm.
\end{align}
We can now build a basis for the vector multiplet fermions by using  the spinors $\eta_\pm,\tilde{\eta}_\pm$ and the scalar spherical harmonics $Y^k_m$.  
Scalar spherical harmonics are labelled by the eigenvalues of the Laplacian and the Cartan generator along the vector $v^\mu$:
\begin{equation}
	\nabla^2 Y^k_m = -4\beta^2 k(k+d-1), \qquad
	v^\mu \nabla_\mu Y^k_m = 2i\beta m Y^k_m .
\end{equation}
The definitions of our spinor harmonics and their eigenvalues under operators $\Gamma^{89}$ and $ 
\tilde{\Gamma}_{0}v^{\tilde{M}} \Gamma_{\tilde{M}}$ are given in \cref{tab:fermbasis}.

\begin{table}[h!]
\centering
\renewcommand{\arraystretch}{1.6}
\setlength{\tabcolsep}{16pt}

\begin{tabular}{l  c  c }
	\toprule 
	
	\textbf{Spinor harmonics} &  
	{\bfseries\boldmath $\Gamma^{89}$-eigenvalue} & 
	{\bfseries\boldmath $\tilde{\Gamma}_{0}v^{\tilde{M}} \Gamma_{\tilde{M}\,}$-eigenvalue} \\ \midrule
	
	$\mathcal{X}^1_\pm \equiv Y^k_m \eta_\pm$ & $\pm i$ & $+1$ \\
	
	$\tilde{\mathcal{X}}^1_\pm \equiv Y^k_m \tilde{\eta}_\pm$ & $\pm i$ & $-1$ \\

	$\mathcal{X}^2_\pm \equiv \tilde{\Gamma}_0\Gamma^{\tilde{M}} \hat{\nabla}_{\tilde{M}} Y^k_m \eta_\pm$, for $m\neq \pm k$ & $\pm i$ & $+1$ \\

	$\tilde{\mathcal{X}}^2_\pm \equiv \tilde{\Gamma}_0\Gamma^{\tilde{M}} \hat{\nabla}_{\tilde{M}} Y^k_m \tilde{\eta}_\pm,$ for $m\neq \mp k$ & $\pm i$ & $-1$ \\ \bottomrule

\end{tabular}

\caption{Spinor harmonics basis and corresponding eigenvalues.}
\label{tab:fermbasis}
\end{table}
\noindent Here $\hat{\nabla}_{\tilde{M}}$ is defined as
\be
\hat{\nabla}_{\tilde{M}} \equiv  \nabla_{\tilde{M}} - v_{\tilde{M}} v\cdot\nabla.
\ee
 $\mathcal{X}_{\pm}^2$ and $\mathcal{X}_{\mp}^2$ vanish identically for $m=\pm k$ (see~\cref{app:vanishing} for a proof).
The set of spinors with a `$+$' subscript is related to the set with a `$-$' subscript via complex conjugation. We take the standard approach~\cite{Pestun:2007rz} that the Euclidean action is an analytical functional in the space of complexified fields and integrate over a certain  half-dimensional subspace in the path integral.  
With this in mind, we will focus on the basis for spinors with $\Gamma^{89}$ eigenvalue $+i$.

Let us show that set of spinors in  \cref{tab:fermbasis} provide a complete set of basis elements for the vector multiplet fermions on $S^d$.  To do so, we compute the action of the Dirac operator, $\tilde{\Gamma}_0\slashed{\nabla}$, on these spinors  using
\begin{equation}\label{eq:Diraceta}
\tilde{\Gamma}_0\slashed{\nabla} \eta_{+}\ =\ + i d \eta_{+}, \qquad \tilde{\Gamma}_0\slashed{\nabla} Y_m^k\ =\ \tilde{\Gamma}_0\Gamma^{\tilde{M}} \hat{\nabla}_{\tilde{M}}Y_m^k + 2 i m \beta Y_m^k.
\ee
This gives
\begin{equation}	
\label{eq:Diracchi1}
	\tilde{\Gamma}_0\Gamma^\mu \nabla_\mu \mathcal{X}^1_+
	  = 2i\beta\left(m + \frac{d}{2}\right)\mathcal{X}^1_+ + \mathcal{X}^2_+.
\end{equation}
Next we note that $\mathcal{X}_{+}^2$ can be written as
\begin{equation}
\mathcal{X}_{+}^2\ =\  \tilde{\Gamma}_0 \slashed{\nabla} Y_m^k \eta_{+} -2 i m \beta \mathcal{X}_{+}^1.
\ee
The action of $\tilde{\Gamma}_0\slashed{\nabla}$ can now be worked out by using~\cref{eq:Diraceta}, \cref{eq:Diracchi1} and the fact that $\bkt{\tilde{\Gamma}_0 \slashed{\nabla}}^2\ =\ \nabla^2$, which gives
	 \be
	 \label{eq:Diracchi2}
	\tilde{\Gamma}_0\Gamma^\mu \nabla_\mu \mathcal{X}^2_+
	  = -4\beta^2\left(k - m\right)\left(k + m + d -1 \right)\mathcal{X}^1_+
	-2i\beta\left(m + \frac{d-2}{2}\right)\mathcal{X}^2_+.
\ee
Similarly, for 
$\tilde{\mathcal{X}}^{1,2}_+$ one finds 
\begin{align}
\label{eq:Diracchi12t}
	\tilde{\Gamma}_0\Gamma^\mu\nabla_\mu\tilde{\mathcal{X}}^1_+
	 & = 2i\beta\left(m - \frac{d}{2}\right)\tilde{\mathcal{X}}^1_+ +\tilde{\mathcal{X}}^2_+, \\
	\tilde{\Gamma}_0\Gamma^\mu\nabla_\mu\tilde{\mathcal{X}}^2_+
	 & = -4\beta^2(k + m)\left(k - m +d-1\right)\tilde{\mathcal{X}}^1_+
	-2i\beta \left(m - \frac{d-2}{2}\right)\tilde{\mathcal{X}}^2_+.
\end{align}
Now we diagonalize the action of $\tilde{\Gamma}_0 \slashed{\nabla}$ on the spinor basis to get the eigenvalues
\begin{equation}\label{eq:eigDirac}
\pm  2 i \beta\bkt{k+\frac{d}{2}},\ \qquad \mp 2 i \beta \bkt{k-1+\frac{d}{2}} \qquad \text{for} \qquad m\neq + k.
\ee
By shifting $k$ in the second set of eigenvalues,  we can arrange the spinor harmonics into two sets of eigenstates of the Dirac operator, with eigenvalues $\pm 2 i \beta \bkt{k+\frac{d}{2}}$ whose degeneracy $\text{deg}_f\bkt{k,d}$, is given by
\begin{equation}
\text{deg}_f\bkt{k,d}\ =\ \mathcal{D}_{k}\bkt{d,0}+\mathcal{D}_{k+1}\bkt{d,0}-N_{k+1,d},
\ee
 where $\mathcal{D}_{k}\bkt{d,r}$ is the total degeneracy of symmetric traceless, divergence-less rank-$r$ tensors defined on $S^d$\cite{Rubin:1984tc}. $N_{m,d}$ is the number of scalar harmonics $Y_m^k$ for the case of eight supersymmetries. The explicit expressions for these degeneracies are given in \cref{app-counting}. Using these expressions we get
 \begin{equation}
\text{deg}_f\bkt{k,d}\ = 4 \frac{\Gamma\bkt{k+d}}{\Gamma\bkt{d}\Gamma\bkt{k+1}}.
\ee
For $d=4,5$ this is equal to the degeneracy of spinor harmonics on $S^d$ \cite{Camporesi:1995fb} and for $d=2,3$ this is twice the degeneracy of spinor harmonics, as expected. Hence, we conclude that the set of spinors defined in \cref{tab:fermbasis} provides a complete basis for the vector multiplet fermions in the case of eight supersymmetries.
 
 Next we use the spinor basis to construct a basis for the fields $A_{\tilde{M}}$. We define
\begin{equation}
	\label{eq:basisVec1}
	\begin{split}
		\mathcal{A}^1_{\tilde{M}} &\equiv\bkt{\eps \Gamma_{\tilde{M}}
			\mathcal{X}_{+}^1}+c^1 \nabla_{\tilde{M}} Y^k_m\ =\ \bkt{\eps \Gamma_{\tilde{M}}
			\mathcal{X}_{-}^1}+c^1 \nabla_{\tilde{M}} Y^k_m\ =\ v_{\tilde{M}}Y^k_m + c^1 \nabla_{\tilde{M}} Y^k_m, \\
		\mathcal{A}^2_{\tilde{M}} &= \frac{i}{2}\bkt{\epsilon \Gamma_{\tilde{M}} \mathcal{X}_+^2- \eps\Gamma_{\tilde{M}} \mathcal{X}_{-}^2
		}+c^2 \nabla_{\tilde{M}} Y^k_m\ =\
		\epsilon{\Gamma_{\tilde{M}}}^\mu\Lambda\epsilon \nabla_\mu Y^k_m + c^2 \nabla_{\tilde{M}} Y^k_m,\\
		\mathcal{A}^3_{\tilde{M}} &\equiv \epsilon {\Gamma_{\tilde{M}}}^\mu \Gamma^{069}\epsilon \nabla_\mu Y^k_m\ =\ \frac{-i}{2} \bkt{\eps \Gamma_{\tilde{M}} \tilde{\mathcal{X}}_{+}^2\ - \eps \Gamma_{\tilde{M}} \tilde{\mathcal{X}}_{-}^2}, \\
		\mathcal{A}^4_{\tilde{M}} &\equiv \epsilon {\Gamma_{\tilde{M}}}^\mu \Gamma^{079}\epsilon \nabla_\mu Y^k_m\ =\ \frac{1}{2} \bkt{\eps \Gamma_{\tilde{M}} \tilde{\mathcal{X}}_{+}^2\ + \eps \Gamma_{\tilde{M}} \tilde{\mathcal{X}}_{-}^2}.
	\end{split}
\end{equation}
Here $c^1, c^2$ are constants which are determined by the condition that $\mathcal{A}^1_\mu$ and $\mathcal{A}^2_\mu$ should be divergenceless:
\begin{align}
	c^1 & = \frac{im}{2\beta k(k+d-1)} , \qquad
	c^2 = \frac{(d-1)im}{k(k+d-1)}.
\end{align}
There is another bilinear involving spinors $\mathcal{X}_{\pm}^{2}$, which is equal to a linear combination of a pure divergence term and $\mathcal{A}_{\tilde{M}}^1$
\be
\eps \Gamma_{\tilde{M}} \mathcal{X}_+^2+ \eps \Gamma_{\tilde{M}} \mathcal{X}_-^2\ =\ 2 \nabla_{\tilde{M}} Y_m^k-4 i m \beta v_{\tilde{M}} Y_m^k.
\ee
Since $\mathcal{X}_{\pm}^2$ vanishes identically for $m=\pm k$, we see that $\mathcal{A}^1$ and $\mathcal{A}^2$ are not linearly independent for $m=\pm k$:
\be
\mathcal{A}^2_{\tilde{M}}\ =\ - 2 k \beta \mathcal{A}^1_{\tilde{M}}, \qquad \text{for }\qquad m=\pm k.
\ee
Similarly, $\mathcal{A}^3$ and $\mathcal{A}^4$ are proportional to each other for $m=\pm k$. 

Let us now show that the bosonic fields defined in \cref{eq:basisVec1} provide a complete 
basis for bosons in the vector multiplet\footnote{Excluding the scalar field $\phi_0$.}. We do so by diagonalizing the action of $\nabla^2$   on $\mathcal{A}_{\tilde{M}}$. 
  It acts on the vector field $v_{\tilde{M}}$ to give
\be
\nabla^2 v_{ \mu}\ = 	-4\beta^2\bkt{d-1} v_{ \mu},  \qquad  \nabla^2 v_{ i}\ = 	-4\beta^2 d v_{ i} .
\ee
Using this along with
\begin{equation}
	\nabla^2\nabla_\mu Y^k_m = -4\beta^2 \left(k (k+d-1) - (d-1)\right) \nabla_\mu Y^k_m,
	\qquad
	\nabla^\mu v_{\tilde{M}} = 2\beta\epsilon\Gamma_{\tilde{M}}{}^\mu\Lambda\epsilon,  
\end{equation}
gives us the action of $\nabla^2$ on the $\mathcal{A}^1_{\tilde{M}}$
\begin{equation}\label{eq:del2A1}
\begin{split}
	\nabla^2 \mathcal{A}_{\mu}^1\  & =
	-4\beta^2\left[k(k+d-1)+d-1\right]\mathcal{A}^1_\mu +4\beta \mathcal{A}^2_\mu, \\
	\nabla^2 \mathcal{A}_{i}^1\    & =
	-4\beta^2\left[k(k+d-1)+d\right]\mathcal{A}^1_i +4\beta \mathcal{A}^2_i.
	\end{split}
\ee
To find the action of $\nabla^2$ on $\mathcal{A}^2_{\tilde{M}}$ we need to know how the operators  $\nabla^\lambda$ and $\nabla^2$ act on more complicated bilinears. Using the Killing spinor equation and the fact that $\tilde{\eps}\ =\ \beta \Lambda \eps$, one gets
\begin{equation}
	\begin{split}
		\nabla^2\epsilon\Gamma_\mu{}^\nu\Lambda\epsilon &= -8\beta^2 \epsilon \Gamma_\mu{}^\nu\Lambda\epsilon,
		\qquad
		\nabla^2\epsilon\Gamma_i{}^\nu\Lambda\epsilon = -4\beta^2\epsilon\Gamma_i{}^\nu\Lambda\epsilon ,
		\\
		\nabla^\lambda \bkt{\eps \Gamma_{\tilde{M}}{}^{\mu} \Lambda \eps} \nabla_\lambda \nabla_\mu Y_{m}^k\  & =\ 8\beta^3 k(k+d-1) v_{\tilde{M}} Y^k_m
		+4\beta^2 \delta_{\tilde{M}}{}^\mu \bkt{ i m \nabla_{\mu}Y^k_m
		+\epsilon\Gamma_\mu{}^\nu\Lambda\epsilon \nabla_\nu Y^k_m}.
	\end{split}
\end{equation}
Using these results we find that
\begin{equation}\label{eq:del2A2}
	\begin{split}
		\nabla^2 \mathcal{A}^2_\mu &=
		16\beta^3 k(k+d-1) \mathcal{A}^1_\mu -4\beta^2\left(k(k+d-1)-(d-1)\right)\mathcal{A}^2_\mu, \\
		\nabla^2 \mathcal{A}^2_i &=
		16\beta^3k(k+d-1) \mathcal{A}^1_i
		-4\beta^2\left(k(k+d-1)-(d-2)\right)\mathcal{A}^2_i.
	\end{split}
\end{equation}
The action of $\nabla^2$ on $\mathcal{A}^{3,4}$ can be computed in a similar way. The following results are necessary for this calculation:
\begin{equation}
	\begin{split}
		\nabla^2\epsilon\Gamma_{\tilde{M}}{}^\nu\Gamma^{0I9}\epsilon &= -4\beta^2(d-2) \epsilon \Gamma_{\tilde{M}}{}^{\nu0I9}\epsilon, \qquad \nabla^\lambda\epsilon\Gamma_{\tilde{M}}{}^\nu\Gamma^{0I9}\epsilon
		\nabla_\lambda\nabla_\nu Y^k_m=0 ,\qquad \text{for}\qquad I=6,7.
	\end{split}
\end{equation}
This gives
\begin{equation}\label{eq:del2A34}
	\begin{split}
		\nabla^2 \mathcal{A}^{3,4}_\mu &=
		-4\beta^2\left(k(k+d-1)-1\right)\mathcal{A}^{3,4}_\mu , \\
		\nabla^2 \mathcal{A}^{3,4}_i &=
		-4\beta^2 k (k+d-1)\mathcal{A}^{3,4}_i .
	\end{split}
\end{equation}
The eigenvalues of $\nabla^2$ acting on the vector and the scalar parts of $\mathcal{A}_{\tilde{M}}$ are given below. The first term in each row corresponds to  $\mathcal{A}_\mu$ and the second to $\mathcal{A}_{i}$:
\begin{equation}
\label{eq:8veceig1}
\begin{array}{l l}
-4\beta^2\bkt{k (k-3) + d (k-1)+1}, \qquad & -4\beta^2\bkt{k-1}\bkt{k+d-2},
 \\
  -4\beta^2\bkt{k\bkt{k+d+1}+d-1},  \qquad & -4\beta^2 \bkt{k+1}\bkt{k+d},
 \\
 -4\beta^2\bkt{k\bkt{k+d-1}-1}, \qquad &-4\beta^2 k\bkt{k+d-1}.
\end{array}
\ee
These eigenvalues correspond to the following linear combinations of the basis
\begin{equation}
\mathcal{A}^1_{ \tilde{M}}+2 \beta\bkt{k+d-1} \mathcal{A}^2_{ \tilde{M}}, \qquad \mathcal{A}^1_{ \tilde{M}}- 2\beta k \mathcal{A}^2_{ \tilde{M}}, 
\qquad \mathcal{A}^3_{ \tilde{M}} \mp i \mathcal{A}^4_{ \tilde{M}}.
\ee
For $m=\pm k$, we use the fact that $\mathcal{A}^2_{ \tilde{M}}\ =\ -2\beta k \mathcal{A}^1_{ \tilde{M}}$ to see that the first eigenvalue in~\cref{eq:8veceig1} does not contribute. Similarly, $\mathcal{A}^3_{ \tilde{M}} \mp i \mathcal{A}^4_{ \tilde{M}}$ vanish identically for $m=\pm k$ so corresponding eigenvalues do not contribute.  By shifting $k$, we can rearrange the basis into vector and scalar harmonics with eigenvalues
\begin{equation}
-4\beta^2 \bkt{k\bkt{k+d-1}-1}, \qquad -4\beta^2 k \bkt{k+d-1},
\ee
respectively. The total number of harmonics is given by
\be
\text{deg}_{\text{b}}\bkt{k,d}\ =  \mathcal{D}_{k+1}\bkt{d,0}+\mathcal{D}_{k-1}\bkt{d,0}+2\mathcal{D}_{k}\bkt{d,0}-2 N_{k+1,d}-2 N_{k,d}.
\ee
Using the explicit expressions for the degeneracies provided in~\cref{app-counting} we get
\begin{equation}
\text{deg}_{\text{b}}\bkt{k,d}\ =\ \bkt{5-d} \mathcal{D}_{k}\bkt{d,0}\ +\mathcal{ D}_{k}\bkt{d,1}.
\ee
 So we deduce that the basis defined in~\cref{eq:basisVec1} provides a complete set of harmonics for the  vector multiplet  in the case of eight supersymmetries.
\subsubsection{One-loop determinant for bosons}
\label{s-1loop8bos}
Let us now compute the one-loop determinant for vector multiplet bosons. We need to diagonalize the action of the operator $\mathcal{O}_{\tilde{M}}{}^{\tilde{N}} $ defined in equation \cref{eq:OpO}
\begin{equation}
	\mathcal{O}_{\tilde{M}}{}^{\tilde{N}} =
	-\delta_{\tilde{M}}^{\tilde{N}} \nabla^2
	+\alpha_{\tilde{M}}{}^{\tilde{N}}
	-2\beta(d-3)\epsilon\Gamma_{\tilde{M}}{}^{\nu\tilde{N}89}\epsilon\nabla_\nu.
\end{equation}
The matrix $\alpha_{\tilde{M}}{}^{\tilde{N}} $ is defined in~\cref{eq:MatAlph}.
 The action of $\nabla^2$ on the basis is given in equations~\eqref{eq:del2A1}, \eqref{eq:del2A2} and \eqref{eq:del2A34}. The next non-trivial part of the operator $\mathcal{O}_{\tilde{M}}{}^{\tilde{N}}$ involves
$
	\epsilon\Gamma_{\tilde{M}}{}^{\lambda\tilde{N}89}\epsilon\nabla_\lambda.
$
For $\mathcal{A}^1$ we have
\begin{equation}
	\epsilon\Gamma_{\tilde{M}}{}^{\lambda\tilde{N}89}\epsilon\nabla_\lambda
	\mathcal{A}^1_{\tilde{N}}\ =\
	2\beta \epsilon\Gamma_{\tilde{M}}{}^{\lambda\tilde{N}89}\epsilon
	\,\epsilon\Gamma_{\tilde{N}\lambda}\Lambda\epsilon Y^k_m
	+
	\epsilon\Gamma_{\tilde{M}}{}^{\lambda\tilde{N}89}\epsilon
	\, v_{\tilde{N}} \nabla_{\lambda} Y^k_m.
\end{equation}
The term multiplying $Y_m^k$ and its derivative can be simplified using triality.
\begin{equation}
	\epsilon\Gamma_{\tilde{\mu}}{}^{\lambda\tilde{N}89}\epsilon
	\,
	\epsilon\Gamma_{\tilde{N}\lambda}\Lambda\epsilon
	=
	-(d-1) v_\mu,
	\qquad
	\epsilon\Gamma_{i}{}^{\lambda\tilde{N}89}\epsilon
	\,
	\epsilon\Gamma_{\tilde{N}\lambda}\Lambda\epsilon
	= -d v_i
	,\qquad
	\epsilon\Gamma_{\tilde{M}}{}^{\lambda\tilde{N}89}\epsilon\,\epsilon\Gamma_{\tilde{N}}\epsilon=    	\epsilon\Gamma_{\tilde{M}}{}^\lambda\Lambda\epsilon.
\end{equation}
Using these relations, we get
\begin{equation}
	\begin{split}
		\epsilon\Gamma_\mu{}^{\lambda\tilde{N}89}\epsilon\nabla_\lambda \mathcal{A}^1_{\tilde{N}}
		&=
		-2\beta(d-1)\mathcal{A}^1_{\mu} + \mathcal{A}^2_\mu ,
		\\
		\epsilon\Gamma_i{}^{\lambda\tilde{N}89}\epsilon\nabla_\lambda \mathcal{A}^1_{\tilde{N}}
		&=
		-2\beta d\mathcal{A}^1_i + \mathcal{A}^2_i.
	\end{split}
\end{equation}
The action on  $\mathcal{A}^2_{\tilde{N}}$ can be computed in a similar manner:
\begin{equation}
	\begin{split}
		\epsilon\Gamma_{\mu}{}^{\lambda\tilde{N}89}\epsilon \nabla_\lambda
		\mathcal{A}^2_{\tilde{N}}\ &=\
		4\beta^2k(k+d-1)\mathcal{A}^1_\mu, \\
		\epsilon\Gamma_{i}{}^{\lambda\tilde{N}89}\epsilon \nabla_\lambda
		\mathcal{A}^2_{\tilde{N}}\ &=\
		4\beta^2k(k+d-1)\mathcal{A}^1_i -2\beta\mathcal{A}^2_i .
	\end{split}
\end{equation}
However, the computation for $\mathcal{A}_{\tilde{M}}^{3,4}$ is slightly different. We have
\begin{equation}\label{eq:A34inter}
	\epsilon\Gamma_{\tilde{M}}{}^{\lambda\tilde{N}89}\epsilon\nabla_\lambda\ \mathcal{A}_{\tilde{N}}^{I-3}\ =\ 	\epsilon\Gamma_{\tilde{M}}{}^{\lambda\tilde{N}89}\epsilon\nabla_\lambda
	\left[\epsilon\Gamma_{\tilde{N}}{}^{\nu0I9}\epsilon\nabla_\nu Y^k_m\right],
\end{equation}
where $I=6,7$ corresponds to $\mathcal{A}^3,\mathcal{A}^4$ respectively. First we note that
\be\begin{split}
	\nabla_\lambda \left(\epsilon\Gamma_{\tilde{M}}{}^{\lambda\tilde{N}89}\epsilon\right)
	 & = 2 \beta d \epsilon\Gamma_{\tilde{M}}{}^{\tilde{N} 0} \epsilon  =0.
	 \end{split}\ee
So we can write the right-hand side of equation~\eqref{eq:A34inter}	as a total derivative. Next we use the following relation due to triality,
\be\begin{split} \qquad   \epsilon\Gamma_{\tilde{M}}{}^{\lambda\tilde{N}89}\epsilon\,
	\epsilon\Gamma_{\tilde{N}}{}^{\nu0I9}\epsilon
	 & =
	-\epsilon\Gamma_{\tilde{M}}{}^{\nu\lambda I8}\epsilon
	-v^\nu \epsilon\Gamma_{\tilde{M}}{}^{\lambda0I8}\epsilon\,,
\end{split}\ee
which allows us to write
\be\begin{split}
	\epsilon\Gamma_{\tilde{M}}{}^{\lambda\tilde{N}89}\epsilon\nabla_\lambda\ \mathcal{A}_{\tilde{N}}^{I-3}\ =\
	-\nabla_\lambda\left(
	\epsilon\Gamma_{\tilde{M}}{}^{\nu\lambda I8}\epsilon\nabla_\nu Y^k_m
	+2im\beta\epsilon\Gamma_{\tilde{M}}{}^{\lambda0I8} Y^k_m\right).
\end{split}\ee
This can be now computed using the Killing spinor equation and the triality identity, resulting in
\begin{equation}
\begin{split}
	\epsilon\Gamma_{\mu}{}^{\lambda\tilde{N}89}\epsilon\nabla_\lambda\ 
	\begin{pmatrix}
	\mathcal{A}_{\tilde{N}}^{3}\\ 
	\mathcal{A}_{\tilde{N}}^{4}
	\end{pmatrix}
	&= - 2 \beta
	\begin{pmatrix}
	d-2 &  i m  \\
	- i m & d-2
	\end{pmatrix}
	\begin{pmatrix}
	\mathcal{A}^3_\mu \\
	\mathcal A^4_\mu
	\end{pmatrix},
 \\
	\epsilon\Gamma_{i}{}^{\lambda\tilde{N}89}\epsilon\nabla_\lambda\ 
	\begin{pmatrix}
	\mathcal{A}_{\tilde{N}}^{3}\\ 
	\mathcal{A}_{\tilde{N}}^{4}
	\end{pmatrix}
	&= -2 \beta
	\begin{pmatrix}
	d-1 &  i m  \\
	-i m & d-1
	\end{pmatrix}
	\begin{pmatrix}
	\mathcal A^3_i \\
	\mathcal A^4_i
	\end{pmatrix}.
	\end{split}
\end{equation}
The action of the complete operator on the set of basis 
vectors
can be written in the following compact form:
\begin{equation}
	\begin{split}
		\left(\mathcal{OA}^1\right)_{\tilde{M}}
		&=
		4\beta^2\left[
			k(k+d-1)+(d-1)^2\right]\mathcal{A}^1_{\tilde{M}}
		-2\beta\left(d-1 \right)\mathcal{A}^2_{\tilde{M}} ,\\
		\left(\mathcal{OA}^2\right)_{\tilde{M}}
		&=
		-8\beta^3 k(d-1)(k+d-1)\mathcal{A}^1_{\tilde{M}}
		+4\beta^2 k(k+d-1)\mathcal{A}^2_{\tilde{M}}, \\
		\bkt{\mathcal{OA}^3}_{\tilde{M}} &=
		4\beta^2\left[
			k(k+d-1+(d-2)^2\right]\mathcal{A}^3_{\tilde{M}}
		+4 i \beta^2 m(d-3)\mathcal{A}^4_{\tilde{M}}, \\
		\bkt{\mathcal{OA}^4}_{\tilde{M}} &=
		4\beta^2\left[
			k(k+d-1)+(d-2)^2\right]\mathcal{A}^4_{\tilde{M}}
		-4 i\beta^2 m (d-3) \mathcal{A}^3_{\tilde{M}} .
	\end{split}
\end{equation}
The corresponding eigenvalues are
\be
4\beta^2 k^2, \qquad  4\beta^2\bkt{k+d-1}^2,\qquad	4\beta^2\left[ k(k+d-1)+(d-2)^2 \pm m (d-3) \right].
\ee
Including the contribution from different roots and taking into account the degeneracy of the basis, we get the one-loop determinant for the bosonic part of the vector multiplet:
\begin{equation}
\begin{split}
\Zv\Big|_{\text{b}} =\ \prod_{\alpha}&\prod_{k=1}^{\infty} \sbkt{4\beta^2\bkt{k^2+\alphi^2}}^{\tfrac{\mathcal{D}_k\bkt{d,0}}{2}-N_{k,d}}
\prod_{k=0}^{\infty} \sbkt{4\beta^2\bkt{\bkt{k+d-1}^2+\alphi^2}}^{\tfrac{\mathcal{D}_k\bkt{d,0}}{2}} \\
&\prod_{k=1}^{k=\infty} \prod_{m=-k}^{m=k-1}\sbkt{4\beta^2\bkt{k\bkt{k+d-1}+\bkt{d-2}^2+\bkt{d-3}m+ \alphi^2}}^{N_{m,d}}.
\end{split}
\ee
\subsubsection{One-loop determinant for fermions}
\label{s-8vfermions}
Next, we calculate the contribution to the one-loop determinant from the vector multiplet fermions. We will use the basis with the `$+$' subscript introduced in Table~\ref{tab:fermbasis}. We need to diagonalize the action of the following operator:
\begin{equation}\label{ovm8}
\mathcal{O}_{\text{v.m}}^{\text{f}} 
=
\tilde{\Gamma}_0\slashed{\nabla}
	-\frac{1}{2}(d-3)\beta v^{\tilde{M}} \tilde{\Gamma}_{\tilde{M}}\Lambda
	-\frac{1}{4}(d-3)\beta\left(\epsilon\tilde{\Gamma}^{\tilde{M}\tilde{N}}\Lambda\epsilon\right) \Gamma_{\tilde{M}\tilde{N}}
	+ \frac{1}{2}\bkt{d-1} \Gamma^{89}.
\ee

The action of $\tilde{\Gamma}_0\slashed{\nabla}$ has been computed in \cref{eq:Diracchi2,eq:Diracchi1,eq:Diracchi12t}. The second operator can be written as 
 \begin{equation}
v^{\tilde{M}}\Gamma_{\tilde{M}}\Lambda\ =\ \Gamma^{89} \bkt{ \tilde{\Gamma}_{0}v^{\tilde{M}}\Gamma_{\tilde{M}}}.
\ee
The spinor basis elements  have definite eigenvalues under the action of  $\Gamma^{89}$ and $ \tilde{\Gamma}_{0}v^{\tilde{M}}\Gamma_{\tilde{M}}$  as given in Table~\ref{tab:fermbasis}. Hence the action of the second and the last operator on the righthand side of \cref{ovm8}  is trivial to evaluate. 

The action of the third term appearing in $\mathcal{O}_{\text{v.m}}^{\text{f}}$  can be obtained using triality.
\begin{equation}
	\begin{split}
		 \epsilon\Gamma^{\tilde{M}\tilde{N}}\Lambda\epsilon \Gamma_{\tilde{M}\tilde{N}}\mathcal{X}^1_+
		&=
		- 4 i \mathcal{X}^1_+ ,\qquad 
		\epsilon\Gamma^{\tilde{M}\tilde{N}}\Lambda\epsilon \Gamma_{\tilde{M}\tilde{N}}\tilde{\mathcal{X}}^1_+
		= + 4 i \tilde{\mathcal{X}}^1_+ , \\
		 \epsilon\Gamma^{\tilde{M}\tilde{N}}\Lambda\epsilon &\Gamma_{\tilde{M}\tilde{N}}\mathcal{X}^2_+
		=
		 \epsilon\Gamma^{\tilde{M}\tilde{N}}\Lambda\epsilon \Gamma_{\tilde{M}\tilde{N}}\tilde{\mathcal{X}}^2_+
		= 0.
	\end{split}
\end{equation}                                                                                                                                                                                 We get the action of the full operator on the spinor basis to be
\begin{equation}
	\begin{split}
	\mathcal{O}_{\text{v.m}}^{\text{f}}\mathcal{X}_{+}^1                        
	 & 
	 =2i\beta\left(m + (d-1)   \right)\mathcal{X}^1_+ + \mathcal{X}^2_+,                            \\
	\mathcal{O}_{\text{v.m}}^{\text{f}}  \mathcal{X}^2_+ & = -4\beta^2(k - m)(k + m + d-1)\mathcal{X}^1_+ -2i \beta m\mathcal{X}^2_+, \\
	\mathcal{O}_{\text{v.m}}^{\text{f}}\tilde{\mathcal{X}}^1_+                                      
	 & = 2i\beta\left(m - (d-2)  \right)\tilde{\mathcal{X}}^1_+ + \tilde{\mathcal{X}}^2_+,\\
	\mathcal{O}_{\text{v.m}}^{\text{f}} \tilde{\mathcal{X}}^2_+ 
	 & = -4\beta^2(k + m)(k - m +d-1)\tilde{\mathcal{X}}^1_+
	-2i\beta\left(m - (d-2) \right)\tilde{\mathcal{X}}^2_+.
	 \end{split}
\end{equation}
For $m\neq \pm  k$, all of the above spinors contribute to the determinant. The contribution from $\mathcal{X}_+^{1,2}$ and $\tilde
{\mathcal{X}}_+^{1,2}$ is
\begin{equation}
\label{eq:contdet}
	4\beta^2\, k \bkt{k+d-1}, \qquad 
 4\beta^2\left[
		k(k+d-1)
		- m(d-3) +(d-2)^2  \right],
\end{equation}
respectively. 
However, as discussed earlier, $\mathcal{X}_{+}^2(\tilde{\mathcal{X}}_{+}^2)$ vanishes identically for $m= k(-k)$.  So for $m=k(-k)$, the first(second) term in \cref{eq:contdet} is replaced by the eigenvalue corresponding to $\mathcal{X}^1_+(\tilde{\mathcal{X}}_{+}^1)$:
\begin{equation}
\begin{split}
	\mathcal{O}_{\text{v.m}}^{\text{f}}\mathcal{X}_{+}^1                        
	 & 
	 =+ 2i\beta\left( k + (d-1)  \right)\mathcal{X}^1_+       ,                    \\      
	 \mathcal{O}_{\text{v.m}}^{\text{f}}\tilde{\mathcal{X}}^1_+                              
	 & = - 2i\beta\left( k  + (d-2)  \right)\tilde{\mathcal{X}}^1_+ .
	 \end{split}
\ee
Including the contribution from different roots, the  one-loop determinant for the fermions is given by
\begin{equation}
\begin{split}
&\Zv\Big|_{\text{f}} = \prod_{\alpha} \prod_{k=0}^{\infty}\sbkt{2 i \beta \bkt{k+d-1-i\alphi}}^{\mathcal{D}_k\bkt{d,0}} \prod_{k=1}^{\infty}\sbkt{-2 i \beta\bkt{k+i\alphi}}^{\mathcal{D}_k\bkt{d,0}-N_{k,d}} 
\\
&\prod_{k=0}^{\infty}
 \sbkt{-2 i \beta\bkt{k+d-2+i \alphi}}^{N_{k,d}}
 \prod_{m=-k}^{k-1}\sbkt{4\beta^2 \bkt{k\bkt{k+d-1}+m \bkt{d-3}+\bkt{d-2}^2+\alphi^2}}^{N_{m,d}}.
\end{split}
\ee

Combining this with the bosonic determinant, we see that  
most
terms cancel and in the end we are left with:
\begin{equation}\label{eq:Zvec8}
\Zv\  \prod_{\alpha} i \alphi   =\ \prod_{\alpha}
 \prod_{k=0}^{\infty}\sbkt{\bkt{k+i\alphi}
 \bkt{k+d-2+i \alphi}}^{N_{k,d}}.
\ee
With $N_{k,d}$ given in ~\cref{eq:Nkk} this matches exactly with the conjecture in \cite{Minahan:2015any}. 
\subsection{Hypermultiplet}
In this section we compute one-loop determinants for a hypermultiplet with eight supersymmetries. We proceed in the same manner as for the vector multiplet by introducing a complete set of states and then computing 
the eigenvalues and degeneracies
 of the quadratic operator.
\subsubsection{One-loop determinant for bosons}
The bosonic part of the quadratic fluctuations about the fixed point locus for the hypermultiplet is given in~\cref{eq:Lhm8bos}. \begin{equation}\label{eq:Lhypb8rev}
\begin{split}
\mathcal{L}_{\text{h.m}}^{\text{b}}\bkt{\mu}\ =& \ \sum_{i=6}^{9} \sbkt{\phi_i\left(
	-\nabla^2
	+\beta^2 (d-2+2i \sigma_i \mu)^2
	\right)\phi_i
	-[\phi^{\text{cl}}_{0},\phi_i][\phi^{\text{cl}}_{0},\phi_i]} \\ &
	+ 4 \beta \bkt{2i\mu-1}\phi_6  v^\mu \nabla_\mu \phi_7 
	+4 \beta \bkt{2i\mu+1}\phi_8  v^\mu \nabla_\mu \phi_9.
	\end{split}
\ee
We see that $\phi_{6,7}$
 and $\phi_{8,9}$ mix under the action of the kinetic operator. We use $Y_m^k$ to diagonalize the action of the operator appearing in \cref{eq:Lhypb8rev}. The eigenvalues for $\phi_{6,7}$ are \begin{equation}
 4\beta^2 \bkt{k\bkt{k+d-1}+\bkt{\frac{d-2}{2}+i \mu}^2 \pm  m \bkt{2 i \mu-1}}.
 \ee
The eigenvalues for $\phi_{8,9}$  are the same as above with $\mu\rightarrow -\mu$. 
 Including the contribution from different roots, the bosonic part of the one-loop determinant is given by
 \begin{equation}
 \begin{split}
\Zh\Big|_{\text{b}} =\ \prod_{\alpha} \prod_{k=0}^{k=\infty} &\prod_{m=-k}^{k}\sbkt{ 4\beta^2 \bkt{k\bkt{k+d-1}+\bkt{\frac{d-2}{2}+i \mu}^2+\alphi^2 +  m \bkt{2 i \mu-1}}}^{N_{m,d}}\\
 & \sbkt{4\beta^2 \bkt{k\bkt{k+d-1}+\bkt{\frac{d-2}{2}-i \mu}^2+\alphi^2 -  m \bkt{2 i \mu+1}}}^{N_{m,d}},
 \end{split}
 \ee
 where we have used the fact that in the product positive and negative values of $m$  come in pairs, so the product is invariant under $m\leftrightarrow -m$.
\subsubsection{One-loop determinant for fermions}
The relevant part of quadratic fluctuations is given in~\cref{eq:Lfermhyp8final}. We need to compute the determinant of the operator
\be\begin{split}
\mathcal{O}_{\text{h.m}}^{\text{f}}\ =\ \tilde{\Gamma_0}\slashed{\nabla}  	  -\frac{1}{2}\beta\left(\epsilon\tilde{\Gamma}^{\tilde{M}\tilde{N}}\Lambda\epsilon\right) \Gamma_{\tilde{M}\tilde{N}} 
	+2  i \mu \beta v^{\tilde{N}} \tilde{\Gamma}_{\tilde{N}} \Lambda 
	.
\end{split}\ee
To diagonalize the action of this operator, we construct a complete basis for the hypermultiplet fermions.  We define the spinors
\begin{equation}
	\begin{split}
		 \lambda_+ &= \left( \Gamma_6 + i\Gamma_7\right) \epsilon, \qquad 
		\tilde{ \lambda}_+ = \left( \Gamma_8 + i \Gamma_9\right) \epsilon,
	\end{split}
\end{equation}
which satisfy
\begin{align}
	\Gamma^{89} \lambda_+         & = + i  \lambda_+ ,        &
\tilde{\Gamma}_0	\Gamma^{\tilde{M}} v_{\tilde{M}}  \lambda_+               & = -  \lambda_+ \, ,               \\
	\Gamma^{89} \tilde{ \lambda}_+ & = + i \tilde{ \lambda}_+ ,&
\tilde{\Gamma}_0	\Gamma^{\tilde{M}} v_{\tilde{M}}  \tilde{ \lambda}_+       & = - \tilde{ \lambda}_+\, .
\end{align}
Now we define the spinor harmonics, using the  spinors $ \lambda_+^{1,2}$ and the scalar spherical harmonics $Y^k_m$:
\begin{equation}\label{eq:8hypbasis}
	\begin{split}
		\mathcal{X}^1_+ &= Y^k_m  \lambda_+,\qquad \mathcal{X}^2_+ = \tilde{\Gamma}^0\Gamma^\mu\left(\hat\nabla_\mu Y^k_m\right)  \lambda_+,  \\
		\tilde{\mathcal{X}}^1_+ &= Y^k_m \tilde{ \lambda}_+,\qquad
		\tilde{\mathcal{X}}^2_+ = \tilde{\Gamma}^0\Gamma^\mu\left(\hat\nabla_\mu Y^k_m\right) \tilde{ \lambda}_+.
	\end{split}
\end{equation}
The spinors
 $\mathcal{X}_{+}^2(\tilde{\mathcal{X}}_{-}^2)$ vanish identically for $m= k(-k)$. An analysis similar to the one in~\cref{s-basis} shows that the basis defined above, provides a complete set of spinor harmonics on $S^d$ for hypermultiplet fermions.
The action of the operator on these basis elements can be computed in an analogous fashion to the vector multiplet fermions, resulting into
\begin{equation}
	\begin{split}
		\mathcal{O}_{\text{h.m}}^{\text{f}} \mathcal{X}^1_+ &=
		-2i\beta\bkt{m + \left(\frac{d-2}{2} + i\mu \right) }\mathcal{X}_+^1 + \mathcal{X}_+^2, \\
		\mathcal{O}_{\text{h.m}}^{\text{f}} \mathcal{X}^2_+
		&=-4\beta^2 (k- m) (k + m + d -1)\mathcal{X}^1_+
		+2i\beta\left(m +\left(\frac{d-2}{2}  + i\mu \right)\right)\mathcal{X}^2_+, \\
		\mathcal{O}_{\text{h.m}}^{\text{f}} \tilde{\mathcal{X}}_+^1	&=
		-2i\beta\left(m - \left(\frac{d-2}{2} - i\mu \right) \right)\tilde{\mathcal{X}}_+^1+\tilde{\mathcal{X}}_+^2, \\
		\mathcal{O}_{\text{h.m}}^{\text{f}} \tilde{\mathcal{X}}_+^2
		&=-4\beta^2 (k + m) (k - m + d -1) \tilde{\mathcal{X}}_+^1
		+2i\beta\left(m -\left(\frac{d-2}{2}  - i\mu \right) \right) \tilde{\mathcal{X}}_+^2.
	\end{split}
\end{equation}
For $m\neq \pm k$, the contribution to the determinant from these basis elements is given by
\be
\label{eq:conthm8}
\begin{split}
&\mathcal{X}_{+}^{1,2}: \qquad 
	4\beta^2\left(k(k+d-1) + m\left(2i\mu -1\right)
	+\left(\frac{d-2}{2}+i\mu\right)^2 \right), 
\\
&\tilde{\mathcal{X}}_{+}^{1,2}: \qquad 
	4\beta^2\left(k(k+d-1) + m\left(2i\mu +1\right)
	+\left(\frac{d-2}{2}-i\mu\right)^2 \right). 
	\end{split}
\ee
For $m=  k\bkt{-k}$,  only  $\mathcal{X}_{+}^1 (\tilde{\mathcal{X}}_{+}^1)$ contributes to the first(second) term in~\cref{eq:conthm8}.
\begin{equation}
\begin{split}
\mathcal{O}_{\text{h.m}}^{\text{f}} \mathcal{X}^1_+ &=
		- 2i\beta\left(k + \left(\frac{d}{2} + i\mu -1\right)  \right)\mathcal{X}_+^1\, ,\\
\mathcal{O}_{\text{h.m}}^{\text{f}} \tilde{\mathcal{X}}_+^1	&=
		-2i\beta\left(k + \left(\frac{d}{2} - i\mu -1\right) \right)\tilde{\mathcal{X}}_+^1\, .
\end{split}
\ee
After including the contribution from roots, the  fermionic  part of the one-loop determinant is given by
\begin{equation}
\begin{split}
\Zh\Big|_{\text{f}} =\ \prod_{\alpha}\prod_{k=0}^{k=\infty}& \prod_{m=-k}^{k-1} \sbkt{4\beta^2\left(k(k+d-1) + m\left(2i\mu -1\right)
	+\left(\frac{d-2}{2}+i\mu\right)^2+ \alphi^2\right)}^{N_{m,d}} \\
	&\sbkt{4\beta^2\left(k(k+d-1) - m\left(2i\mu +1\right)
	+\left(\frac{d-2}{2}-i\mu\right)^2+ \alphi^2\right)}^{N_{m,d}}\\
	&\sbkt{4 \beta^2  \bkt{\bkt{k+\frac{d-2}{2}+i \mu}- i \alphi} \bkt{\bkt{k+\frac{d-2}{2}-i \mu}+i \alphi}}^{N_{k,d}}.
\end{split}
\ee
Combining this with the bosonic determinant and after many cancellations, we are left with:
\begin{equation}\label{eq:Zhyp8}
\Zh\ =\ \prod_{\alpha}\prod_{k=0}^{k=\infty}
\sbkt{\bkt{k+\frac{d-2}{2}+i \mu+i\alphi} \bkt{k+\frac{d-2}{2}-i \mu-i\alphi }}^{-N_{k,d}}.
\ee
This matches with the conjectured form in \cite{Minahan:2015any}.
\section{Determinants for four supersymmetries}
\label{s-Det4}
In this section we will compute one-loop determinants for theories with four supersymmetries. Most of the computation is similar to the case of eight supersymmetries. However there is an additional subtlety in the construction of complete sets of basis elements. 
\subsection{The complete set of basis elements} 
\label{ss-4susybasis}
One can verify that only the first two of the spinors defined in Table~\ref{tab:fermbasis}  have $+1$ eigenvalue for $\Gamma'$ and hence belong to the vector multiplet of theories with four supersymmetries. However, they do not provide a complete set of basis elements for spinor harmonics. To see this, recall that eigenvalues of the Dirac operator acting on $\mathcal{X}_+^{1,2}$  are given in \cref{eq:eigDirac}. By shifting the value of $k$, one can arrange them into spinor harmonics with eigenvalues $\pm 2i \beta\bkt{k+\frac{d}{2}}$. However  the degeneracy of positive and negative eigenvalues is not the same.
\begin{equation}
\text{deg}_{+}\ =\ \mathcal{D}_k\bkt{d,0}, \qquad \text{deg}_{-}\ =\ \mathcal{D}_{k+1}\bkt{d,0}-n_{k+1,d}.
\ee
Here $n_{m,d}$ denotes the number of scalar harmonics $Y_{m}^k$ for the case of four supersymmetries. This  differs from $N_{m,k}$, as the vector field $v_\mu$ now vanishes only on an $S^{2-d}$. An explicit expression for $n_{k,d}$ is provided in \cref{app-counting}. Using that, we get
\begin{equation}
\deg_-\ =\  2\frac{\Gamma\bkt{k+d}}{\Gamma\bkt{d}\Gamma\bkt{k+1}},
\ee
which is equal to the degeneracy of spinor harmonics on $S^d$ for $d=2,3$. Clearly $\text{deg}_+$ is different. Moreover one can show that:
\begin{equation}
\text{deg}_-\bkt{k,d}\ -\text{deg}_+\bkt{k,d}\ =\ n_{k,d}.
\ee
So $\mathcal{X}^{1}_+$ and $\mathcal{X}^{2}_{+}$  do not provide a complete basis for the spinor harmonics. This can be fixed by including another spinor $\mathcal{X}_+^{1'}$, which has the 
correct eigenvalue and degeneracy,
\be
\label{eq:altFermVec}
		\mathcal{X}^{1'}_\pm = \Gamma^{0579} \eta_{\mp} Y_{\mp k}^k, \qquad 
\tilde{\Gamma}_0 \slashed{\nabla} \mathcal{X}^{1'}_{\pm}\ =\ \pm 2 i \beta\bkt{k+\frac{d}{2}} \mathcal{X}^{1'}_{\pm}.
\ee
So a complete basis for spinor harmonics is provided by $\mathcal{X}_{+}^{1},\mathcal{X}^{2}_+$ and $\mathcal{X}_{+}^{1'}$.

For the vector multiplet bosons we use the following basis\begin{equation}
\label{eq:vecbasis4}
	\begin{split}
		\mathcal{A}^1_{\tilde{M}} &= v_{\tilde{M}}Y^k_m + c^1 \nabla_{\tilde{M}} Y^k_m \\
		\mathcal{A}^2_{\tilde{M}} &=
		\epsilon{\Gamma_{\tilde{M}}}^\mu\Lambda\epsilon \nabla_\mu Y^k_m + c^2 \nabla_{\tilde{M}} Y^k_m
	\end{split}
\end{equation}
These are the first two basis elements that we used for theories with eight supersymmetries as defined in~\cref{eq:basisVec1}.  As discussed in~\cref{s-basis}, these basis elements can be arranged into vector and scalar harmonics on $S^d$ with the total number given by
\be
\text{deg}_{\text{b}}\bkt{k,d}\ =\ \mathcal{ D}_{k+1}\bkt{d,0}+\mathcal{D}_{k-1}\bkt{d,0}-2n_{k+1,d}.
\ee
Using explicit values one can show that
\begin{equation}
\text{deg}_{\text{b}}\bkt{k,d}-\mathcal{D}_k\bkt{d,1}- \bkt{3-d} \mathcal{D}_k\bkt{d,0}\ =\ -2 n_{k-1,d}\neq 0.
\ee
Hence the above basis is not complete. We can complete it by including
\begin{equation}
\mathcal{A}^{\pm}_{\tilde{M}}\ =\ \eps \Gamma_{\tilde{M}} \mathcal{X}^{1'}_{\pm}.
\ee
 The elements $\mathcal{A}^{\pm}_{\mu}$ defined above are divergenceless. Their eigenvalues under action of Laplacian are
\begin{equation}
\nabla^2 \mathcal{A}^{\pm}_{\mu}\ =\ -4\beta^2\bkt{k\bkt{k+d+1}+d-1} \mathcal{A}^{\pm}_{\mu}, \qquad \nabla^2 \mathcal{A}^{\pm}_{i}\ =\ -4\beta^2 \bkt{k+1}\bkt{k+d}\mathcal{A}^{\pm}_{i}.
\ee
By shifting $k\to k-1$, we can put eigenvalues in the canonical form with the total number of harmonics given by $2n_{k-1,d}$, precisely what is needed to complete the basis. 
\subsection{Vector multiplet}
\subsubsection{One-loop determinant for bosons}

To compute the one loop determinant we need the action of the operator $\mathcal{O}_{\tilde{M}}{}^{\tilde{N}}$  on the basis elements $\mathcal{A}_{\tilde{M}}^{1,2}$ and $\mathcal{A}_{\tilde{M}}^{\pm}$. The computation for $\mathcal{A}_{\tilde{M}}^{1,2}$ was performed in detail in ~\cref{s-1loop8bos}. Their contribution to the one-loop determinant is given by
\be
\prod_\alpha \prod_{k=1}^{\infty} \sbkt{ 4\beta^2\bkt{k^2+\alphi^2}}^{\mathcal{D}_k\bkt{d,0}-2n_{k,d}} \prod_{k=0}^{\infty} \sbkt{ 4\beta^2 \bkt{\bkt{k+d-1}^2+\alphi^2}}^{\mathcal{D}_k\bkt{d,0}}.\nonumber \\
\ee
The action of $\mathcal{O}_{\tilde{M}}{}^{\tilde{N}}$ on $\mathcal{A}^{\pm}_{\tilde{M}}$ can be calculated using the same techniques as were employed in~\cref{s-1loop8bos}. 
\be
\mathcal{O}_{\tilde{M}}{}^{\tilde{N}} \mathcal{A}^{\pm}_{\tilde{N}}\ =\ 4\beta^2\bkt{k+d-1}^2 \mathcal{A}^{\pm}_{ {\tilde{M}}}.
\ee
Including the contributions from all basis elements, we get the bosonic part of the one-loop determinant:

\be
\begin{split}
\Zv\Big|_{\text{b}}
	 =\ \prod_\alpha & \prod_{k=1}^{\infty} 
	 \sbkt{ 4\beta^2\bkt{k^2+\alphi^2}}^{\frac{\mathcal{D}_k\bkt{d,0}}{2}-n_{k,d}} 
	 \prod_{k=0}^{\infty} \sbkt{ 4\beta^2 \bkt{\bkt{k+d-1}^2+\alphi^2}}^{\frac{\mathcal{D}_k\bkt{d,0}}{2}+2n_{k,d}}. 
\end{split}
\ee
\subsubsection{One-loop determinant for fermions}
The quadratic fluctuations for the vector multiplet fermions for the case of four supersymmetries are given in~\cref{eq:LvmFermFinal}. We need to diagonalize the operator
\begin{equation}
\mathcal{O}_{\text{v.m}}^{\text{f}}\ =\  \tilde{\Gamma}_0\Gamma^\nu\nabla_\nu
	-\frac{1}{2}(d-3)\beta v^{\tilde{M}}\Gamma_{\tilde{M}}\Lambda
	-\frac{1}{4}\beta(d-3)\epsilon\Gamma^{\tilde{M}\tilde{N}}\Lambda\epsilon\Gamma_{\tilde{M}\tilde{N}}
	+\bkt{d-2}\beta \Gamma^{89} 
\ee
 acting on $\mathcal{X}_{+}^{1,2}$ and $\mathcal{X}_{+}^{1'}$. The details of this computation are similar to the case of eight supersymmetries. One gets
\be
\begin{split}
	\mathcal{O}_{\text{v.m}}^{\text{f}} \mathcal{X}^1_+ 
	&=2i\beta\left(m + (d-1) \right)\  \mathcal{X}^1_+ + \mathcal{X}^2_+, \\
	\mathcal{O}_{\text{v.m}}^{\text{f}} \mathcal{X}^2_+
	&= -4\beta^2(k - m)(k + m + d-1)\mathcal{X}^1_+ -2i\beta  m\mathcal{X}^2_+, \\
	\mathcal{O}_{\text{v.m}}^{\text{f}}\mathcal{X}_{+}^{1'}&=\ + 2 i\beta \bkt{k+d-1}  \mathcal{X}_{+}^{1'}.
\end{split}
\ee
From this we get the one-loop determinant
\be
\begin{split}
	\Zv\Big|_{\text{f}}\ =\ \prod_{\alpha}& \prod_{k=1}^{k=\infty} \sbkt{-2 i \beta \bkt{k+  i \alphi }}^{\mathcal{D}_k\bkt{d,0}-n_{k,d}}
	\prod_{k=0}^{\infty} \sbkt{2 i \beta \bkt{k+d-1-i\alphi}}^{\mathcal{D}_k\bkt{d,0}} \\
	&\prod_{k=0}^{k=\infty} \sbkt{-2i \beta\bkt{k+d-1-i\alphi}}^{   n_{k,d}}.  \end{split}
\ee
Combining this result with the bosonic determinant, we get the the full one-loop determinant for the vector multiplet:
\be
\label{4ssvec}
 \Zv \prod_{\alpha} i \alphi =\prod_{\alpha}  \prod_{k=0}^{k=\infty}
\sbkt{
	\frac{\bkt{k+i \alphi}}{ \bkt{k+d-1-i\alphi}}}^{n_{k,d}}.  
\ee
One can check that for $d=3$, this gives the correct one-loop determinant  which matches with the results in \cite{Kapustin:2009kz}. One can also check that~\cref{4ssvec} agrees with the perturbative result for two dimensional theories with $ (2,2) $  supersymmetry \cite{Benini:2012ui,Doroud:2012xw}.
\subsection{Chiral multiplet}
Let us now compute the one-loop determinants for the chiral multiplet. For the case of four supersymmetries, the 
mass-deformed  
Lagrangian contains three chiral multiplets.
\subsubsection{One-loop determinant for bosons}
Let us consider the chiral multiplet  containing the scalar fields $\phi_4,\phi_5$.
The relevant bosonic part of the quadratic fluctuations is given by
\be
\begin{split}
  \ \sum_{i=4,5} \sbkt{\phi_i\left(
	-\nabla^2
	+\beta^2 (d-2+2i\mu_1)^2
	\right)\phi_i
	 }\  - 4 \beta \bkt{1-2 i \mu_1}\phi_4  v^\mu \nabla_\mu \phi_5.
\end{split}
\ee
Using the scalar spherical harmonics, the action of the kinetic operator can be diagonalized to obtain the one-loop determinant
\be
\label{eq:dethypbos45}
\Zc\bkt{\mu_1}\Big|_{\text{b}}=\ \prod_{\alpha} \prod_{k=0}^{\infty} \prod_{m=-k}^{k} \sbkt{4\beta^2\bkt{k\bkt{k+d-1}+\bkt{\frac{d-2}{2}+i \mu_1}^2+\alphi^2  + m \bkt{1-2i \mu_1}}}^{n_{m,d}}.
\ee
The determinant for scalar fields $\phi_{6,7} \bkt{\phi_{8,9}}$ is the same as the above expression, but with $\mu_1\to \mu_2 (-\mu_3)$.
\subsubsection{One-loop determinant for fermions}
To compute the one loop determinant, we introduce a basis for the spinor harmonics as before. We introduce three sets of basis elements for three types of  chiral multiplets:
\be
\begin{split}
\mathcal{X}_{+ \ell}^1\ &\equiv \ Y_m^k  \lambda_{+ \ell}, \qquad {\mathcal{X}_{{+ \ell}}^2} \equiv \Gamma_0 \Gamma^{\tilde{M}} \hat{\nabla}_{\tilde{M}}\ Y_m^k\lambda_{+ \ell}, \\
 \mathcal{X}_{+ \ell}^{1'}\ & \equiv\ \Gamma^{0579} Y_{m}^{k} \lambda_{-\ell}, \qquad\text{for}\qquad  m = - \bkt{-1}^{\beta_1\bkt{\ell}\beta_{2}\bkt{\ell}} k,
\end{split}
\ee
where $\lambda_{\pm \ell}$ is defined as  
\be
\lambda_{\pm \ell}\ =\ \Gamma_0\bkt{\Gamma_{2\ell+2}\pm \Gamma_{2\ell+3}} \epsilon.
\ee
The index $\ell=1,2,3$ corresponds to the three chiral multiplets. Now, we need to diagonalize the action of the following operator 
\be 
\begin{split}
\mathcal{O}_{\text{c.m}}^{\text{f}} & =  \sum_{\ell=1}^{3} \mathcal{O}_{\text{c.m}, \ell}^{\text{f}} , \\
 \mathcal{O}_{\text{c.m}, \ell}^{\text{f}}& = \tilde{\Gamma}_0 \slashed{\nabla} 
-\frac{1}{2}\beta\left(\epsilon\tilde{\Gamma}^{\tilde{M}\tilde{N}}\Lambda\epsilon\right) \Gamma_{\tilde{M}\tilde{N}}
	+\sigma_{(\ell)} \beta \bkt{2  i \mu_{\ell} v^{\tilde{N}}   \tilde{\Gamma}_{\tilde{N}} \Lambda  + \Gamma^{89}} .
	\end{split}
\ee

Let us give the result for the $\ell=1$ explicitly:
\be
\begin{split}
	&\mathcal{O}_{\text{c.m},  1}^{\text{f}} \mathcal{X}_{+1}^1\ =\  -2 i  \beta \bkt{m+ \frac{d-2}{2} + i \mu_1 } \mathcal{X}_{+1}^1\ + \mathcal{X}_{+1}^2\ , \\
	& \mathcal{O}_{\text{c.m},  1}^{\text{f}}\mathcal{X}_{+1}^2\ =\ -4 \beta^2\bkt{k- m}\bkt{k+ m +d-1} \mathcal{X}_{+1}^1\ +2 i\beta \bkt{m+ \frac{d-2}{2} + i\mu_1  } \mathcal{X}_{+1}^2 ,
	\\
	&  \mathcal{O}_{\text{c.m},  1}^{\text{f}}\mathcal{X}_{+1}^{1'}\ =\ + 2 i \beta\bkt{k+\frac{d}{2} -i \mu_1 }\mathcal{X}_{+1}^{1'}.
\end{split}
\ee
From this, one gets the one-loop determinant for fermions:
\be
\begin{split}
	\Zc\bkt{\mu_1}& \Big|_{\text{f}}= \\
	& \prod_{\alpha}\prod_{k=0}^{k=\infty}  \prod_{m=- k}^{m=k-1}\sbkt{4 \beta^2\bkt{\bkt{\frac{d-2}{2}+i\mu_1}^2+ m \bkt{2i\mu_1-1}+\alphi^2+k\bkt{k+d-1}}}^{n_{m,d}} \\
	&\qquad \prod_{k=0}^{k=\infty} \sbkt{-2 i\beta\bkt{k+\frac{d-2}{2}+i\mu_1+i\alphi}}^{n_{k,d}}  \sbkt{2 i\beta\bkt{k+\frac{d}{2}-i\mu_1-i\alphi}}^{n_{k,d}}. 	
\end{split}
\ee
Combining this with the bosonic determinant, we get the full one-loop determinant for  the chiral multiplet:
\be
\label{4sschi}
\Zc\ \bkt{\mu_1}=\ \prod_{\alpha} \prod_{k=0}^{k=\infty} \sbkt{\frac{k+\frac{d}{2}-i \mu_1-i \alphi }{k+\frac{d-2}{2}+i \mu_1+i \alphi}}^{n_{k,d}}.
\ee
$n_{k,d}$ is given in ~\cref{eq:nkk}
The one-loop determinant for $\chi_{2\bkt{3}}$ can be obtained by simply replacing $\mu_1$ with $\mu_2\bkt{-\mu_3}$. Hence, the full one-loop determinant for the chiral multiplet part is given by
\begin{equation}
\Zc\bkt{\mu_1,\mu_2,\mu_3}\ =\ \Zc\bkt{\mu_1} \Zc\bkt{\mu_2} \Zc\bkt{-\mu_3}.
\ee
\section{Analytic continuation to  $d=4$ with four supersymmetries}
\label{s-4danalytic}
Now that we have obtained expressions for partition functions with eight supersymmetries in $d\le5$ dimensions and four supersymmetries in $d\le3$ dimensions, it is tempting to continue the results to higher dimensions.  In \cite{Minahan:2017wkz} this was done for eight supersymmetries where it was shown that the results were 
consistent with the
one-loop running of coupling constants in  flat space.  In this section we consider continuing theories with four supersymmetries up to $d=4$ using the expressions in \cref{s-Det4}.

\subsection{Consistency checks of analytic continuation}
\label{s-consistency}
In this subsection we perform consistency checks on the  analytic continuation with four supersymmetries.
We will show that in the $g_{YM}\to 0$ limit, the analytic continuation gives the correct partition function for a free vector and free chiral multiplets on $S^4$. We also show that the analytic continuation gives the correct one-loop divergence for theories with four supersymmetries in the decompactification limit. 
\subsubsection*{Partition function of $U(1)$ theory on $ S^{4}$}
A $U(1)$ gauge theory with four supersymmetries and massless adjoint matter in four dimensions is free and conformal. Hence it can be conformally coupled to $S^{4}$ and the partition function can be explicitly computed. This  matches with the result of our analytical continuation as we demonstrate now.

Consider the chiral multiplet in the adjoint representation of the $U(1)$ gauge group. Our expressions for the one-loop determinants can then be simplified to take the form
\be
\label{eq:ZhZvu1}
\Zc\ =\ \prod_{k=0}^{\infty}  \bkt{\frac{k+2}
{k+1}}
^{\frac{\bkt{k+1}\bkt{k+2}}{2}}, \qquad \Zv\ =\ \prod_{k=0}^{\infty} \bkt{k+1}^{3\bkt{k+1}}. 
\ee
The full partition function in this case is equal to the product of the one-loop determinants up to an overall constant.  
 
 The chiral multiplet of $\NN=1$ supersymmetry in four dimensions contains a two component Weyl fermion and two real scalars. The conformally coupled action for a free chiral multiplet on the sphere takes the following form:
 \begin{equation}
 S_{U(1)}^{\text{chi}} \ = \ \int d^{4} x \sqrt{g}\ \bkt{
\frac{1}{2}\sbkt{\phi_{1}\bkt{-\nabla^{2}+ 8\beta^{2}}\phi_{1} +\phi_{2}\bkt{-\nabla^{2}+ 8\beta^{2}}\phi_{2} }  
 -\psi \slashed{\nabla} \psi }.
 \ee
 The partition function for the matter part is then given by
 \begin{equation}
 \ZZ_{U(1)}^{\text{chi}}\ =\ \frac{\text{det} \slashed{\nabla} }{\text{det} \bkt{-\nabla^{2}+ 8\beta^{2}}}.
 \ee
 The eigenvalues and the degeneracies of these operators are given in appendix~\ref{app-counting}. Using these we get
 \begin{equation}
 \begin{split}
\text{det} \slashed{\nabla}\ 
&=\ \prod_{k=0}^{\infty}\ \sbkt{4\beta^{2}\bkt{k+2}^{2}}^{\frac{ \bkt{k+1}\bkt{k+2}\bkt{k+3}}{3}}
\\
& =\ 
\prod_{k=0}^{\infty}\ \sbkt{2\beta \bkt{k+2}}^{\frac{\bkt{k+1}\bkt{k+2}\bkt{k+3}}{3}}
\sbkt{2\beta\bkt{k+1}}^{\frac{ k \bkt{k+1}\bkt{k+2}}{3}},
\end{split}
 \ee
 where the last equality follows by splitting the product into two parts and shifting $k\to k-1$ in one of the parts. 
 Similarly, we have
 \begin{equation}
 \text{det} \bkt{-\nabla^{2}+ 8\beta^{2}}\ =\ \prod_{k=0}^{\infty}\sbkt{4\beta^{2}\bkt{k+1}\bkt{k+2}}^{\frac{\bkt{2k+3}\bkt{k+2}\bkt{k+1}}{6}}.
 \ee
Combing the the two factors of determinants, one gets
\begin{equation}
 \ZZ_{U(1)}^{\text{chi}}\ =\ \Zc\ =\ \prod_{k=0}^{\infty}  \bkt{\frac{k+2}
{k+1}}
^{\frac{\bkt{k+1}\bkt{k+2}}{2}}\,,
\ee
which matches  the analytic continuation. 

Next let us compute the partition function for the vector multiplet.  The $\NN =1$ vector multiplet in four dimensions contains a gauge field and a two-component Weyl fermion. The relevant action on $S^{4}$, with the gauge fixing term included is given by
\begin{equation}
\begin{split}
S_{U(1)}^{\text{vec}}=\int d^{4} x\sqrt{g}  \Bigl(&
A'^{\nu}\sbkt{\delta_{\nu}{}^{\mu}\bkt {-\nabla^{2}+12\beta^{2}} +\nabla_{\nu}\nabla^{\mu}} A'_{\mu}
-\psi\slashed{\nabla}\psi \\
&+ b \nabla_{\mu}A'^{\mu} - \bar{c} \nabla^{\mu} \partial_{\mu} c
\Bigr).
\end{split}
\ee
We split the vector field as follows
\begin{equation}
A'_{\mu} = A_{\mu}+ \nabla_{\mu} \phi, \qquad \text{such that} \qquad \nabla_{\mu} A^{\mu}\ = 0.
\ee
By using the fact that $\mathcal{D}\bkt{\nabla_{\mu} \phi}\ =\ \mathcal{D}' \phi \sqrt{\text{det}\bkt{- \nabla^{2}}}$, we can write the partition function as follows
\begin{equation}
\ZZ_{U(1)}^{\text{vec}}
 = \int \mathcal{D}A\mathcal{D}\psi\mathcal{D}' \phi\mathcal{D} b\mathcal{D} c \mathcal{D} \bar{c}  \sqrt{\text{det}\bkt{- \nabla^{2}}} \exp\bkt{-S_{U(1),\text{v.m}}}.
\ee
Integration over $b$ gives a factor of $\delta\bkt{-\nabla^{2} \phi}$. This, upon integrating over $\phi$ gives a factor of  $\sbkt{\text{det}\bkt{- \nabla^{2}}}^{-1} $which cancels against the contribution coming from integrating over ghosts. Hence the partition function becomes
\begin{equation}
\ZZ_{U(1)}^{ \text{vec}}
 =
 \frac{\sqrt{\text{det}'\bkt{-\nabla^{2}}}
 \text{det}\bkt{\slashed{\nabla}}}{\sqrt{\text{det}\bkt{-\nabla^{2}+12\beta^{2}}}},
\ee
where the operator in the denominator acts on divergence less vector fields. Using the formulae for eigenvalues and degeneracies of the operators, the above expression reduces to the following infinite product:
\begin{equation}
\ZZ_{U(1)}^{\text{vec}}\ =\ \frac{1}{\sqrt{3}} \prod_{k=0}^{\infty} \bkt{k+1}^{3\bkt{k+1}}.
\ee
This is the same as the analytically continued $\Zv$ up to an overall finite constant. 
\subsubsection*{Beta function from analytic continuation}
The one loop beta function for a gauge theory in four dimensions with $N_{\text{f}}$ Dirac fermions in the representation $\bf{R_{\text{f}}}$
 and $N_{\text{s}}$ complex scalars in the representation $R_{\text{s}}$ of the gauge group is given by:
 \begin{equation}
 \beta\bkt{g}\ = -\frac{g^{3}}{16\pi^{2}}\ \bkt{\frac{11}{3} C_{2}\bkt{{\bf Adj}} - \frac{4}{3} N_{\text{f}} C_{2} \bkt{\bf R_{\text{f}}}
 -\frac{1}{3} N_{\text{s}} C_{2}\bkt{\bf R_{\text{s}}} 
 }.
 \ee
 For an $\mathcal{N}=1$ supersymmetric theory with a vector multiplet and $N_{{c}}$ chiral multiplets in the representation $R_{c}$ of the gauge group the above expression for beta function reduces to:
 \begin{equation}
 \beta\bkt{g}\ =\    -\frac{g^{3}}{16\pi^{2}}\ \bkt{3  C_{{2}}\bkt{{\bf Adj}} - N_{c} C_{2}\bkt{\bf R_{c}}
 }.
  \ee
  We will reproduce this result by dimensional regularization of the analytically continued expression. To do so, we need to determine the $\mathcal{O}\bkt{\sigma^2}$ terms appearing in the one-loop determinants. We proceed as in~\cite{Minahan:2017wkz}, by replacing $\sigma \to t\sigma$ in the expressions for the one-loop determinants. The parameter $t$ keeps track of the order of $\sigma$. Focusing only on the vector multiplet,  one can easily find that
\be
\begin{split}
\frac{d\log \Zv}{dt^2}\ + \sum_{\alpha>0} \frac{1}{t^2} =  
 \sum_{\alpha>0}\tv\bkt{ \mathcal{F}\bkt{d-1,0,t\ \langle \alpha,\sigma\rangle}\ + \mathcal{F}\bkt{d-1,d-1,t\ \langle \alpha,\sigma\rangle}},
\end{split}
\ee
where
\begin{equation}
\mathcal{F}\bkt{x,y,z}\equiv\sum_{n=0}^{\infty} \frac{\Gamma\bkt{n+x}}{\Gamma\bkt{n+1}\Gamma\bkt{x}}\ \frac{1}{\bkt{n+y}^2 +z^2}\ =\ \frac{i}{2 z} \bkt{\frac{1}{y+i z} {}_2F_1 \bkt{x, y+i z; y+ iz+1;1}-c.c}. 
\ee
For $d=4-\epsilon$, we expand the R.H.S in powers of $t$ and $\epsilon$.  Keeping only the leading terms, we find
\begin{equation}
\frac{d\log \Zv}{dt^2}\ = \frac{3}{\epsilon} C_2\bkt{\text{\bf Adj}} \sigma^2+\cdots.
\ee
From this we can easily obtain
\begin{equation}
\log \Zv\ =\ \frac{3}{\epsilon} C_2\bkt{{\bf Adj }} \sigma^2+\cdots. 
\ee
A completely analogous calculation for a chiral multiplet in the representation ${\bf R_{{c}}}$ of the gauge group gives
\begin{equation}
\log \Zc= -\frac{1}{\epsilon} \sigma^2 C_2\bkt{{\bf R_c}}+\cdots.
\ee
 We combine the $\mathcal{O}\bkt{\sigma^2}$ contribution from one-loop determinants  with the $\mathcal{O}\bkt{\sigma^2}$ term in the fixed point action as given in equation (\ref{fpaction}), to get
\begin{equation}
\frac{8\pi^2}{g^2\bkt{{\Lambda}}} = \bkt{ \frac{8\pi^2}{g_{0}^2}- \frac{3}{\epsilon} C_2\bkt{{\bf Adj}}+\frac{1}{\epsilon} N_c C_2\bkt{\bf R_c} }{\Lambda}^{-\epsilon},
\ee
where $\Lambda$ is the renormalization scale and $g_0$ is the bare coupling. Differentiating the above equation w.r.t the $\log \Lambda$, one obtains the beta function
\begin{equation}
\begin{split}
 \beta\bkt{g}\ & =\  -\frac{g^3}{16\pi^2} \bkt{\ 3 C_2\bkt{\text{\bf Adj}}
 -N_{c}C_2\bkt{\bf R_{c}}},
\end{split}
\ee 
which is exactly what we wanted to show.

\subsection{Free energy of mass-deformed $\NN=4$ SYM}
\label{s-NN1star}

In this subsection we compare results from analytic continuation to a recent
holographic analysis for $\NN=1^*$ super Yang-Mills  \cite{Bobev:2016nua}.
 There are some caveats which we explain below, but to the extent that we can make a comparison our results are consistent with the 
holographic results.

The  $\NN=4$ super Yang-Mills  multiplet decomposes into an $\NN=1$ vector multiplet and three massless adjoint chiral  multiplets.  The  superpotential also has a cubic term which is the product of all three chiral fields.  We can give masses $m^{j}$, $j=1\dots3$, to the three chiral multiplets and still preserve $\NN=1$ supersymmetry.  If we choose $m^{(1)}=0$ and $m^{(2)}=m^{(3)}$ then we preserve $\NN=2$ supersymmetry, with the massless chiral multiplet joining with the $\NN=1$ vector multiplet to form an $\NN=2$ vector multiplet, while the two massive chiral multiplets combine into a hypermultiplet.  The cubic term in the superpotential  remains unchanged.   
The supersymmetry is broken to $\NN=1$ if the third chiral multiplet is given a mass or the first two multiplets have unequal masses.  The theory is called $\NN=1^*$ if the cubic term in the superpotential is left  unchanged.

It was shown explicitly in \cite{Festuccia:2011ws} how to put an $\NN=1$ theory on $S^4$,  and the $\NN=1^*$ theory is no exception.  However, there are some subtleties.  First for a Lorentzian $\NN=1$ theory, every chiral superfield $\Phi$ has a complex conjugate superfield $\bar\Phi$.  In Euclidean space, these fields should be considered independent.  Likewise, for a flat Lorentzian $\NN=1$ theory, a mass term would appear in the superpotential, $W_m=\frac{1}{2}m\Phi^2$.  The conjugate fields would have a complex conjugate mass $\bar m$.  In Euclidean space these masses are independent. 
In the 
holographic analysis in \cite{Bobev:2016nua}  $m^{(j)}$ is set equal to $\tilde m^{(j)}$.

There is no known localization procedure for $\NN=1^*$ on $S^4$.  Instead we propose analytically continuing the mass deformed theory in $d\le3$ up to $d=4$.  There is an important warning in doing this.  If we consider  $\NN=1^*$  on flat space and compactify down to three dimensions,  the resulting three-dimensional chiral multiplets have complex masses.  As  explained in \cref{s-dimred}, the mass deformed theory we use in the analytic continuation  has real masses. 
 Hence, it is not obvious that the analytic continuation of the perturbative mass-deformed partition function  actually equals the perturbative partition function for $\NN=1^*$ on $S^4$, 
  where the continuation of each real mass is set equal to the mass, or its negative, of the corresponding $\NN=1^*$ chiral multiplet\footnote{Note that these concerns do not apply to $\NN=2^*$ theories, which correspond to $\NN=4$ in three dimensions.  In decomposing the three dimensional $\NN=4$ vector multiplet into an $\NN=2$ vector and chiral multiplet, one can choose to have the scalar field $\phi_0$ be part of the vector multiplet, which leads to real mass terms.  However, we could have also chosen $\phi_4$ to be part of the vector multiplet and $\phi_0$ to pair up with $\phi_5$ in the chiral multiplet.  If at the same time one changes the pairings of the other four scalar fields, then the mass terms and the cubic term proportional to the mass in  (\ref{3phi}) would come from the superpotential.}. 
 Perhaps there is a more involved relation between the two sets of the mass parameters for which the analytically continued partition function equals that of the $\NN=1^*$. We leave this question for future work. Here we simply explore the consequences of analytically continuing to $d=4$ and find that the general form of the real part of the  
 free energy at large $N$ is consistent with the holographic results.

In three dimensions the mass parameters that appear in the partition function are written as $\mu_j^{(3)}=i\,\Delta_j+r\,m^R_j$ where $m^R_j$ is the real three dimensional mass and $\Delta_j$ is a charge under a corresponding flavor symmetry.  When continuing up to four dimensions we assume that this becomes $\mu_j^{(3)}\to \s_{(j)}\mu_j$ where $\s_{(j)}$ is defined in (\ref{al9}) and $\mu_j$ the four-dimensional complex mass multiplied by $r$.  If we then set $d=4$ in \cref{4ssvec} and \cref{4sschi} for three massive adjoint chiral multiplets, we find the perturbative partition function
 \be
 \bes
 \label{4d4sspert}
\ZZ_{\rm pert}&=\int d\sigma_i e^{-\frac{8\pi^2}{g_{YM}^2}\Tr\,\sigma^2}
\prod_\alpha\prod_{k=0}^\infty\left[\frac{(k\ms i\langle \alpha,\sigma\rangle)}{(k\ps i\langle \alpha,\sigma\rangle\ps 3)}\prod_{j=1}^3\frac{(k\ms i\langle \alpha,\sigma\rangle\ms i\s_{(j)}\mu_j+2)}{(k\ps i\langle \alpha,\sigma\rangle\ps i\s_{(j)}\mu_j \ps1)}\right]^{\frac{(k+1)(k+2)}{2}}\\
&=\int d\sigma_i {{e^{-\frac{8\pi^2}{g_{YM}^2}\Tr\sigma^2}\prod_\alpha i\langle \alpha,\sigma\rangle}}\Zm,
\end{split}
\ee
where $\Zm$ is the mass correction to the $\NN=4$ partition function,
\be
\Zm=
{{\prod_\alpha\prod_{k=0}^\infty\prod_{j=1}^3\left[\frac{(k\ms i\langle \alpha,\sigma\rangle\ms i\s_{(j)}\mu_j+2)(k\ps i\langle \alpha,\sigma\rangle \ps1)}{(k\ps i\langle \alpha,\sigma\rangle\ps i\s_{(j)}\mu_j \ps1)(k\ms i\langle \alpha,\sigma\rangle+2)}\right]^{\frac{(k+1)(k+2)}{2}}}}\,.
\ee
This last expression collapses to $\Zm=1$ if all $\mu_j=0$.  In deriving the second line in \cref{4d4sspert} we used the identity
\be
\prod_{k=0}^\infty\left[\frac{(k\ps i\langle \alpha,\sigma\rangle)(k\ps i\langle \alpha,\sigma\rangle\ps2)^3}{(k\ps i\langle \alpha,\sigma\rangle\ps 3)(k\ps i\langle \alpha,\sigma\rangle \ps1)^3}\right]^{\frac{(k+1)(k+2)}{2}}=i\langle \alpha,\sigma\rangle 
\ee
and that every root in the product comes with its negative.
 The $\s$ are $N\times N$ matrices and the root vectors are all possible combinations $\s_i-\s_j$, $i\ne j$ where $\s_i$ are the $N$ eigenvalues of $\s$.

This term is divergent if any $\mu_j\ne0$ and  needs to be regularized.  To this end we define
\be
Z_k(\sigma\ms\sigma',\mu)\equiv \left[\frac{(k\ms i(\sigma-\sigma')\ms i\mu+2)(k\ps i(\sigma-\sigma') \ps1)}{(k\ps i(\sigma-\sigma')\ps i\mu \ps1)(k\ms i(\sigma-\sigma')+2)}\right]^{\frac{(k+1)(k+2)}{2}}.
\ee
For $k\gg1$ we  expand $\log[Z_k(\sigma\ms\sigma',\mu)]$ in $1/k$, where we find
\be
 \log(Z_k(\sigma\ms\sigma',\mu))=-i\left(k\ps\frac{1}{2}\ps\frac{(\sigma\ms\sigma')^2}{k}\right)\mu \ms\frac{1}{2k}\mu^2\ps\frac{i}{3k}\mu^3\ps{\rm O}\left(\frac{1}{k^2}\right)\,. 
\ee
Hence, if we expand $\log\Zm$ in powers of $\mu_j$, the terms up to cubic order in the masses will be divergent.  The term linear in $\mu$ can be dropped as it eventually will cancel because of the mass condition (\ref{masscond}), which in terms of the $\mu_j$ is
 \be
\label{0cond}
\mu_1\ps\mu_2\ms\mu_3=0\,.
\ee
   The remaining divergent terms are independent of $\s-\s'$ and can be removed by adding constant local counterterms to the Lagrangian.   

In the large $N$-limit the free energy can be found by saddle point.  We are particularly interested in the behavior at strong coupling, where the 't Hooft coupling $\lambda\equiv g_{YM}^2N\gg1$.  In this case, the saddle point will  have the separation between two generic eigenvalues $|\s_i-\s_j|$ to  be much greater than $1$.  One can then check that for $|\s-\s'|\gg1$,
\be
\bes
\label{Zkasymp}
\sum_{j=1}^3\sum_{k=0}^{\infty} \log(Z_k(\sigma\ms\sigma',\s_{(j)}\mu_j)_{\rm reg})&\sim +\frac{1}{4}\log(\s\ms\s')^2(\mu_1^2\ps\mu_2^2\ps\mu_3^2)\\
&-\frac{i}{6}\log(\s\ms\s')^2(\mu_1^3\ps\mu_2^3\ms\mu_3^3).
\end{split}
\ee
Using (\ref{0cond}) we can reexpress the cubic term as
 \be
 \mu_1^3\ps\mu_2^3\ms\mu_3^3=-3\,\mu_1\mu_2\mu_3\,.
 \ee
 Then, when  \cref{Zkasymp} is combined with the $\NN=4$ part of the partition function, the saddle point equation reduces to
\be
\label{sadpt}
\frac{16\pi^2}{\lambda}\s\approx2\pint d\s'\rho(\s')\frac{1+\frac{1}{2}(\mu_1^2\ps\mu_2^2\ps\mu_3^2)+i\,\mu_1\mu_2\mu_3}{\s-\s'},
\ee
where $\rho(\s')$ is the eigenvalue density.  Notice that \cref{sadpt} is similar to the $\NN=2^*$ saddle point equation \cite{Russo:2012ay,Chen:2014vka} which has the same form as the saddle point equation for a Gaussian matrix model.  One then solves for $\rho(\s)$ in the standard way, where one finds the Wigner semi-circle distribution, 
\be
\rho(\s)=\frac{2}{\pi A^2}\sqrt{A^2-\s^2}\,,
\ee
with 
\begin{equation}\label{Aeq}
A^2=\frac{\lambda(1+\frac{1}{2}(\mu_1^2\ps\mu_2^2\ps\mu_3^2)+i\,\mu_1\mu_2\mu_3)}{8\pi^2}\,.
\ee
Because of the imaginary part in \cref{Aeq} the eigenvalue distribution runs at an angle off of the real axis.  One then substitutes $\rho(\s)$  back into the free energy, where the dominant part is given by
\be
\bes
\label{FE}
 F\approx& -\frac{N^2}{2}\int d\s d\s'\log(\s-\s')^2\\
\approx& -\frac{N^2}{2}\left(1+\frac{1}{2}(\mu_1^2\ps\mu_2^2\ps\mu_3^2)+i\,\mu_1\mu_2\mu_3\right)\\
&\qquad\qquad\times \log\left(\lambda\left(1+\frac{1}{2}(\mu_1^2\ps\mu_2^2\ps\mu_3^2)+i\,\mu_1\mu_2\mu_3\right)\right)\,,
\end{split}
\ee
Expanding about small $\mu_i$ and dropping terms up to cubic order which are not universal \cite{Gerchkovitz:2014gta,Bobev:2016nua}, \cref{FE} becomes
\be
\bes
\label{FE2}
F\approx&  -N^2\bigg(\frac{1}{16}(\mu_1^2\ps\mu_2^2\ps\mu_3^2)^2+\frac{i}{4}(\mu_1^2\ps\mu_2^2\ps\mu_3^2)\mu_1\mu_2\mu_3\\
&\qquad\qquad\qquad-\frac{1}{96}(\mu_1^2\ps\mu_2^2\ps\mu_3^2)^3-\frac{1}{4}(\mu_1\mu_2\mu_3)^2+{\rm O}(\mu^7)\bigg)\,,
\end{split}
\ee

In \cite{Bobev:2016nua} it was argued that  the terms in the free energy could only come with factors of $m^{(1)}m^{(2)}m^{(3)}$, $\tilde m^{(1)}\tilde m^{(2)}\tilde m^{(3)}$, or $\sum_j(m^{(j)}\tilde m^{(j)})^n$ where $n$ is a positive integer in order to  be consistent with supersymmetry.  If $m^{(j)}=\tilde m^{(j)}$ then this translates into terms of the form $\mu_1\mu_2\mu_3$ or $\mu_1^{2n}\ps\mu_2^{2n}\ps\mu_3^{2n}$.  Equation (\ref{FE2}) is consistent with this observation.
One should also note that the regularization should preserve the supersymmetry.  If equation (\ref{0cond}) had not been in effect, we would have had to add counterterms linear in $\mu_j$, which  violates this supersymmetry prescription. 

Assuming that a regularization can be performed, one expects the free energy for a general choice of $\mu_j$ to have the form \cite{Bobev:2016nua}
\begin{equation}\label{freeN1}
\begin{split}
F&=  -N^2\bigg(A_1(\mu_1^4\ps\mu_2^4\ps\mu_3^4)\ps A_2(\mu_1^2\ps\mu_2^2\ps\mu_3^2)^2+i\,B_1(\mu_1^2\ps\mu_2^2\ps\mu_3^2)\mu_1\mu_2\mu_3\\
&\qquad\qquad\qquad-C_1(\mu_1^6\ps\mu_2^6\ps\mu_3^6)\ms C_2(\mu_1^2\ps\mu_2^2\ps\mu_3^2)^3\ms C_3(\mu_1\mu_2\mu_3)^2+{\rm O}(\mu^7)\bigg)
\end{split}
\end{equation}
Comparing with \cref{FE2} and using \cref{0cond}, we find that 
\begin{equation}
\bes
A_1+2A_2=\frac{1}{8}\,,\quad\
B_1=-\frac{1}{4}\,,\quad
C_1+C_2=\frac{1}{24}\,,\quad
-12C_2+C_3=\frac{1}{8}\,.
\end{split}
\end{equation}
The first and third relations were derived in  \cite{Bobev:2016nua} using the $\NN=2^*$ results, where one has $\mu_1=0$.  The second relation differs from \cite{Bobev:2016nua} since their free energy is real.  The fourth relation is a new prediction.  

One feature that is different here compared to the holographic dual  is that the free energy in (\ref{freeN1}) has an imaginary piece, while the 
holographic result has a real free energy \cite{Bobev:2016nua}.  Since the theory is  Euclidean and nonconformal it is not reflection positive \cite{Festuccia:2011ws}, so it is not obvious on general grounds why the 
supergravity dual gives a real free energy.  This issue deserves further investigation.   

One further issue is that  a gaugino condensate appears in the 
holographic analysis if all three chiral multiplet masses are nonzero \cite{Bobev:2016nua}.  It is not clear how one sees the condensate in the analytic continuation.

\section{Summary and discussion}
\label{s-summary}

In this article we computed perturbative partition functions for theories with eight and four supersymmetries on spheres, with the dimension of the sphere being a continuous parameter. This proved the conjecture in \cite{Minahan:2015any} for eight supersymmetries and provided a new result for the case of four supersymmetries. 
We analytically continued our result for four supersymmetries to $d=4$ and  performed non-trivial consistency checks. We showed that in the limit of zero coupling the analytic continuation gives the correct partition function for the free conformal theories on $S^4$.  We also showed that the analytic continuation is consistent with the one-loop running of the coupling in four dimensions.
Then we used our results to study the free energy for mass-deformed theories with four supercharges on $S^4$ and compared these to the holographic  results for $\NN=1^*$ theory.

For eight supersymmetries our analysis can be straightforwardly generalized to  hypermultiplets in other representations of the gauge group.  For four supersymmetries one should also be able to extend the chiral multiplets to other representations, with possible restraints on the masses to be consistent with supersymmetry.  

At the same time it would be desirable to weaken any constraints on the masses so that one could obtain 
determinants with independent masses for three adjoint chiral multiplets. 
Despite the constraint, our work provides a way forward for localizing minimally supersymmetric theories on $S^4$.


Our work opens up various directions for  exploring the dynamics of $\NN=1$ theories on $S^4$.  One obvious possibility is to apply the analytic continuation to $\NN=1$ superconformal theories, analogously to the work in \cite{Fei:2014yja,Giombi:2014xxa,Fei:2014xta,Fei:2015kta,Fei:2015oha}.  We have already shown that it works for free theories.  These theories would also have the advantage of not having any ambiguities about real versus complex masses.

 A natural extension of our results is to include  instanton contributions. These contributions  have only been studied  for $d=4,5$ for supersymmetric theories on spheres.  It would be interesting to revisit those computations and investigate if they admit an analytic continuation in  dimensions.  If the continuation exists it must be nontrivial as instantons themselves do not exist below four dimensions.  The analytic continuation would have to flow to some other non-perturbative behavior.

From a more formal perspective, it would be  instructive to derive our results using index theorem techniques. In our computations with non-integer $d$ we witnessed large cancellations between bosonic and fermionic contributions. This hints that our results may be derived from some underlying index theorem for non-integer $d$. It would be interesting to explore this issue further.

Another avenue for future work is  to consider the analytical continuation for  $\NN=1$ theories in other dimensions where it is not known how to localize explicitly.  Theories with eight supersymmetries on $S^6$ and with 16 supersymmetries on $S^{8,9}$ can be constructed \cite{inprogress}. However just like the case of $\NN=1$ on $S^4$, it is not known how to localize these theories. We hope to explore these issues in future.
\section*{Acknowledgements}
J.A.M  thanks the CTP at MIT  for
hospitality during the course of this work.  He also thanks the Leinweber Center for Theoretical Physics at the University of Michigan for the invitation to present  this research at the workshop ``Supersymmetric Localization and Holography".  We thank H. Elvang, G. Festuccia, S. Pufu for useful discussions.  We also thank H. Elvang, S. Pufu and B. Zwiebach for comments on a draft of this paper.
The research of A.G. and J.A.M.  is supported in part by
Vetenskapsr{\aa}det under grants \#2012-3269 and \#2016-03503 and by the Knut and Alice Wallenberg Foundation under grant Dnr KAW 2015.0083.
Work of U.N is supported by the U.S. Department of Energy under grant Contract Number de-sc0012567.

\appendix

\appendix

\section{Conventions and useful properties}
\label{s-useful}
We use 10-dimensional Majorana-Weyl spinors $\epsilon_\alpha$ and $\Psi_\alpha$, etc. The 10-dimensional $\Gamma$-matrices are chosen to be real and symmetric:
\begin{equation}
	\Gamma^{M\alpha\beta} = \Gamma^{M\beta\alpha}, \qquad
	\tilde{\Gamma}^{M\alpha\beta} = \tilde{\Gamma}^{M\beta\alpha}.
\end{equation}
Products of $\Gamma$-matrices are given by:
\be\begin{split}
	\Gamma^{MN}   \equiv \tilde{\Gamma}^{[M}\Gamma^{N]},
\qquad	              \tilde{\Gamma}^{MN}                          & \equiv  \Gamma^{[M}\tilde{\Gamma}^{N]}                                                \\
	\Gamma^{MNP}  \equiv\Gamma^{[M}\tilde{\Gamma}^N\Gamma^{P]},
\qquad	              \tilde{\Gamma}^{MNP}                         & \equiv \tilde{\Gamma}^{[M}\Gamma^N\tilde{\Gamma}^{P]}, \qquad \text{etc.} 
\end{split}\ee
we also have that $\Gamma^{MNP\alpha\beta}= -\Gamma^{MNP\beta\alpha}$, hence:
\begin{equation}
	\epsilon\Gamma^{MNP}\epsilon = 0
\end{equation}
for any bosonic spinor $\epsilon$. We also introduce:
\begin{equation}
	\tilde{\epsilon}=\beta\Lambda\epsilon,
\end{equation}
where $\beta=\frac{1}{2r}$ and $\Lambda= \Gamma^{089}$.
A very useful relation is the triality condition,
\begin{equation}\label{triality}
	\Gamma^M_{\alpha\beta}\Gamma_{M\gamma\delta}
	+\Gamma^M_{\beta\delta}\Gamma_{M\gamma\alpha}
	+\Gamma^M_{\delta\alpha}\Gamma_{M\gamma\beta} = 0.
\end{equation}
Using \cref{triality} one can show
\begin{equation}
	\epsilon\Gamma^M\epsilon\,\epsilon\Gamma_M\chi = 0,
\end{equation}
where $\chi$ is any spinor. It immediately follows that $v^M v_M = 0$, where $v^M$ is the vector field
\begin{equation}
	v^M \equiv \epsilon \Gamma^M \epsilon.
\end{equation}

We define  another  set of  bosonic spinors, $\nu_m$ for $m=1,2,\cdots, 7$. They satisfy the following properties.
\begin{equation} \label{purespinors}
	\begin{split}
		\nu_m\Gamma^M\epsilon &= 0, \\
		\nu_m\Gamma^M\nu_n &= \delta_{mn}v^M, \\
		\nu^m_\alpha\nu^n_\beta + \epsilon_\alpha\epsilon_\beta &= \frac{1}{2} v^M \tilde{\Gamma}_{M\alpha\beta}.
	\end{split}
\end{equation}
They are invariant under an internal $SO(7)$ symmetry, which can be enlarged to $SO(8)$ by including $\epsilon$.

To reduce to eight supersymmetries  we impose the condition $\epsilon= +\Gamma^{6789}\epsilon$. Furthermore, for $d\leq5$, $\Psi$ can be split up into even and odd eigenstates of $\Gamma^{6789}$. The even eigenstates, $\psi = \frac{1}{2}\bkt{1+\Gamma^{6789}}\Psi$, make up the fermions in the vector multiplet, while the odd eigenstates, $\chi =\frac{1}{2}\bkt{1 -\Gamma^{6789}}\Psi$, make up the fermions in the hypermultiplet. The scalars $\phi^I$, $I=6,\ldots9$ constitute the bosonic fields of the hyper multiplet. The gauge fields $A^\mu$ and the rest of the scalars $\phi_I$, $I=0,d+1,\ldots5$ make up the bosonic fields in the vector multiplet. Finally, the auxiliary fields split up, with $K^m$, $m=1,2,3$, being in the vector multiplet, and $K_m$, $m=4,5,6,7$, being in the hypermultiplet. The same is true for the pure-spinors $\nu^m$. Reduction to four supersymmetries can be done similarly by imposing $\eps = +\Gamma^{4589} \eps$.

\section{Quadratic fluctuations about the fixed point locus}
\label{ap:fluct}
In this appendix we give details of the computation of  quadratic fluctuations about the fixed point locus. We focus on bosonic and fermionic parts separately.
\subsection{Bosonic part}
The bosonic part of fluctuations about the fixed point locus is equal to~\cite{Minahan:2015any}:
\be\label{eq:bosMain}
\begin{split}
	\mathcal{L}^{\text{b}}& =\delta_\epsilon\Psi \overline{\delta_\epsilon\Psi}\phantom{a}                                     \\
	                    & =\frac{1}{2}F_{MN}F^{MN}-\frac{1}{4}F_{MN}F_{M'N'}\left( \epsilon \Gamma^{MNM'N'0}\epsilon \right) 
	                     +\frac{\beta d\alpha_I}{4}F_{MN}\phi_I
	\left(\epsilon\Lambda\left(\tilde{\Gamma}^I \tilde{\Gamma}^{MN}\Gamma^0-
		\tilde{\Gamma}^0\Gamma^I\Gamma^{MN}
		\right)\epsilon
	\right)                                                                                                                  \\
	                    &-K^m K_m v^0 -\beta d\alpha_0\phi_0 K^m(\nu_m\Lambda\epsilon)
	+\frac{\beta^2 d^2}{4}\sum_I (\alpha_I)^2 \phi_I\phi^Iv^0. \end{split}
\ee
Expanding the first term in~\cref{eq:bosMain} we get
\be
\begin{split}	
\frac{1}{2}F_{MN}F^{MN} & =
	\frac{1}{2}\,F_{\mu\nu}F^{\mu\nu}+F_{\mu 0}F^{\mu 0}
	-[\phi_0^{\text{cl}},\phi_J][\phi^{\text{cl}}_0,\phi^J]
	+\nabla_\mu\phi_J\nabla^\mu\phi^J                                       \\
	                        & = \nabla_\mu A_\nu\nabla^\mu A^\nu
	-\nabla_\mu A_\nu \nabla^\nu A^\mu
	+\nabla_\mu\phi_{0}\nabla^\mu\phi^{\,0}
	+2\nabla_\mu\phi_{0} [A^\mu,\phi^{0}_{cl}]                               \\ 
	                        & \qquad-[A_\mu,\phi^{\text{cl}}_0][A^\mu,\phi^{\text{cl}}_0]
	-[\phi_0^{\text{cl}},\phi_J][\phi^{\text{cl}}_0,\phi^J]+ \nabla_\mu\phi_J\nabla^\mu\phi^J, 
\end{split}
\ee
where $J=d+1,\ldots,9$. The second term in~\cref{eq:bosMain} can be expanded to get:
\be
\begin{split}
	-\frac{1}{4}F_{MN}F_{M'N'} \left(\epsilon \Gamma^{MNM'N'0}\epsilon\right) =
	 & -\nabla_\mu A_\nu \nabla_{\mu'} A_{\nu'}
	\left(\epsilon \Gamma^{\mu \nu \mu' \nu'0}\epsilon \right)
	-2\nabla_\mu A_\nu \nabla_{\mu'} \phi_{J}
	\left(\epsilon \Gamma^{\mu \nu \mu' J 0}\epsilon\right) \\
	 & \quad-\nabla_\mu \phi_J \nabla_{\mu'} \phi_{J'}
	\left(\epsilon \Gamma^{\mu J \mu' J' 0}\epsilon\right). 
	\end{split}
\ee
The third term in~\cref{eq:bosMain} is
\be
\begin{split}
	\frac{\beta d \alpha_I}{4}\, F_{MN} \phi_I \left( \epsilon \Lambda
	\left(\tilde{\Gamma}^I\tilde{\Gamma}^{MN} \Gamma^0 - \tilde{\Gamma}^0\Gamma^I\Gamma^{MN} \right) \epsilon\right)
	 & =
	\frac{\beta d \alpha_J}{2}\,
	\left(\nabla_\mu A_\nu -\nabla_\nu A_\mu \right) \phi_J
	\left(\epsilon\Lambda \tilde{\Gamma}^J\Gamma^0\tilde{\Gamma}^\mu\Gamma^\nu \epsilon\right) \\
	 & \,\,\,+\beta d \alpha_{J'}\, \nabla_\mu \phi_J \phi_{J'} \left(
	\epsilon\Lambda \tilde{\Gamma}^{J'}\Gamma^0\tilde{\Gamma}^\mu\Gamma^J \epsilon
	\right). 
	\end{split}
	\ee
	Collecting our results, we find that the bosonic part is:
\be
\begin{split}
\mathcal{L}^{\text{b}} & =
	\nabla_\mu A_\nu\nabla^\mu A^\nu
	-\nabla_\mu A_\nu \nabla^\nu A^\mu
	-\nabla_\mu\phi_{\,0}\nabla^\mu\phi_{\,0}
	-2\nabla_\mu\phi_{\,0} [A^\mu,\phi_{0}^{\text{cl}}]
	-[A_\mu,\phi^{\text{cl}}_0][A^\mu,\phi^{\text{cl}}_0]                               \\
	                    & -[\phi^{\text{cl}}_0,\phi_J][\phi^{\text{cl}}_0,\phi^J]
	+\nabla_\mu\phi_J\nabla^\mu\phi^J                                     \\
	                    & -\nabla_\mu A_\nu \nabla_{\mu'} A_{\nu'}
	\left( \epsilon \Gamma^{\mu\nu\mu'\nu'0}\epsilon \right)
	-2\nabla_\mu A_\nu \nabla_{\mu'} \phi_J
	\left( \epsilon \Gamma^{\mu\nu\mu'J0}\epsilon \right)
	-\nabla_\mu \phi_J \nabla_{\mu'} \phi_{J'}
	\left( \epsilon \Gamma^{\mu J\mu' J'0}\epsilon \right)                \\
	                    & +\frac{\beta d \alpha_J}{2}\,
	\left(\nabla_\mu A_\nu -\nabla_\nu A_\mu \right) \phi_J
	\left(\epsilon\Lambda \tilde{\Gamma}^J\Gamma^0\tilde{\Gamma}^\mu\Gamma^\nu \epsilon\right)
	+\beta d \alpha_{J'}\, \nabla_\mu \phi_J \phi_{J'} \left(
	\epsilon\Lambda \tilde{\Gamma}^{J'}\Gamma^0\tilde{\Gamma}^\mu\Gamma^J \epsilon
	\right)
	\\
	                    & -v^0 K^m K_m
	-\beta d\alpha_0 \phi_{0} K^{\,m} (\nu_m\Lambda\epsilon)
	+\frac{\beta^2 d^2}{4}v^0\sum_I (\alpha_I)^2\phi_I \phi^{\,I}. 
\end{split}
\ee
Next, we rewrite this expression as a quadratic form:
\be
\begin{split}
	\mathcal{L}^{\text{b}}&= A^\mu \left(-\delta_\mu^\nu\nabla^2
	+\nabla^\nu\nabla_\mu
	-\left(\epsilon\Gamma_\mu{}^{\mu'\nu'\nu0}\epsilon\right)\nabla_{\mu'}\nabla_{\nu'}
	-2\beta(d-3)\left(\epsilon\Lambda
		\Gamma_\mu{}^{\mu'\nu0}\epsilon\right)\nabla_{\mu'}
	\right) A_\nu
	-[A_\mu,\phi^{\text{cl}}_0][A^\mu,\phi^{\text{cl}}_0]                            \\
	 & +\phi^J\left(
	-\nabla^2 \delta_J^{J'}
	-2\beta(d-1)
	\left( \epsilon \Lambda \Gamma_J{}^{\mu J'0} \epsilon \right)\nabla_\mu
	-\beta d \alpha_{J'}\,  \left(
	\epsilon\Lambda \tilde{\Gamma}^{J'}\Gamma^0\tilde{\Gamma}^\mu\Gamma_J \epsilon
	\right)	\nabla_\mu
	+\frac{\beta^2 d^2}{4}\, (\alpha_J)^2\delta_J^{J'}
	\right)\phi_{J'}                                                   \\
	 & -[\phi^{\text{cl}}_0,\phi_J][\phi^{\text{cl}}_0,\phi^J]+\phi_{0}\left(
	\nabla^2 -\frac{\beta^2 d^2}{4}\, \alpha_0^2 \right)\phi_{0}
	-4 \beta (d-2) A_\nu \nabla_\mu \phi_{J}
	\left( \epsilon \Lambda \Gamma^{\nu\mu}\Gamma^{J0}\epsilon \right) \\
	 & +\beta d \alpha_J\, \phi_J \nabla_\mu A_\nu
	\left(\epsilon\Lambda \tilde{\Gamma}^J\Gamma^0 \Gamma^{\mu\nu} \epsilon\right)
	-K^m K_m -\beta d\alpha_0 \phi_{0} K^m (\nu_m\Lambda\epsilon),
\end{split}
\ee
where we have used the Lorenz gauge condition and the relation $\tilde{\Gamma}_\mu\Gamma^{\mu\nu\mu'}=(d-2) \Gamma^{\nu\mu'}$. Now, note that the third term in the first row vanishes, and that for the second term, we can exchange the order of the covariant derivatives to get a term which is zero due to the Lorenz gauge condition and another one which contains a Ricci tensor, which on spheres is proportional to a Kronecker delta. Furthermore, we can combine the two terms which are proportional to $\nabla_\mu\phi_{J'}$ into one, and finally get:
\be\begin{split}
	\mathcal{L}^{\text{b}} &= A^\mu \left(-\delta_\mu^\nu\nabla^2
	+4\beta^2(d-1)\delta^\nu_\mu
	-2\beta(d-3)\left(\epsilon\Lambda
	\Gamma_\mu{}^{\mu'\nu0}\epsilon\right)\nabla_{\mu'}
	\right) A_\nu
	-[A_\mu,\phi^{\text{cl}}_0][A^\mu,\phi^{\text{cl}}_0]
	\\
	 & +\phi^J\left(
	-\nabla^2 \delta_J^{J'}
	+\beta\left(-2(d-1)+ d \alpha_{J'} \right)
	\left( \epsilon \Lambda \Gamma^{J'} \Gamma_J \Gamma^{\mu 0} \epsilon \right)\nabla_\mu
	+\frac{\beta^2 d^2}{4}\, (\alpha_J)^2\delta_J^{J'}
	\right)\phi_{J'}                                                     \\
	 & -[\phi^{\text{cl}}_0,\phi_J][\phi^{\text{cl}}_0,\phi^J]
	+\phi_{0}\left( \nabla^2 -\frac{\beta^2 d^2}{4}\, \alpha_0^2
	\right)\phi_0
	+\beta\left(-4(d-2) +d\alpha_J\right)A_\nu \nabla_\mu\phi_J
	\left(\epsilon\Lambda \Gamma^{\nu\mu} \Gamma^{J0}  \epsilon\right)   \\
	 & -K^m K_m -\beta d\alpha_0 \phi_{0} K^m (\nu_m\Lambda\epsilon).
	 \label{general.bosons}
\end{split}\ee
This general result includes both the vector multiplet and the hypermultiplet bosons. Let us now specialize to the vector multiplet.

\subsubsection{Vector multiplet}
The vector multiplet contains the vector field $A_\mu$ and the scalar fields $\phi_0,\phi_i$, where the index $i$ takes values $i=d+1,\cdots D$ and  $D=5$ for eight supersymmetries and $D=3$ for four supersymmetries.  We use
\begin{equation}
	\alpha_0 = \frac{4(d-3)}{d},\ \ \ \
	\alpha_i = \frac{4}{d},\ \ \ \ \text{for} \ \ \ i=d+1,\ldots , D.
\end{equation}
We also combine $\mu$ and $i $ indices into $\tilde{M}\ =\{\mu, i\}$ to write the bosonic part of the vector-multiplet Lagrangian from equation~\eqref{general.bosons} in the following compact form:
\begin{equation}\label{eq:ALvmBos}
\begin{split}
	\mathcal{L}_{\text{v.m}}^{\text{b}}\ & =\
	A^{\tilde{M}}
	\ \mathcal{O}_{\tilde{M}}{}^{\tilde{N}} \
	A_{\tilde{N}}
	-[A_{\tilde{M}},\phi^{\text{cl}}_0][A^{\tilde{M}},\phi^{\text{cl}}_0]
	\\&
	-K^m K_m -4\beta (d-3) \phi_{0} K^m (\nu_m\Lambda\epsilon)
	+\phi_{0}\left(
	\nabla^2
	-4\beta^2 (d-3)^2
	\right)\phi_0.
\end{split}
\ee
The operator $\mathcal{O}_{\tilde{M}}{}^{\tilde{N}} $ is defined as follows:
\begin{equation}\label{eq:AOpO}
	\mathcal{O}_{\tilde{M}}{}^{\tilde{N}} =
	-\delta_{\tilde{M}}^{\tilde{N}} \nabla^2
	+\alpha_{\tilde{M}}{}^{\tilde{N}}
	-2\beta(d-3)\epsilon\Gamma_{\tilde{M}}{}^{\nu\tilde{N}89}\epsilon\nabla_\nu.
\end{equation} and $\alpha_{\tilde{M}}{}^{\tilde{N}} $ is the diagonal matrix given by:
\begin{equation}\label{eq:AMatAlph}
\alpha_{\tilde{M}}{}^{\tilde{N}} \ =\  4\beta^2
\begin{pmatrix}
	\bkt{d-1} \delta_\mu^\nu & 0          \\
	0                        & \delta_i^j
\end{pmatrix}.
\ee
\subsubsection{Hyper/chiral-multiplet}
The scalars $\phi^I$, $I=D+1,\ldots,9$ are part of the hypermultiplet. For eight supersymmetries we get a single hypermultiplet and for four supersymmetries we get three hypermultiplets by reduction of $10$-d theory. Let us first focus on four supersymmetries:
\begin{equation}
\begin{split}
	\mathcal{L}_{\text{c.m}}^{\text{b}}\ =\ & \phi^J\left(
	-\nabla^2 \delta_J^{J'}
	+\beta\left(-2(d-1)+ d \alpha_{J'} \right)
	\left( \epsilon \Lambda \Gamma^{J'} \Gamma_J \Gamma^{\mu 0} \epsilon \right)\nabla_\mu
	+\frac{\beta^2 d^2}{4}\, \alpha_J^2\delta_J^{J'}
	\right)\phi_{J'} \\
	&
	-[\phi^{\text{cl}}_0,\phi_J][\phi^{\text{cl}}_0,\phi^J]. \label{Ahyper.bosons.1}
	\end{split}
\end{equation}
For four supersymmetries, the values of $\alpha_I$  are given in~\eqref{al9}. The Lagrangian of  equation~\eqref{Ahyper.bosons.1} splits up in three decoupled parts which take the form:
\begin{equation}
\begin{split}
	\label{eq:Lab}
	\mathcal{L}_{\text{c.m}}^{\text{b}}  = \sum_{\ell =1}^{3} & \phi^{I_\ell} \left(
	-\nabla^2 \delta_{I_\ell}^{L_\ell}
	-2\beta(1-2i\sigma_{(\ell)}\mu_\ell)
	\left(\epsilon \Lambda \Gamma^{J_\ell} \Gamma_{I_\ell} \Gamma^{\mu 0} \epsilon \right)\nabla_\mu
	+\beta^2 (d-2+2i\sigma_{(\ell)}\mu_\ell)^2\delta_{I_\ell}{}^{J_\ell}
	\right)\phi_{J_\ell} \\
	&
	-[\phi^{\text{cl}}_0,\phi_{I_\ell}][\phi^{\text{cl}}_0,\phi^{I_\ell}].
	\end{split}
\end{equation}
This can be simplified by noting that 
\be
\qquad \eps\Lambda \Gamma^{7 6 \mu 0} \eps \ = v^\mu, \qquad \eps\Lambda \Gamma^{9 8 \mu 0} \eps\ =\ -v^\mu, \qquad
\eps\Lambda \Gamma^{5 4 \mu 0} \eps= v^\mu.
\ee
This gives the following form of the chiral multiplet Lagrangian.
\be
\begin{split}
	\mathcal{L}_{\text{c.m}}^{\text{b}}
	\  =
	 \ \sum_{\ell=1}^{3}  & \sbkt{\phi_{I_\ell}\left(
	-\nabla^2
	+\beta^2 (d-2+2i \sigma_{(\ell)}\mu_\ell)^2
	\right)\phi^{I_\ell}
	-[\phi^{\text{cl}}_{0},\phi_{I_\ell}][\phi^{\text{cl}}_{0}\phi^{I_\ell}]} \\
	& +4 \beta \bkt{2i\mu_{\ell}-\sigma_{(\ell)}} \phi_{2\ell+2}  v^\mu \nabla_\mu \phi_{2\ell+3}.
\end{split}
\ee

For the case of eight supersymmetries the Lagrangian for hypermultiplet bosons can be obtained from the above expressions by ignoring $\phi_4, \phi_5$ and setting $\mu_2=\mu_3$:
\be 
\begin{split}
\mathcal{L}_{\text{h.m}}^{\text{b}}=& \ \sum_{i=6}^{9} \sbkt{\phi_i\left(
	-\nabla^2
	+\beta^2 (d-2+2i \sigma_i \mu)^2
	\right)\phi_i
	-[\phi^{\text{cl}}_{0},\phi_i][\phi^{\text{cl}}_{0},\phi_i]} \\ &
	+ 4 \beta \bkt{2i\mu-1}\phi_6  v^\mu \nabla_\mu \phi_7 
	+4 \beta \bkt{2i\mu+1}\phi_8  v^\mu \nabla_\mu \phi_9.
\end{split}
\ee
\subsection{Fermionic part}
The fermionic part of the fluctuations around the fixed point locus is given by:
\begin{equation}
	\label{eq:Lfermfull1}
	\mathcal{L}^{\text{f}} =
	\Psi\delta_\epsilon\left(\overline{\delta_\epsilon\Psi}\right) =
	\Psi\Gamma^0\delta_\epsilon^2\Psi
	-\Psi\Gamma^0\left[
		2\Gamma^{M'0} \delta_\epsilon F_{M'0} \epsilon
		+\alpha_0\Gamma^{\mu0}\delta_\epsilon\phi_0\nabla_\mu\epsilon
		+2\delta_\epsilon K^m\nu_m
		\right].
\end{equation}
Let us focus on the first term, involving two variations of the fermion.  
\be\begin{split}
	\delta_\epsilon^2\Psi & =
	\left(\epsilon\Gamma^N D^M\Psi\right)\Gamma_{MN}\epsilon
	-\beta\left(\epsilon\Lambda\tilde{\Gamma}^\mu\Gamma^N\Psi\right)\Gamma_{\mu N}\epsilon
	-\frac{\alpha_I \beta d}{2} \left(\epsilon\Gamma_I\Psi\right)\tilde{\Gamma}^I \Lambda\epsilon \\
	                      & \ \ + \left(\epsilon\slashed{D}\Psi\right)\epsilon
	-\frac{1}{2}v^M\tilde{\Gamma}_M\slashed{D}\Psi + \Delta K^m \nu_m.	 \label{hyper.fermions.1}
\end{split}\ee
This expression can be brought into the desired form by using triality and other identities. Using triality the first term in~\cref{hyper.fermions.1}
\be\begin{split}
	 & 
	-\left(\epsilon\slashed{D}\Psi\right)\epsilon
	-\left(\epsilon\Gamma^N\epsilon\right)D_N\Psi
	+\frac{1}{2}\left(\epsilon\Gamma^N\epsilon\right)\tilde{\Gamma}_N\slashed{D}\Psi.  
\end{split}
\ee
Using triality, the second term in \cref{hyper.fermions.1} becomes
\be\begin{split}
	&\beta d\left(\epsilon\Lambda\Psi\right)\epsilon
	+\frac{1}{2}\beta\left(\epsilon\Lambda\Gamma_{MN}\epsilon\right)\Gamma^{MN}\Psi
	-\frac{1}{2}\beta\left(\epsilon\Lambda\Gamma_{IJ}\epsilon\right)\Gamma^{IJ}\Psi           \\ &       
	 +\frac{1}{2}\beta\left(\epsilon\Lambda\Gamma_{\mu\nu}\epsilon\right)\Gamma^{\mu\nu}\Psi
	-2\beta\left(\epsilon\Gamma^\mu\Psi\right)\tilde{\Gamma}_\mu\Lambda\epsilon
	+d\beta\left(\epsilon\Gamma^N\Psi\right)\tilde{\Gamma}_N\Lambda\epsilon.
\end{split}\ee
The second term in the above expression can be simplified using the following Fierz identity~\cite{Pestun:2007rz, Kennedy_clifford_1981}:
\begin{equation}
	\label{eq:Fierz}
	-\frac{1}{2}\left(\tilde{\epsilon}\Gamma_{MN}\epsilon\right)\Gamma^{MN}\Psi
	-4\left(\Psi\tilde{\epsilon}\right)\epsilon
	+2\left(\epsilon\Gamma_N\Psi\right)\tilde{\Gamma}^N\tilde{\epsilon} = 0,
\end{equation}
Combining all these pieces we get
\be\begin{split}
	\delta_\epsilon^2\Psi & =
	-v^N D_N\Psi
	-\frac{1}{4}\nabla_{[\mu}v_{\nu]} \Gamma^{\mu\nu}\Psi
	-\frac{1}{2}\beta\left(\epsilon\Lambda\Gamma_{IJ}\epsilon\right)\Gamma^{IJ}\Psi
	+\beta \bkt{2	-\frac{\alpha_I d}{2}} \left(\epsilon\Gamma_I\Psi\right)\tilde{\Gamma}^I \Lambda\epsilon                                                       \\
	                      & \qquad+\beta (d-4)\bkt{\left(\epsilon\Lambda\Psi\right)\epsilon + \left(\epsilon\Gamma_N\Psi\right)\tilde{\Gamma}^N\Lambda\epsilon}
	+\Delta K^m \nu_m.   \label{hyper.fermions.3}
\end{split}\ee
Let's focus on the second term in~\eqref{eq:Lfermfull1} and simplify all three terms appearing there. The first one is
\begin{equation}
	-2 \left( \Psi\Gamma^0\Gamma^{M'0}\epsilon \right)\delta_\epsilon F_{M'0}
	=
	2 \left(\Psi\Gamma^0\epsilon \right)\left(\epsilon\slashed{\nabla}\Psi\right)
	-2 \beta d\left(\Psi\Gamma^0\epsilon\right) \left(\epsilon\Lambda\Psi\right)
	+2 \left( \Psi\Gamma^{M'}\epsilon \right)\left(\epsilon\Gamma_{M'}D_0\Psi \right).
\end{equation}
The second term is:
\begin{equation}
	-\alpha_0 \Psi\Gamma^0 \Gamma^{\mu0}\delta_\epsilon\phi_0\nabla_\mu\epsilon
	=\alpha_0 d \beta\left(\Psi\Gamma^0\epsilon\right)\left(\epsilon\Lambda\Psi\right). 
\end{equation}
The third term is:
\be\begin{split}
	-2\left(\Psi\Gamma^0\nu_m\right)\delta_\epsilon K^m & =
	-2\left(\Psi\Gamma^0\epsilon\right)\left(\epsilon\slashed{\nabla}\Psi\right)
	+2\left(\Psi\Gamma^0\epsilon\right)\left(\epsilon\Gamma_0 D_0\Psi\right)
	+v^M\left(\Psi\Gamma^0\tilde{\Gamma}_M\slashed{D}\Psi\right)                                        \\
	                                                    & \ \  -2 \bkt{\Psi \Gamma^0 \nu_m} \Delta K^m.
\end{split}\ee
Collecting all the terms, we get:
\be\begin{split}
	v^M\left(\Psi\Gamma^0\tilde{\Gamma}_M\slashed{D}\Psi\right)
	+ d \beta\bkt{\alpha_0-2} \left(\Psi\Gamma^0\epsilon\right)\left(\epsilon\Lambda\Psi\right)
	+2 \left( \Psi\Gamma^M\epsilon \right)\left(\epsilon\Gamma_M D_0\Psi \right)  -2 \bkt{\Psi \Gamma^0 \nu_m} \Delta K^m.	 
\end{split}\ee
The first term in the above expression can be rewritten using the identity $\tilde{\Gamma}^M\Gamma^N = g^{MN} + \Gamma^{MN}$, and the third one  can be manipulated using the triality identity. The result is
\be\begin{split}
	\bkt{\Psi \slashed{\nabla} \Psi} + v^\mu \left(\Psi\Gamma^0\Gamma_{\mu \nu} \nabla^\nu\Psi\right)+  \sum_{I=d+1}^{9} v^I\left(\Psi\Gamma^0\Gamma_{I \nu} \nabla^\nu\Psi\right)
	+v^\mu\left(\Psi\Gamma^0 \nabla_\mu\Psi\right) \\
	+ 2 \bkt{\Psi \Gamma^0 D_0 \Psi}
	+ d \beta\bkt{\alpha_0-2} \left(\Psi\Gamma^0\epsilon\right)\left(\epsilon\Lambda\Psi\right)-2 \bkt{\Psi \Gamma^0 \nu_m} \Delta K^m.  
\end{split}\ee
Using integration by parts, the second and third terms can be modified to give
\be\begin{split}
	\bkt{\Psi \slashed{\nabla} \Psi} -\beta\bkt{\eps \tilde{\Gamma}^{\mu\nu} \Lambda \eps}  \bkt{\Psi \Gamma^0 \Gamma_{\mu\nu} \Psi}\ -\beta\bkt{\eps \tilde{\Gamma}^{J \nu} \Lambda \eps}  \bkt{\Psi \Gamma^0 \Gamma_{J \nu} \Psi}\  	+v^\mu\left(\Psi\Gamma^0 \nabla_\mu\Psi\right) \\
	+ 2 \bkt{\Psi \Gamma^0 D_0 \Psi}
	+ d \beta\bkt{\alpha_0-2} \left(\Psi\Gamma^0\epsilon\right)\left(\epsilon\Lambda\Psi\right)-2 \bkt{\Psi \Gamma^0 \nu_m} \Delta K^m.  
\end{split}\ee
Now, combining this with the result for $\Psi\Gamma^0\delta_\epsilon^2\Psi$, we get the complete expression for the fermionic part
\be\begin{split}
	\mathcal{L}^{\text{f}} = &
	\bkt{\Psi \slashed{\nabla} \Psi}	+ \bkt{\Psi \Gamma^0  D_0\Psi}+ \beta\bkt{ 3 d - 16} \left(\Psi\Gamma^0\epsilon\right)\left(\epsilon\Lambda\Psi\right) -\frac{1}{2}\beta\left(\epsilon\tilde{\Gamma}^{MN}\Lambda\epsilon\right)\left( \Psi \Gamma^0 \Gamma_{MN}\Psi \right) \\ &
	+\beta \bkt{2-\frac{\alpha_I d}{2}}  \bkt{\Psi \Gamma^0 \tilde{\Gamma}^I \Lambda\epsilon}\left(\epsilon\Gamma_I\Psi\right)
	+\beta (d-4)\bkt{\Psi \Gamma^0 \tilde{\Gamma}^N\Lambda\epsilon} \left(\epsilon\Gamma_N\Psi\right)
	\\
	                         &
	- \bkt{\Psi \Gamma^0 \nu_m} \Delta K^m.   
\end{split}\ee
The terms on the second line can be modified by using the following identity:
\be
\tilde{\Gamma}^N\Lambda\epsilon \left(\epsilon\Gamma_N\Psi\right) \ - 2  \tilde{\Gamma}^A\Lambda\epsilon \left(\epsilon\Gamma_A\Psi\right)\ =\ \frac{1}{2} v^N \tilde{\Lambda } \Gamma_N \Psi .
\ee
So the quadratic part becomes
\be\begin{split}
	\mathcal{L}^{\text{f}} = &
	\bkt{\Psi \slashed{\nabla} \Psi}	+ \bkt{\Psi \Gamma^0  D_0\Psi}+ \beta\bkt{ 3 d - 16} \left(\Psi\Gamma^0\epsilon\right)\left(\epsilon\Lambda\Psi\right) -\frac{1}{2}\beta\left(\epsilon\tilde{\Gamma}^{MN}\Lambda\epsilon\right)\left( \Psi \Gamma^0 \Gamma_{MN}\Psi \right) \\ &
	+ \beta\mathcal{C}_I   \bkt{\Psi \Gamma^0 \tilde{\Gamma}^I \Lambda\epsilon}\left(\epsilon\Gamma_I \Psi\right)\ 	+\beta \frac{ d-4}{2} v^N \bkt{\Psi \Gamma^0 \tilde{\Lambda} \Gamma_N \Psi}
	- \bkt{\Psi \Gamma^0 \nu_m} \Delta K^m,   \label{eq:LfermFulFinal}
\end{split}\ee
where the coefficient $\mathcal{C}_I$ which appear in the first term in second line is given by:
\be
\mathcal{C}_A\ =\ 2d-6-\frac{\alpha_A d}{2}\ \ \ ,\ \qquad \mathcal{C}_i\ =\ 2-\frac{\alpha_i d}{2}.
\ee
Let us now specialize to vector and hypermultiplets separately.
\subsubsection{Vector multiplet}
The vector multiplet fermions have same eigenvalues under the projection operators $\Gamma, \Gamma'$ as the Killing spinor.  We denote the vector multiplet fermion by $\psi$.
For a fermion in the vector multiplet, the first term on the second line of~\cref{eq:LfermFulFinal} does not contribute.  It is easy to verify that for this term, either $\mathcal{C}_I$ vanishes or $\bkt{\epsilon \Gamma_I \psi}\ =\ 0$. Furthermore, for the last term in~\cref{eq:LfermFulFinal}, we take pure spinors $\nu^m, m=1,2,\cdots D-2$ to have the same eigenvalues under projection operators as the Killing spinor and the vector multiplet fermion, while the rest of the pure spinors have the same eigenvalues as the hypermultiplet fermions. We use
\be
\begin{split}
	\Delta K^m\ &=\ \beta \bkt{d-4} \nu^m  \Lambda \psi,\ \ \qquad \text{for}\qquad m=1,2,\cdots D-2,
\end{split}
\ee
to write:
\be
\begin{split}
	- \bkt{\psi \Gamma^0 \nu_m} \Delta K^m\ &=\ -\beta\bkt{d-4}\sum_{m=1}^{m=D-2 }\bkt{\psi \Gamma^0 \nu_m}\bkt{\nu^m \Lambda \psi }, \\
	& =-\beta\bkt{d-4}\sum_{m=1}^7\ \bkt{\psi \Gamma^0 \nu_m}\bkt{\nu^m \Lambda \psi }\ , \\
	&\ =\ \beta\bkt{d-4} \bkt{\psi \Gamma^0 \epsilon} \bkt{\epsilon \Lambda \psi} -\frac{1}{2} \beta\bkt{d-4} v^N \bkt{\psi \Gamma^0 \tilde{\Gamma}_N \Lambda \psi},
\end{split}
\ee
where in the second equality we have used the fact that for rest of the pure spinors  $\bkt{\psi \Gamma^0\nu^m}\ =\ 0$. The last equality follows by using the completeness relation of the pure spinors and the Killing spinor. Next we use the fact that:
\be
v^N \bkt{\psi \Gamma^0 \tilde{ \Lambda}\Gamma_N \psi}-v^N \bkt{\psi \Gamma^0 \tilde{\Gamma}_N \Lambda \psi}\ =\ -2 v^{\tilde{N}} \bkt{\psi \Gamma^0 \tilde{\Gamma}_{\tilde{N}} \Lambda \psi}.
\ee
Using all this information in equation ({\ref{eq:LfermFulFinal}}), we get the quadratic Lagrangian for vector multiplet fermions to be:

\be\begin{split}
	\mathcal{L}^{\text{f}}_{\text{v.m}} = &
	\bkt{\psi \slashed{\nabla} \psi}	+ \bkt{\psi \Gamma^0  D_0\psi}+ 4\beta\bkt{ d-5} \left(\psi\Gamma^0\epsilon\right)\left(\epsilon\Lambda\psi\right) -\frac{1}{2}\beta\left(\epsilon\tilde{\Gamma}^{MN}\Lambda\epsilon\right)\left( \psi \Gamma^0 \Gamma_{MN}\psi \right) \\ &
	- \beta \bkt{d-4}  v^{\tilde{N}} \bkt{\psi \Gamma^0 \tilde{\Gamma}_{\tilde{N}} \Lambda \psi}	.   \label{eq:Lfermvm1}
\end{split}\ee
We use a few relations to simplify the Lagrangian further. First  we can use the Fierz identity quoted in~\cref{eq:Fierz} and triality to bring the third term above in the desired form:
\begin{equation}
	4\beta(d-5)\left(\psi\Gamma^0\epsilon\right)\left(\epsilon\Lambda\psi\right)  =
	-\frac{1}{4}\beta(d-5)\left(\psi\Gamma^0\Gamma^{MN}\psi\right) \left(\epsilon\Lambda\Gamma_{MN}\epsilon\right)
	-\frac{1}{2}\beta(d-5) v_N\left(\psi\Gamma^0\tilde{\Lambda}\Gamma^N\psi\right).
\end{equation}
Secondly, we rewrite the last term of this equation as
\begin{equation}
	\begin{split}\label{eq:vnsplitid}
		v_N \left(\psi\Gamma^0\tilde{\Lambda}\Gamma^N\psi\right) & =
		\left(\psi\Lambda\psi\right)
		-v^{\tilde{M}}\left(\psi\Gamma^0\tilde{\Gamma}_{\tilde{M}}\Lambda\psi\right).
	\end{split}
\end{equation}
Further, we note that for the vector multiplet fermions, we have:
\begin{equation}
	\begin{split}
		-\frac{1}{2} (d-4)\beta \left(\epsilon \tilde{\Gamma}^{AB} \Lambda \epsilon\right) \left( \psi \Gamma^0\Gamma_{AB} \psi\right) &=
		\beta(d-4) \left( \psi \Lambda \psi\right) \\
		\left(\epsilon\tilde{\Gamma}^{MN}\Lambda\epsilon\right)\left( \psi \Gamma^0 \Gamma_{MN}\psi \right) & =
		\left(\epsilon\tilde{\Gamma}^{\tilde{M}\tilde{N}}\Lambda\epsilon\right)\left( \psi \Gamma^0 \Gamma_{\tilde{M}\tilde{N}}\psi \right)
		- (9-D) \left(\psi \Lambda \psi\right).
	\end{split}
\end{equation}
Combining these results in the general Lagrangian \cref{eq:Lfermvm1} we get finally get the following expression for Lagrangian of vector multiplet fermions:
\be
\label{eq:aLvmFermFinal}
\begin{split}
	\mathcal{L}_{\text{v.m}}^{\text{f}} =&
	\left(\psi\slashed{\nabla}\psi\right)
	-\frac{1}{2}(d-3)\beta v^{\tilde{M}}\left(\psi\Gamma^0\tilde{\Gamma}_{\tilde{M}}\Lambda\psi\right)
	+v^0 \left( \psi \Gamma^0 D_0 \psi \right) \\
	&-\frac{1}{4}(d-3)\beta\left(\epsilon\tilde{\Gamma}^{\tilde{M}\tilde{N}}\Lambda\epsilon\right)\left( \psi \Gamma^0 \Gamma_{\tilde{M}\tilde{N}}\psi \right)
	+ m_{\psi} \left(\psi\Lambda\psi\right).
\end{split}
\ee
Here $m_{\psi}\ =\ \frac{d-1}{2}$ for eight supersymmetries and $m_\psi\ =\  \ \bkt{d-2} $ for four supersymmetries.
\subsubsection{Hyper/chiral-multiplet}

Let us treat eight and four supersymmetries separately. For eight supersymmetries, we have a single fermion in the hypermultiplet. Let us denote it as $\chi\ =\ - \Gamma \chi$.  For the hypermultiplet fermion $\bkt{\epsilon \Lambda \chi}\ =\ 0$. We have $\mathcal{C}_6\ =\ \mathcal{C}_7\ =\ -\mathcal{C}_8\ = -\mathcal{C}_9\ = -\bkt{d-4+2 i \mu}$. Also using the fact that $\epsilon \Gamma_M \chi\ =\ 0$ for $M=0, \tilde{M}$, we see that the first term on second line of \cref{eq:LfermFulFinal} can be written as
\be
- \beta\mathcal{C}_6   \bkt{\chi\Gamma^0 \tilde {\Lambda} \Gamma^N\epsilon}\left(\epsilon\Gamma_N \chi \right)\ =\ \beta \frac{ \mathcal{C}_6}{2} v^N \bkt{\chi \Gamma^0 \tilde{\Lambda}\Gamma^N \chi}.
\ee
Using this, we get the following expression for the hypermultiplet fermion's Lagrangian
\be\begin{split}
	\mathcal{L}^{\text{f}} = &
	\bkt{\chi \slashed{\nabla} \chi}	+ \bkt{\chi \Gamma^0  D_0\chi} -\frac{1}{2}\beta\left(\epsilon\tilde{\Gamma}^{MN}\Lambda\epsilon\right)\left( \chi \Gamma^0 \Gamma_{MN}\chi \right) \\ &
	- i \mu \beta  v^N \bkt{\chi \Gamma^0 \tilde{\Lambda} \Gamma_N \chi}
	- \bkt{\chi \Gamma^0 \nu_m} \Delta K^m.   \label{eq:LfermHyp1}
\end{split}\ee
It is easy to verify that the contribution of third term in \cref{eq:LfermHyp1} is
\be
\left(\epsilon\tilde{\Gamma}^{MN}\Lambda\epsilon\right)\left( \chi \Gamma^0 \Gamma_{MN}\chi \right)\ = \left(\epsilon\tilde{\Gamma}^{\tilde{M}\tilde{N}}\Lambda\epsilon\right)\left( \chi \Gamma^0 \Gamma_{\tilde{M}\tilde{N}}\chi \right).
\ee
The last term in~\cref{eq:LfermHyp1} gets contributions from
\be
\Delta K^m\ =\ - 2 i\mu  \nu^m  \Lambda \chi, \ \ \ \text{for }\ \ \ m\ =\ 4,5,6,7.
\ee
Using the completeness property for pure spinors and the fact that $\epsilon \Lambda \chi\ =\ 0$, we get:
\be
-\bkt{\chi \Gamma^0 \nu ^m} \Delta K^m\ =\ i \mu \beta v^N \bkt{\chi \Gamma^0 \tilde{\Gamma}_N \Lambda \chi}.
\ee
This and the second to last term in \cref{eq:LfermHyp1} can be combined using the identity \cref{eq:vnsplitid}. After all the simplifications, we obtain the following form for the Lagrangian of the hypermultiplet fermion with eight supersymmetries:
\be\begin{split}
	\mathcal{L}^{\text{f}}_{\text{h.m}} = &
	\bkt{\chi \slashed{\nabla} \chi}	+ \bkt{\chi \Gamma^0  D_0\chi} -\frac{1}{2}\beta\left(\epsilon\tilde{\Gamma}^{\tilde{M}\tilde{N}}\Lambda\epsilon\right)\left( \chi \Gamma^0 \Gamma_{\tilde{M}\tilde{N}}\chi \right)
	+2  i \mu \beta v^{\tilde{N}} \bkt{\chi \Gamma^0 \tilde{\Gamma}_{\tilde{N}} \Lambda \chi}
	. 
\end{split}\ee
The chiral multiplet fermionic part with four can be obtained by similar computation:
\be
\begin{split}
\mathcal{L}^{\text{f}}_{\text{h.m}}\ & =\sum_{\ell=1}^3 \bkt{\chi_{\ell} \slashed{\nabla} \chi_{\ell}}	
+ \bkt{\chi_{\ell} \Gamma^0  [\phi^{\text{cl}}_{0},\chi_{\ell}]} 
-\frac{1}{2}\beta\left(\epsilon\tilde{\Gamma}^{\tilde{M}\tilde{N}}\Lambda\epsilon\right)\left( \chi_{\ell} \Gamma^0 \Gamma_{\tilde{M}\tilde{N}}\chi_{\ell} \right)
\\
	&
	+\sigma_{(\ell)} \beta \bkt{2  i \mu_{\ell} v^{\tilde{N}} \bkt{\chi_i \Gamma^0 \tilde{\Gamma}_{\tilde{N}} \Lambda \chi_\ell}+ \chi_i\Lambda \chi_\ell} .
	\end{split}
\ee
\section{Degeneracy of harmonics on $S^{d}$}
\label{app-counting}
The spectrum of the Laplacian and the degeneracy of symmetric traceless tensors on $S^{d}$ is given in \cite{Rubin:1984tc}. We summarize the results for scalar and divergence-less vectors here for completeness.  Scalar harmonics are labelled by the eigenvalues of the Laplacian on the sphere, with the eigenvalues and the degeneracy given by:
\begin{equation}
\label{eq:scalarDeg}
\nabla^{2} Y_{m}^{k}\ =\ -4\beta^2 k\bkt{k+d-1} Y_{m}^{k}, \qquad \mathcal{D}_{k}\bkt{d,0}\ =\ 
\frac{\bkt{2k+d-1}{\Gamma\bkt{k+d-1}}}{\Gamma\bkt{d} \Gamma\bkt{k+1}}.
\ee
Divergence-less vector harmonics are also labelled by eigenvalues of the Laplacian on the sphere which are different than scalars. Their degeneracy is given by:
\begin{equation}
\label{eq:vecDeg}
\begin{split}
\nabla^{2} A_{\mu}^{k}\  &=\ -4\beta^{2}\bkt{k\bkt{k+d-1}-1} A_{\mu}^{k}, 
 \qquad  \nabla\cdot A^{k}\ =\ 0, \\
 \mathcal{D}_{k}\bkt{d,1}\ & =\ \frac{k\bkt{k+d-1}\bkt{2k+d-1}\Gamma\bkt{k+d-2}}{\Gamma\bkt{d-1}\Gamma\bkt{k+2}}.
 \end{split}
\ee
 Spinor harmonics on $S^{d}$ are labelled by the eigenvalues of the Dirac operator. We summarize results of \cite{Camporesi:1995fb} here:
 \begin{equation}
 \slashed{\nabla} \psi^{k}_{\pm}\ =\ \pm i\bkt{k+\frac{d}{2}}\psi^{k}_{\pm}, \qquad \mathcal{D}_{k}\bkt{d, +}=\mathcal{D}_{k}\bkt{d, -}= \frac{2^{\lfloor \frac{d}{2}\rfloor } \Gamma\bkt{k+d}}{\Gamma\bkt{d}\Gamma\bkt{k+1}}.
 \ee
 
 An important degeneracy factor that appears in the computation of the one-loop determinant is the number of spherical harmonics $Y_{\pm k}^{k}$. Since the spin is labelled by the Cartan generator along the direction of the vector field $v^{\tilde{M}}$, the degeneracy is different for the case of eight and four supersymmetries. Let us derive this degeneracy for the case of eight supersymmetries now.
 
Consider an $S^{d}$ parameterized as follows:
\begin{equation}
        |z_i|^2 + x_j^2 = 1,  \label{sphere}
\end{equation}
where $z_i\in\mathbb{C}$, $x_j\in\mathbb{R}$ and the indices $i,j$ range in $i=1,\ldots , \frac{d-k'+1}{2}$ and $j=1,\ldots k'$. Let us consider the vector field $v^{\tilde{M}}$, which acts on the sphere coordinates as
\begin{equation}
\begin{split}
        z_i &\to z_i e^{i\phi} , \qquad
        x_j \to x_j.
\end{split}
\end{equation}
The fixed point locus of this vector field is given by the equation $z_i = 0$, $i= 1, \ldots k$, which, when substituted in equation~\eqref{sphere}, leaves a $(k'-1)$-sphere fixed. For example, in the case of eight supersymmetries we have
\begin{itemize}
        \item $S^5$: $|z_1|^2 + |z_2|^2 + |z_3|^2 = 1$ has a fixed $S^{-1}$,
        \item $S^4$: $|z_1|^2 + |z_2|^2 + x_1^2 = 1$ has a fixed $S^0$ (two points on the poles),
        \item $S^3$: $|z_1|^2 + x_1^2 + x_2^2 = 1$ has a fixed $S^1$,
        \item $S^2$: $x_1^2 + x_2^2 + x_3^2 = 1$ has a fixed $S^2$.
\end{itemize}
So in the case of eight supersymmetries the action of the vector field  leaves an $S^{4-d}$ fixed. 
In this parametrization, the scalar spherical harmonics $Y^k_m$ can be written as polynomials in the variables $z_i,\bar{z}_i$ and $x_j$. To construct a spherical harmonic of level $k$ and ``charge'' $m$, we assign charge $+1$ to $z_i$, $-1$ to $\bar{z}_i$ and $0$ to $x_j$. Thus, the top spherical harmonics can be written as:
\begin{equation}
        Y^k_k \sim z_{i_1} z_{i_2} \ldots z_{i_k},
\end{equation}
with the degeneracy given by:
\begin{equation}
\label{eq:Nkk}
        N_{k, d} = {{k+\frac{d-k'+1}{2}-1}\choose{k} }\ =\ \frac{\Gamma(k+d-2)}{\Gamma(k+1)\Gamma(d-2)}.
\end{equation}

In the case of four supersymmetries, $v^{\tilde{M}}$  leaves an $S^{{2-d}}$ fixed, so the degeneracy of the top level harmonics is:
\begin{equation}
\label{eq:nkk}
       n_{k,d} = \frac{\Gamma(k+d-1)}{\Gamma(k+1)\Gamma(d-1)}.
\end{equation}
\section{Vanishing of top spinor modes}
\label{app:vanishing}
Certain elements of the basis for spinor harmonics vanish identically for $m=\pm k$. Here we will demonstrate explicitly that for $m=k$
 \begin{equation}
        \mathcal{X}^2_+
\ = \Gamma^{\tilde{M}} \hat{\nabla}_{\tilde{M}} Y^k_{+ k}\eta_+
        = \Gamma^\mu\nabla_\mu Y^k_{+ k} \eta_+ + 2i\beta k Y^k_{+ k}\eta_+
        = 0. \label{spherharm1}
\end{equation}
We will take the top and the bottom modes of the scalar spherical harmonics  to be given by:
\begin{equation}
        Y^k_k = z^k \qquad \text{and} \qquad
        Y^k_{-k} = \bar{z}^k,
\end{equation}
where
\begin{equation}
    z = 2\beta \, \frac{x^1 + ix^2}{1+x^2\beta^2}.
\end{equation} 
We will also use the relation between the gamma matrices with the flat and curved indices, given by:
\begin{equation}
	\Gamma^\mu = (1+x^2 \beta^2) \Gamma^{\hat{\mu}}.
\end{equation}
 Now, the first term in equation~\eqref{spherharm1} becomes:
\begin{equation}
        \Gamma^\mu\nabla_\mu Y^k_k \eta_+ =
        \left[ 2\beta k(\Gamma^1 + i\Gamma^2)z^{k-1}
        -2\beta^2 k x\cdot\Gamma z^k \right]\eta_+, \label{spherharm2}
\end{equation}
whose first term can be expanded to:
\begin{equation}
        (\Gamma^1 + i\Gamma^2)\eta_+ =
        \Gamma_0\left(\tilde{\Gamma}_6 + i\tilde{\Gamma}_7\right)
        \left[(\Gamma_1+i\Gamma_2) + \beta(\Gamma_1+i\Gamma_2)x\cdot\tilde{\Gamma}\Lambda\right]\frac{\epsilon_s}{\sqrt{1+\beta^2x^2}}.
\end{equation}
Note however that:
\be\begin{split}
        (\tilde{\Gamma}_6 + i\tilde{\Gamma}_7)
        (\Gamma_1 + i\Gamma_2)\epsilon_s
        &=
        \left[ \tilde{\Gamma}_6\Gamma_1 - \tilde{\Gamma}_7\Gamma_2
        +i(\tilde{\Gamma}_6\Gamma_2 + \tilde{\Gamma}_7\Gamma_1) \right] \epsilon_s \\
        &=
         \left( \Gamma_{61} - \Gamma_{72}\Gamma^{1267}
        +i(\Gamma_{62} + \Gamma_{71}\Gamma^{1267})\right]\epsilon_s \\
        &= 0,  
\end{split}\ee
where we have used the fact that $\Gamma^{1267}\epsilon_s =+\epsilon_s$. This result implies that:
\begin{equation}
        (\tilde{\Gamma}_6 + i\tilde{\Gamma}_7)
        (\Gamma_1 + i\Gamma_2)\tilde{\Gamma}_M\Lambda\epsilon_s =0,
        \quad
        \text{for } M \neq 1,2.
\end{equation}
Thus:
\be\begin{split}
        (\Gamma^1 + i\Gamma^2) \eta_+ &=
        \Gamma_0\left(\tilde{\Gamma}_6 + i\tilde{\Gamma}_7\right)\beta
        \left[ (x^1 +ix^2)\Lambda\epsilon_s + i(x^1+ix^2)\Gamma^0\epsilon_s\right]
        \frac{1}{\sqrt{1+\beta^2 x^2}} \\
        &=
        -2i\beta(x^1 + ix^2)(\Gamma_6 + i\Gamma_7) \frac{\epsilon_s}{\sqrt{1+\beta^2 x^2}}.  
\end{split}\ee
Let's proceed to second term of equation~\eqref{spherharm2}:
\be\begin{split}
        x\cdot\Gamma\eta_+ &= x\cdot\Gamma \, \tilde{\Gamma}_0(\Gamma_6 +i \Gamma_7)
        (1+\beta x\cdot\tilde{\Gamma}\Lambda) \frac{\epsilon_s}{\sqrt{1+\beta^2 x^2}} \\
        &=
        -\frac{1}{\beta}\Gamma_0 (\tilde{\Gamma}_6 + i\tilde{\Gamma}_7) \Lambda
        (\beta x\cdot\tilde{\Gamma}\Lambda - \beta^2 x^2) 
        \frac{\epsilon_s}{\sqrt{1+\beta^2 x^2}} \\
        &=
        +\frac{1}{\beta}\,i (\Gamma_6 + i\Gamma_7)(\beta x\cdot\tilde{\Gamma}\Lambda
        -\beta^2 x^2)\frac{\epsilon_s}{\sqrt{1+\beta^2 x^2}}.  
\end{split}\ee
Combining our results for the two terms of~\eqref{spherharm2}, we get:
\be\begin{split}
        k\left[-2i\beta(1+x^2\beta^2)(\Gamma_6 + i\Gamma_7)
        -2i\beta(\Gamma_6 +i\Gamma_7)(\beta x\cdot\tilde{\Gamma}\Lambda - \beta^2 x^2)
        \right]
        \frac{\epsilon_s}{\sqrt{1+\beta^2 x^2}}
        = -2i\beta k (\Gamma_6 + i\Gamma_7) \epsilon  
\end{split}\ee
which finally implies that
\begin{equation}
        \Gamma^{\tilde{M}} \hat{\nabla}_{\tilde{M}} Y^k_k \eta_+ = 0.
\end{equation}
\bibliographystyle{JHEP}
\bibliography{refs}

\end{document}